\documentclass{amsart}
\usepackage{amsfonts}
\usepackage{amsmath}
\usepackage{amsxtra}
\usepackage{amssymb}
\usepackage{graphicx}

\theoremstyle{plain}
\newtheorem{theorem}{Theorem}[section]
\newtheorem{lemma}[theorem]{Lemma}

\newtheorem{remark}{Remark}

\input{tcilatex}

\begin{document}
\title[Stability Criterion for Vlasov-Maxwell ]{A sharp stability criterion
for the Vlasov-Maxwell system}
\author{Zhiwu Lin}
\address{Mathematics Department\\
University of Missouri\\
Columbia, MO 65211}
\email{lin@math.missouri.edu}
\author{Walter A. Strauss}
\address{Department of Mathematics and Lefschetz Center for Dynamical
Systems, Brown University, Providence, RI 02912 }
\email{wstrauss@math.brown.edu}

\begin{abstract}
We consider the linear stability problem for a 3D cylindrically symmetric
equilibrium of the relativistic Vlasov-Maxwell system that describes a
collisionless plasma. For an equilibrium whose distribution function
decreases monotonically with the particle energy, we obtained a linear
stability criterion in our previous paper \cite{lw-linear}. Here we prove
that this criterion is sharp; that is, there would otherwise be an
exponentially growing solution to the linearized system. 
%Therefore for the class of symmetric Vlasov-Maxwell equilibria, we
%establish an energy principle for linear stability.
We also treat the considerably simpler periodic $1\frac{1}{2}$D case. The
new formulation introduced here is applicable as well to the nonrelativistic
case, to other symmetries, and to general equilibria.
\end{abstract}

\maketitle

\section{Introduction}

We consider a plasma at such high temperature or low density that collisions
can be ignored compared with the electromagnetic forces. Such a
collisionless plasma is modeled by the relativistic Vlasov-Maxwell (RVM)
system. We assume all physical constants like the speed of light $c$ and the
mass of particles $m$ to be $1$, for the sole purpose of simplifying our
notation. All the results we obtain below can be modified straightforwardly
to apply to the true physical situations with general masses, charges, etc.
In the physical literature, the nonrelativistic version of the
Vlasov-Maxwell system is more commonly considered but our results easily
extend to that case. Our notation is as follows. Let $f^{\pm }\left(
t,x,v\right) $ be the ion and electron distribution functions, $\mathbf{E}%
(t,x)$ and $\mathbf{B}(t,x)$ be the electric and magnetic fields and $%
\mathbf{E}^{ext},\mathbf{B}^{ext}$ be the external fields. Then the RVM
system is 
\begin{subequations}
\label{RVM-3D}
\begin{equation}
\partial _{t}f^{\pm }+\hat{v}\cdot \nabla _{x}f^{\pm }\pm \left( \mathbf{E}+%
\mathbf{E}^{ext}+\hat{v}\times \left( \mathbf{B}+\mathbf{B}^{ext}\right)
\right) \cdot \nabla _{v}f^{\pm }=0,
\end{equation}%
\begin{equation}
\partial _{t}\mathbf{E}=\nabla \times \mathbf{B}-\vec{j},\quad \nabla \cdot 
\mathbf{E}=\rho ,\quad \rho =\int \left( f^{+}-f^{-}\right) dv,
\end{equation}%
\begin{equation}
\partial _{t}\mathbf{B}=-\nabla \times \mathbf{E},\quad \nabla \cdot \mathbf{%
B}=0,\quad \vec{j}=\int \hat{v}\left( f^{+}-f^{-}\right) dv,
\end{equation}%
where $\hat{v}=v/\left\langle v\right\rangle $and $\left\langle
v\right\rangle =\sqrt{1+\left\vert v\right\vert ^{2}}$. Alternatively, in
many physical problems (\cite{davidson}, \cite{davidson-qin}), a nonneutral
plasma is also \ considered, where there is only a single species of
particle.

One of the central problems in the theory of plasmas is to understand plasma
stability and instability (\cite{nicholson-plasma}, \cite{trievelpiece}).
For example, to control the plasma instability in a fusion device is a key
issue for the nuclear fusion program. Many other examples occur in
astrophysical contexts. So far, most studies on plasma stability are based
on macroscopic MHD models. For such fluid models, the famous \textit{energy
principle} was discovered by Bernstein, Frieman, Kruskal and Kulsrud (\cite%
{energy58}) in the 1950s, first for static equilibria and later for
symmetry-preserving perturbations of symmetric equilibria (\cite{newcomb}).
These energy principles allow one to reduce the study of linear stability to
checking the positivity of a certain relatively simple quadratic energy form 
$W\left( \xi ,\xi \right) $. They have been widely used in the plasma
physics community (\cite{friedberg-mhd}, \cite{goedbload}) to study many
types of important plasma instabilities. However, the collision-dominant
assumption required in deriving these MHD models from kinetic models is
seriously violated in many almost collisionless situations in nuclear fusion
(\cite{friedberg-mhd}) and space plasmas (\cite{parks}). This puts into
question the applicability of such energy principles in physical situations
where collisions are infrequent. While energy principles have been derived
for some simple approximate models, such as collisionless MHD and guiding
center models (\cite{kruskal} \cite{kulsrud-energy} \cite{grad}), there have
been very few such studies on the more accurate but more complicated
microscopic Vlasov-Maxwell models. A good understanding of stability for
Vlasov systems could provide a theoretical basis to compare and check the
validity of stability results for various approximate plasma models like
MHD. Moreover, many plasma instability phenomena have an essentially
microscopic nature, for which kinetic models like Vlasov-Maxwell are
required (\cite{parks}).

Combining the results of this paper with \cite{lw-linear}, we have
established an energy principle for a large class of symmetric equilibria of
various Vlasov-Maxwell systems. More precisely, for a large class of
equilibria that enjoy certain kinds of symmetry, the study of linear
stability of symmetry-preserving perturbations is reduced to simply checking
the positivity of a self-adjoint operator $\mathcal{L}^{0}$, or equivalently
the positivity of the quadratic form $\left( \mathcal{L}^{0}\xi ,\xi \right) 
$. Compared with the usual MHD energy principle, our energy principle has
several new features and advantages. In the MHD case, the quadratic energy
form $W\left( \xi ,\xi \right) $ can be written as $\left( F\xi ,\xi \right) 
$ where the force operator $F$ has a complicated spectral structure such as
gaps in its essential spectrum (\cite{goedbload}). It is difficult to
analyze, especially in higher dimensions and in nontrivial magnetic field
geometries. Our operator $\mathcal{L}^{0}$ for RVM is essentially an
elliptic operator plus a bounded nonlocal term and thus has a relatively
simple spectral structure. This structure allows us to obtain important
additional information about the linear instability. For example, we show
that the maximal growth rate is controlled by the lowest negative eigenvalue
of $\mathcal{L}^{0}$ and that the number of growing modes equals the number
of negative eigenvalues of $\mathcal{L}^{0}$.

Linear stability under the condition $\mathcal{L}^{0}\geq 0$ was proven in (%
\cite{lw-linear}). The main result of the present paper is to prove the
converse; that is, to construct a growing mode if $\mathcal{L}^{0}\ngeq 0$.
As in \cite{lw-linear} we specifically consider two RVM models, the simpler $%
1\frac{1}{2}$D periodic case with $x\in \mathbb{R},v\in \mathbb{R}^{2}$, and
the full 3D case in the whole space $\mathbb{R}^3$ with cylindrical
symmetry. However, our methods are also applicable to Vlasov-Maxwell models
with other symmetries, with boundary conditions, or in a nonrelativistic
setting, and will yield similar results.

%%%%%%%%%%%%%%%%%  3D Case  %%%%%%%%%%%%%%%%%%%%%%%%%%
Now we state our main result for the cylindrically symmetric 3D case. As
remarked in (\cite{lw-linear}), the existence of a plasma equilibrium of the
3D RVM model in the whole space requires an external field. To simplify
notation we consider a 3D nonneutral \textit{electron} plasma with an
external field. This scenario does indeed does occur in many physical
situations (\cite{davidson}). So $f^+=0$, and instead of $f^{-}$ we use the
notation $f$ for the electrons. Our equilibrium is cylindrically symmetric
with electron distribution $f^{0}=\mu \left( e,p\right) ,$ where

\end{subequations}
\begin{equation*}
e=\sqrt{1+\left\vert v\right\vert ^{2}}-\phi ^{0}\left( r,z\right) -\phi
^{ext}\left( r,z\right) ,
\end{equation*}%
\begin{equation*}
p=r\left( v_{\theta }-A_{\theta }^{0}\left( r,z\right) -A_{\theta
}^{ext}\left( r,z\right) \right)
\end{equation*}%
and with equilibrium fields 
\begin{equation*}
\mathbf{E}^{0}=-\partial _{r}\phi ^{0}\mathbf{e}_{r}-\partial _{z}\phi ^{0}%
\vec{e}_{z},\text{ \ }\mathbf{B}^{0}=-\partial _{z}A_{\theta }^{0}\mathbf{e}%
_{r}+\frac{1}{r}\partial _{r}\left( rA_{\theta }^{0}\right) \mathbf{e}_{z}.
\end{equation*}
In order to be an equilibrium, $\left( A_{\theta }^{0},\phi ^{0}\right) $
must satisfy the elliptic system 
\begin{equation}
\Delta \phi ^{0}=\partial _{zz}\phi ^{0}+\partial _{rr}\phi ^{0}+\tfrac{1}{r}%
\partial _{r}\phi ^{0}=\int \mu dv  \label{eqn-equi-3d-ele}
\end{equation}%
\begin{equation}
\left( \Delta -\tfrac{1}{r^{2}}\right) A_{\theta }^{0}=\partial
_{zz}A_{\theta }^{0}+\partial _{rr}A_{\theta }^{0}+\tfrac{1}{r}\partial
_{r}A_{\theta }^{0}-\tfrac{1}{r^{2}}A_{\theta }^{0}=\int \hat{v}_{\theta
}\mu dv.  \label{eqn-equi-3d-mag}
\end{equation}%
Here we use cylindrical coordinates $\left( r,\theta ,z\right) $ and denote
by $\left( \mathbf{e}_{r},\mathbf{e}_{\theta },\mathbf{e}_{z}\right) $ the
standard basis. We also assume axisymmetry of the external fields in the
form 
\begin{align*}
\mathbf{E}^{ext}& =-\partial _{r}\phi ^{ext}\left( r,z\right) \mathbf{e}%
_{r}-\partial _{z}\phi ^{ext}\mathbf{e}_{z}, \\
\mathbf{B}^{ext}& =-\partial _{z}A_{\theta }^{ext}\left( r,z\right) \mathbf{e%
}_{r}+\tfrac{1}{r}\partial _{r}\left( rA_{\theta }^{ext}\right) \mathbf{e}%
_{z}.
\end{align*}%
We assume that the equilibrium is confined, namely that $f^{0}$ has compact
support $S$ in phase space. Having compact support is a realistic assumption
for a confined plasma. We make the further assumption that $f^{0}$ and $%
\mathbf{E}^{0},\mathbf{B}^{0}$ are continuous everywhere, including on the
boundary of the support. In \cite{lw-linear}, with properly chosen external
fields, an example of a continuous nonneutral plasma equilibrium with
support in a torus was constructed. We also assume that $\partial \mu
/\partial e=\mu _{e}<0$ inside $S$. This condition is widely believed to
make the equilibrium more likely to be stable (\cite{davidson}, \cite{grad}, 
\cite{taylor 74}). We study the stability of such an equilibrium under
perturbations that preserve cylindrical symmetry.

In order to state our main results, we define certain linear operators
acting on cylindrically symmetric scalar functions $h\in L^{2}(\mathbb{R}%
^{3})$ by 
\begin{equation}
\mathcal{A}_{1}^{0}h=-\partial_{zz}h-\partial_{rr}h-\frac{1}{r}\partial
_{r}h-\int\mu_{e}dvh+\int\mu_{e}\mathcal{P}\left( h\right) dv,
\label{defn-a1-3d}
\end{equation}%
\begin{equation}
\mathcal{A}_{2}^{0}h=-\partial_{zz}h-\partial_{rr}h-\frac{1}{r}\partial
_{r}h+\frac{1}{r^{2}}h-\int\hat{v}_{\theta}\mu_{p}dv\text{ }rh-\int\hat {v}%
_{\theta}\mu_{e}\mathcal{P}\left( \hat{v}_{\theta}h\right) dv,
\label{defn-a2-3d}
\end{equation}%
\begin{equation}
\mathcal{B}^{0}h=\int\mu_{e}\mathcal{P}\left( \hat{v}_{\theta}h\right)
dv-\int\hat{v}_{\theta}\mu_{e}dv\text{ }h,  \label{defn-b0-3d}
\end{equation}
and 
\begin{equation}
\mathcal{L}^{0}=\left( \mathcal{B}^{0}\right) ^{\ast}\left( \mathcal{A}%
_{1}^{0}\right) ^{-1}\mathcal{B}^{0}+\mathcal{A}_{2}^{0}.  \label{defn-l0-3d}
\end{equation}
where $\mathcal{P}$ is the projection operator of $L_{\left\vert \mu
_{e}\right\vert }^{2}$ onto $\ker D$. Here $D$ denotes the transport
operator associated with the steady fields, namely 
\begin{equation*}
D=\hat{v}\cdot\nabla_{x}+\left( \mathbf{E}^{0}+\mathbf{E}^{ext}+\hat{v}%
\times\left( \mathbf{B}^{0}+\mathbf{B}^{ext}\right) \right) \cdot\nabla_{v}
\end{equation*}
and $L_{\left\vert \mu_{e}\right\vert }^{2}$ denotes the $\left\vert \mu
_{e}\right\vert $-weighted $L_{x,v}^{2}$ space. It was proven in \cite%
{lw-linear} that these operators are well-defined and that $\mathcal{L}^{0}$
is self-adjoint. First we recall our previous result in \cite{lw-linear}.

%%%%%%%%%%%%%%%  Theorem 1.1  %%%%%%%%%%%%%%%%%%%%%%%%

\begin{theorem}[\protect\cite{lw-linear}]
\label{3d theorem} Consider a nonnegative axisymmetric equilibrium $\left(
f^{0} ,\mathbf{E}^{0},\mathbf{B}^{0}\right) $ as above with compact support $%
S$ in phase space. Assume $\mu_{e}<0$ inside $S$. For axisymmetric
perturbations, we have following results.

(i) $\mathcal{L}^{0}\geq0$ implies spectral stability. That is, if $\mathcal{%
L}^{0}\geq0$ then there does not exist a growing mode.

(ii) Any growing mode must be purely growing. That is, if 
\begin{equation*}
\left[ e^{\lambda t}f(x,v),e^{\lambda t}\mathbf{E}(x),e^{\lambda t}\mathbf{B}%
(x)\right] \quad \left( \text{Re}\lambda >0\right)
\end{equation*}
with $\mathbf{E}, \mathbf{B }\in L^2,\ f\in L^1\cap L^\infty$ is a solution
of the linearized system, then $\lambda $ is real.

(iii) If $\mathcal{L}^{0}\ngeq 0$ and $-\alpha ^{2}$ denotes the lowest
negative eigenvalue of the operator $\mathcal{L}^{0}$, then the maximal
growth rate $\lambda $ cannot exceed $\alpha $.
\end{theorem}

Theorem \ref{3d theorem} asserts the linear stability if $\mathcal{L}%
^{0}\geq 0$ and it also estimates the maximal growth rate if $\mathcal{L}%
^{0}\ngeq 0$. However, it leaves open the converse, namely the question of
the existence of a growing mode when $\mathcal{L}^{0}\ngeq 0$. In this
paper, we fill this gap by showing that there indeed always exists a growing
mode if $\mathcal{L}^{0}\ngeq 0$. This is the main result of the present
paper.

%%%%%%%%%%%%%%%  Theorem 1.2  %%%%%%%%%%%%%%%%%%%%%%%%

\begin{theorem}
\label{3d growing} Under the same assumptions as in Theorem \ref{3d theorem},

(i) if $\mathcal{L}^{0}\ngeq 0$, there exists a growing mode; that is, an
exponentially growing weak solution 
\begin{equation*}
\lbrack e^{\lambda t}f(x,v),e^{\lambda t}\mathbf{E}(x),e^{\lambda t}\mathbf{B%
}(x)]\quad \left( \lambda >0\right)
\end{equation*}%
of the linearized problem with $f\in L^{1}\cap L^{\infty }$ and $\mathbf{E},%
\mathbf{B}\in H^{1}$.

(ii) The dimension of the space of symmetry-preserving growing modes equals
the dimension of the negative eigenspace of $\mathcal{L}^{0}$.
\end{theorem}

The combination of Theorems \ref{3d theorem} and \ref{3d growing},
establishes an \textquotedblleft energy principle\textquotedblright\ for
this class of equilibria, in terms of the relatively simple operator $%
\mathcal{L}^{0}$. Thus this operator $\mathcal{L}^{0}$ not only provides the
sharp stability criterion, but also contains information about the number of
unstable modes and their maximal growth rate. The projection $\mathcal{P}$
that occurs in the definition of $\mathcal{L}^{0}$ is a highly nonlocal
operator since $\mathcal{P}h\left( x,v\right) $ turns out to be essentially
the average of $h$ in the phase space occupied by the particle trajectory
with the steady field $\left( \mathbf{E}^{0}+\mathbf{E}^{ext},\mathbf{B}^{0}+%
\mathbf{B}^{ext}\right) $ starting at $\left( x,v\right) $. So our sharp
stability criterion $\mathcal{L}^{0}\geq 0$ is also highly \textit{non-local}%
, which reflects the collective nature of plasma stability. Because of the
condition $\mu _{e}<0$, it turns out that all the nonlocal terms are
stabilizing.

In \cite{guo2}, Y. Guo investigated the stability of a two-species plasma
satisfying 3D RVM without external fields, in a bounded domain with the
perfectly conducting boundary condition. In a similar setting to ours, a
sufficient condition for stability was obtained in \cite{guo2} by the
energy-Casimir method. Extending the calculations in \cite{guo2} to the
whole space case, we would obtain the stability condition that $L^{0}>0$,
where $L^{0}$ is the differential operator 
\begin{equation}
L^{0}=-\partial _{zz}-\partial _{rr}-\tfrac{1}{r}\partial _{r}+\tfrac{1}{%
r^{2}}-r\int \hat{v}_{\theta }\mu _{p}dv,  \label{defn-L0-3d}
\end{equation}
the last two terms being multiplication operators. However, since $\mathcal{L%
}^{0}>L^{0}$, the stability criterion $\mathcal{L}^{0}\geq 0 $ in our
Theorem \ref{3d theorem} is a significant improvement because of the
additional stabilizing effects that come from the non-local terms in $%
\mathcal{L}^{0}$. More importantly, in the $1\frac{1}{2}$D case discussed
below, we showed in \cite{lw-linear} that these nonlocal stabilizing terms
are indispensable to prove the stability of any equilibrium, even a
homogeneous one. We believe that the non-local stabilizing terms must play
an important role in plasma stability in the $3$D case as well.

The simplest case that permits a magnetic field is the so-called $1\frac{1}{2%
}$ dimensional case. In this case, physical space is one-dimensional $x\in {%
\mathbb{R}}$ and momentum space is two-dimensional $v=(v_{1},v_{2})\in {%
\mathbb{R}}^{2}$. Moreover, $\mathbf{E}=(E_{1},E_{2},0)$ and $\mathbf{B}%
=(0,0,B)$. We consider solutions that are periodic in $x$ and we may assume
that there is no external field. In Section 2, before going on to the proofs
of Theorems \ref{3d theorem} and \ref{3d growing} in 3D, we prove precise
analogues of these theorems for this much simpler case. Various particular
stable and unstable $1\frac{1}{2}$D examples were constructed in \cite%
{lw-linear}. In \cite{lw-nonlinear} we also proved the validity of these
linear stability and instability results on the \textit{nonlinear} dynamical
level.

We now sketch the main ideas in the proofs of Theorems \ref{3d growing} and
its $1\frac{1}{2}$ dimensional analogue, which are concerned with the
construction of growing modes provided that $\mathcal{L}^{0}\ngeq 0$. We
begin with some brief historical comments on linear instability for Vlasov
systems. One of the main difficulties in studying Vlasov instability is its
collective and thus highly nonlocal nature. In the physics literature, most
classical studies (\cite{nicholson-plasma}, \cite{trievelpiece}) treat
homogeneous equilibria with vanishing electric and magnetic fields, in which
case an explicit algebraic dispersion relation is usually available.
However, any nontrivial electromagnetic field will make the dispersion
relations much more difficult to analyze because they depend upon some
complicated trajectory integrals. In \cite{gs2} and later publications \cite%
{gs5}, \cite{gs3}, \cite{gs4}, Guo and Strauss developed a perturbation
approach to prove the instability of weakly inhomogeneous equilibria of
Vlasov-Poisson systems. They proved the instability of various electrostatic
structures that are close to an unstable homogeneous equilibrium. In \cite%
{lin01} Lin developed a new non-perturbative approach to find purely growing
modes for highly inhomogeneous equilibria of 1D Vlasov-Poisson. This
approach has recently been used (\cite{guo-lin}) as well for galaxy models
satisfying 3D Vlasov-Poisson. There are two elements in this approach. One
is to formulate a family of dispersion operators $A_{\lambda }$ for the
electric potential, depending on a positive parameter $\lambda $. The other
is to prove the existence of a purely growing mode by finding a parameter $%
\lambda _{0}$ such that the $A_{\lambda _{0}}$ has a kernel. The key
observation is that these dispersion operators are self-adjoint due to the
reversibility of the particle trajectories. A continuation argument is
applied to find the parameter $\lambda _{0}$ corresponding to a growing
mode, by comparing the spectra of $A_{\lambda }$ for very small and large
values of $\lambda $.

Let us explain the difficulties in extending this approach to the
electromagnetic case. We first recall the method in \cite{lw-linear} in the
periodic $1\frac{1}{2}$D case. Assuming that the growing mode has periodic
electromagnetic potentials $\left( \phi ,\psi \right) $ such that $%
E_{1}=-\partial _{x}\phi ,\ B=\partial _{x}\psi ,E_{2}=-\partial _{t}\psi $,
we express $f$ in term of them by integrating along the trajectories.
Plugging $f$ into the Maxwell system and using the condition $\mu _{e}<0$ to
eliminate $\phi $, we get a self-adjoint dispersion operator for $\psi $
alone. Then we apply the continuation argument as in \cite{lin01}. The
difficulty with this approach is that the equation 
\begin{equation}
\partial _{t}E_{1}=-j_{1}  \label{eqn-current}
\end{equation}
(the first current equation of Maxwell) does not follow from the dispersion
operator. Under additional evenness assumptions in the variable $x $, we
proved in \cite{lw-linear} by means of a parity argument that $j_{1}$ has
zero mean. Then (\ref{eqn-current}) does follow from the Poisson equation 
\begin{equation*}
\partial _{x}E_{1}=\rho
\end{equation*}%
and we get a growing mode.

In order to make this construction without any evenness assumption, we need
a new formulation. To do this, we express $E_{1}=-\partial _{x}\phi -\lambda
b$ where the scalar $b$ is introduced to account for the possible nonzero
spatial average of $E_{1}$ and $\lambda $ is the exponential growth rate.
Once again we express $f$ in terms of $\left( \phi ,\psi ,b\right) $ by
integration over the trajectories and plug it into the Maxwell system. The
equation (\ref{eqn-current}) can now be handled by means of this additional
number $b$. Again we eliminate $\phi $ using the the condition $\mu _{e}<0$
and the resulting equations for $\psi $ and $b$ can be written in a
self-adjoint matrix operator form. We then apply the continuation argument
to this new dispersion matrix by keeping track of its negative spectrum.

The proof of Theorem \ref{3d growing} in the 3D case is much more subtle. We
start with the electric potential $\phi $ and the magnetic potential $%
\mathbf{A}=\left( A_{r},A_{\theta },A_{z}\right) $. Of course we define $%
\mathbf{E}=-\nabla \phi -\partial _{t}\mathbf{A}$ and $\mathbf{B}=\nabla
\times \mathbf{A}$. Our strategy is to represent $f$ in terms of $\left(
\phi ,\mathbf{A}\right) $ and plug it into the Maxwell system to get a 
\textit{self-adjoint} formulation for the electromagnetic potentials. To
achieve this goal as well as to satisfy the current equation, our key
observation is to impose the Coulomb gauge condition $\nabla \cdot \mathbf{A}%
=0$ and use the cylindrical symmetry to define a \textquotedblleft
super-potential" $\pi \left( r,z\right) $ such that $\nabla \times \left(
\pi \mathbf{e}_{\theta }\right) =\left( A_{r},0,A_{z}\right) $. We then
express $f$ in terms of $\left( \phi ,A_{\theta },\pi \right) $ by
integrating along the trajectories of the equilibrium and the external field
and then plugging it into the Maxwell system to derive a system of equations
for the three unknowns $\phi ,A_{\theta }$ and $\pi $. The introduction of
this \textquotedblleft super-potential" $\pi $ allows us to separate the $%
\theta $ and $\left( r,z\right) $ components of the current equation (\ref%
{lin-maxwell-current}). The resulting system for $\left( \phi ,A_{\theta
},\pi \right) $ indeed turns out to be self-adjoint. We then eliminate $\phi 
$ using the the condition $\mu _{e}<0$ to get a $2\times 2$ self-adjoint
matrix operator $\mathcal{M}^{\lambda }$ for $\left( A_{\theta },\pi \right) 
$, depending on a positive parameter $\lambda $.

However, this matrix operator is bounded neither from below nor from above
so the continuation argument cannot be applied directly. To handle this new
difficulty, we perform an $n$-dimensional truncation in the function space
for $\pi $. The truncated matrix operator $\mathcal{M}_{n}^{\lambda }$ has
entries that are high-order integro-differential operators. It is bounded
from below (by a bound depending on $n$), which allows us to apply the
continuation argument to get an approximate growing mode. We then let $n$ go
to infinity. The limit of this approximate growing mode is shown to satisfy
the original linearized Vlasov-Maxwell system weakly.

However, it is still very subtle to show that this limit indeed gives us a
true growing mode. There are two issues to clarify. The first is to show
that the limit does not vanish, for which we need a uniform bound of the
approximate growing modes. The second issue is to show that the growth rate
does not tend to zero as $n$ go to infinity. For this, we need to get
uniform control of the spectrum of $\mathcal{M}_{n}^{\lambda }$ for small $%
\lambda $ and large $n$. This turns out to be quite delicate since the
operators involved merely converge to their limits weakly as $\lambda
\searrow 0$. In our proof the compactness of the support of the confined
plasma equilibria plays a crucial role, allowing us to get some compactness
of the operators.

As for Theorem \ref{3d growing} (ii), the lower bound on the number of
growing modes is a corollary of the continuation argument. To get the upper
bound, the key observation is that any two growing modes are orthogonal in
some sense due to a certain invariance property proven in \cite{lw-linear}.
We note that such counting formulae are unknown for the standard energy
principles (\cite{energy58}, \cite{kulsrud-energy}) for approximate plasma
models like MHD. In our case the simple spectral structure of the operator $%
\mathcal{L}^{0}$ is essential.

The new formulation and techniques developed in this paper can also be used
to detect linear instability of general Vlasov-Maxwell equilibria without
the monotone assumption $\mu _{e}<0$. The idea is to formulate the growing
mode problem as a $3\times 3$ indefinite matrix dispersion operator
including $\phi $ and then to use the truncation and continuation arguments
to study it. In this way we find a sufficient instability criterion by
utilizing the difference of the signatures of the matrix operators at small
and large parameters. We illustrate this idea in Section 9 by getting a
instability criterion in $1\frac{1}{2}$D purely magnetic case that
generalizes the sharp criterion in the monotone case.

The methods of this paper and of \cite{lw-linear} can also be used for 
\textit{nonrelativistic} Vlasov-Maxwell systems and also for other
symmetries, for example, the $2\frac{1}{2}$D Vlasov-Maxwell system with its $%
z$-symmetry \cite{gsc2.5}. For such cases, but still assuming that the
distribution function depends monotonically on the particle energy, we can
establish similar energy principles in terms of a certain self-adjoint
operator $\mathcal{L}^{0}$. For the nonrelativistic case the operator $%
\mathcal{L}^{0}$ is formally obtained from its relativistic version by
dropping the hat in $\hat{v}$. Since the results and the proofs are similar
to the cases we treat here, we do not elaborate any further.

The paper is organized as follows. In Section 2, we treat the easier $1\frac{%
1}{2}$D case. The proof for the 3D case is split into six sections. In
Section 3, we formulate the problem using $\left( \phi ,A_{\theta },\pi
\right) $ and derive the dispersion matrix operator $\mathcal{M}^\lambda$
for $\left( A_{\theta },\pi \right) $. In Section 4, we present the key
mapping and spectral properties of the operators appearing in the
formulation. In Section 5, we study their behavior for small $\lambda $ and
introduce the finite-dimensional truncation. In Section 6, we find the
approximate growing mode for each $n$. In Section 7, we take the limit of
the approximate growing modes. In Section 8, we check that this limit is
indeed a true growing mode. In Section 9, we extend our formulation to
equilibria that are not monotone in the energy $e$.

%%%%%%%%%%%%%%%  Section 2  %%%%%%%%%%%%%%%%%%%%%%%%%%%%%

\section{$1\frac{1}{2}$ dimensional case}

In this case, physical space is one-dimensional $x\in {\mathbb{R}}$ and
momentum space is two-dimensional $v=(v_{1},v_{2})\in {\mathbb{R}}^{2}$.
Moreover, $\mathbf{E}=(E_{1},E_{2},0)$ and $\mathbf{B}=(0,0,B)$. Assuming no
external field and setting all physical constants to be $1$, system (\ref%
{RVM-3D}) reduces to the following $1\frac{1}{2}$D RVM system 
\begin{subequations}
\label{1.5RVM}
\begin{equation}
\partial _{t}f^{\pm }+\hat{v}_{1}\partial _{x}f^{\pm }\pm (E_{1}+\hat{v}%
_{2}B)\partial _{v_{1}}f^{\pm }\pm (E_{2}-\hat{v}_{1}B)\partial
_{v_{2}}f^{\pm }=0  \label{M}
\end{equation}%
\begin{equation}
\partial _{t}E_{1}=-j_{1},\quad \partial _{t}E_{2}+\partial _{x}B=-j_{2}
\end{equation}%
\begin{equation}
\partial _{t}B=-\partial _{x}E_{2},\quad \partial _{x}E_{1}=\rho
\end{equation}%
with 
\end{subequations}
\begin{equation*}
\rho =\int (f^{+}-f^{-})dv,\quad j_{i}=\int \hat{v}_{i}(f^{+}-f^{-})dv\quad
\left( i=1,2\right) .
\end{equation*}%
The main reason to consider $1\frac{1}{2}$D RVM is its simplicity and yet it
preserves many of the essential features of $3$D RVM. We refer to \cite%
{parks} for astrophysical applications of this model and to \cite{gsc1.5}
for a proof of global well-posedness. We will consider solutions of the
system (\ref{1.5RVM}) that are periodic in the variable $x$ with a given
period $P$.

First we take a $P$-periodic equilibrium 
\begin{equation}
f^{0,\pm }=\mu ^{\pm }(e^{\pm },p^{\pm })=\mu ^{\pm }(\langle v\rangle \pm
\phi ^{0}(x),v_{2}\pm \psi ^{0}(x)),  \label{equilibria}
\end{equation}%
\begin{equation*}
E_{1}^{0}=-\partial _{x}\phi ^{0},\quad E_{2}^{0}=0,\quad B^{0}=\partial
_{x}\psi ^{0},
\end{equation*}%
where the pair $\left( \phi ^{0},\psi ^{0}\right) $ satisfies the ODE system
\ 
\begin{equation}
\partial _{x}^{2}\phi ^{0}=-\rho ^{0}=-\int (f^{0,+}-f^{0,-})dv,\quad
\partial _{x}^{2}\psi ^{0}=-j_{2}^{0}=-\int \hat{v}_{2}(f^{0,+}-f^{0,-})dv.
\label{equi-fields}
\end{equation}%
We assume that 
\begin{equation}
\mu ^{\pm }\geq 0,\ \ \mu ^{\pm }\in C^{1},\ \ \mu _{e}^{\pm }<0,\ \ |\mu
_{e}^{\pm }|+|\mu _{p}^{\pm }|\leq c(1+\left\vert e\right\vert )^{-\alpha }
\label{assumption-mu+-}
\end{equation}%
for some $\alpha >2$. In \cite{lw-linear} we proved that there exist
infinitely many periodic electromagnetic equilibria of the above form. Now
we denote 
\begin{equation*}
D^{\pm }=\hat{v}_{1}\partial _{x}\pm (E_{1}^{0}+\hat{v}_{2}B^{0})\partial
_{v_{1}}\mp \hat{v}_{1}B^{0}\partial _{v_{2}}\ ,
\end{equation*}%
\begin{equation*}
L_{\left\vert \mu _{e}^{\pm }\right\vert }^{2}=\left\{ f\ \Big |\ f\ \text{%
periodic in }x,\ \Vert f\Vert _{\pm }^{2}\equiv \int_{0}^{P}\int_{-\infty
}^{\infty }|f|^{2}|\mu _{e}^{\pm }|dvdx<\infty \right\} \ ,
\end{equation*}%
and $\mathcal{P}^{\pm }=$ the projection operator of $L_{\left\vert \mu
_{e}^{\pm }\right\vert }^{2}$ onto $\ker D^{\pm }$. We define the following
linear operators acting on $L_{P}^{2}(\mathbb{R})$, where the subscript $P$
refers to the periodicity. 
\begin{equation}
\mathcal{A}_{1}^{0}h=-\partial _{x}^{2}h-\left( \sum_{\pm }\int \mu
_{e}dv\right) h+\sum_{\pm }\int \mu _{e}^{\pm }\ \mathcal{P}^{\pm }h\ dv.
\label{defn-a1-1.5d}
\end{equation}%
\begin{equation}
\mathcal{A}_{2}^{0}h=-\partial _{x}^{2}h-\left( \sum_{\pm }\int \hat{v}%
_{2}\mu _{p}^{\pm }dv\ \right) h-\sum_{\pm }\int \mu _{e}^{\pm }\hat{v}_{2}\ 
\mathcal{P}^{\pm }(\hat{v}_{2}h)dv.  \label{defn-a2-1.5d}
\end{equation}%
\begin{equation}
\mathcal{B}^{0}h=\left( \sum_{\pm }\int \mu _{p}^{\pm }dv\right) h+\sum_{\pm
}\int \mu _{e}^{\pm }\ \mathcal{P}^{\pm }(\hat{v}_{2}h)\ dv
\label{defn-b0-1.5d}
\end{equation}%
and 
\begin{equation}
\mathcal{L}^{0}=(\mathcal{B}^{0})^{\ast }(\mathcal{A}_{1}^{0})^{-1}\mathcal{B%
}^{0}+\mathcal{A}_{2}^{0}.  \label{defn-L0-1.5d}
\end{equation}

Similarly to the $3$D case, we proved in \cite{lw-linear} the following
theorem.

%%%%%%%%%%%%%%%%%  Theorem 2.1  %%%%%%%%%%%%%%%%%%

\begin{theorem}
\label{thm-1.5d-2s-sta} Consider periodic perturbations of any equilibrium
satisfying the conditions given above. Then

(i) $\mathcal{L}^{0}\geq0$ implies spectral stability. \ 

(ii) Any growing mode must be purely growing.

(iii) If $-\alpha ^{2}$ denotes the lowest eigenvalue of the operator $%
\mathcal{L}^{0}$, then the maximal growth rate cannot exceed $\alpha $.
\end{theorem}

Moreover, it was shown in \cite{lw-linear} that if $\psi ^{0},\phi ^{0}$ are
even functions of $x$ around $x=P/2$ and if $\mathcal{L}^{0}$ has an \textit{%
even} eigenfunction corresponding to a negative eigenvalue, then there
exists a growing mode. In the following theorem proven in this section, we
assert that $\mathcal{L}^{0}\ngeq 0$ always implies the existence of a
growing mode, without any additional evenness assumptions.

\begin{theorem}
\label{1.5d growing} Under the same assumptions,

(i) If $\mathcal{L}^{0}\ngeq 0$, then there exists a real periodic growing
mode $[e^{\lambda t}f(x,v),e^{\lambda t}E(x)$, $e^{\lambda t}B(x)]$ with $%
f,E,B\in W^{1,1}_P$ and $\lambda >0$.

(ii) The dimension of the space of growing modes equals the dimension of the
negative eigenspace of $\mathcal{L}^{0}$.
\end{theorem}

The combination of Theorems \ref{thm-1.5d-2s-sta} and \ref{1.5d growing}
provides an energy principle for the $1\frac{1}{2}$D case, in terms of the
operator $\mathcal{L}^{0}.$

With the sole purpose of simplifying our notation, we present the proof in
the case of a constant ion background $n_{0}$. (For the more general
two-species case, the proofs remain almost the same except for the more
cumbersome notation.) Then the $1\frac{1}{2}$D RVM for one species becomes 
\begin{subequations}
\label{1.5D-one species}
\begin{equation}
\partial _{t}f+\hat{v}_{1}\partial _{x}f-\left( E_{1}+\hat{v}_{2}B\right)
\partial _{v_{1}}f-\left( E_{2}-\hat{v}_{1}B\right) \partial _{v_{2}}f=0
\label{rvm-1}
\end{equation}%
\begin{equation}
\partial _{t}E_{1}=-j_{1}=\int \hat{v}_{1}f\text{ }dv,\quad
\partial_{t}B=-\partial _{x}E_{2}
\end{equation}%
\begin{equation}
\partial _{t}E_{2}+\partial _{x}B=-j_{2}=\int \hat{v}_{2}f\text{ }dv
\end{equation}%
with the constraint 
\end{subequations}
\begin{equation}
\partial _{x}E_{1}=n_{0}-\int f\text{ }dv.
\end{equation}%
Fixing any such equilibrium with a period $P$, we will consider the system
(21) with periodic boundary conditions of the same period $P$.

The equilibrium is assumed to have the form $f^{0}=\mu (e,p),$ $%
E_{1}^{0}=-\partial _{x}\phi ^{0},E_{2}^{0}=0,B^{0}=\partial _{x}\psi ^{0},$
where the electromagnetic potentials $\left( \phi ^{0},\psi ^{0}\right) $
satisfy the ODE system%
\begin{equation*}
\partial _{x}^{2}\phi ^{0}=n_{0}-\int \mu (e,p)dv,\quad \partial
_{x}^{2}\psi ^{0}=\int \hat{v}_{2}\mu (e,p)dv
\end{equation*}%
with the electron energy and the \textquotedblleft angular
momentum\textquotedblright\ defined by 
\begin{equation}
e=\langle v\rangle -\phi ^{0}(x),\quad p=v_{2}-\psi ^{0}(x).  \label{mu-defn}
\end{equation}%
(The $e$ is distinguished from the exponential $e$ in context.) The only
assumptions we make on $\mu $ are 
\begin{equation}
\mu \geq 0,\quad \mu \in C^{1},\ \ \mu _{e}\equiv \frac{\partial \mu }{%
\partial e}<0  \label{mu-assumption}
\end{equation}%
and, in order for $\int \left( |\mu _{e}|+\left\vert \mu _{p}\right\vert
\right) dv$ to be finite, 
\begin{equation}
\left( |\mu _{e}|+\left\vert \mu _{p}\right\vert \right) \left( e,p\right)
\leq c(1+\left\vert e\right\vert )^{-\alpha }\text{ for some }\alpha >2.
\label{mu-assumption1}
\end{equation}

Hence the linearized evolution equations are 
\begin{equation}
(\partial_{t}+D)f=\mu_{e}\hat{v}_{1}E_{1}-\mu_{p}\hat{v}_{1}B+(\mu_{e}\hat {v%
}_{2}+\mu_{p})E_{2},  \label{linearized-V}
\end{equation}
where $D$ is the transport operator associated with the steady fields, 
\begin{equation}
D=\hat{v}_{1}\partial_{x}-\left( E_{1}^{0}+\hat{v}_{2}B^{0}\right)
\partial_{v_{1}}+\hat{v}_{1}B^{0}\partial_{v_{2}}  \notag
\end{equation}
together with 
\begin{equation}
\partial_{x}E_{1}=-\int fdv,\ \partial_{t}E_{1}=\int\hat{v}_{1}fdv,\
\partial _{t}E_{2}+\partial_{x}B=\int\hat{v}_{2}fdv,\
\partial_{t}B+\partial_{x}E_{2}=0.  \label{linearized-M}
\end{equation}
%For convenience, we may introduce the magnetic potential $\psi(t,x)$ such that
%$B=\pa_x\psi$ and $E_2=-\pa_t\psi$.  (It is not necessarily true that $\psi$ is
%periodic.  It is periodic if and only if
%$B$ has zero mean.)
%The magnetic potential is a scalar $\psi$ and the magnetic field is
%$[0,0,B]$ with $B=\partial_{x}\psi$. The electric potential is $\phi$ and the
%electric field is $[E_{1},E_{2},0]$ with $E_{1}=-\partial_{x}\phi$ and
%$E_{2}=-\partial_{t}\psi$.
We define the Hilbert space 
\begin{equation*}
L_{\left\vert \mu_{e}\right\vert }^{2} = \left\{ f\left( x,v\right) \ \Big|\
\ f\ \text{periodic in }x,\ \Vert f\Vert_{_{\left\vert \mu _{e}\right\vert
}}^{2}\equiv\int_{0}^{P}\int_{\mathbb{R}^2}|f|^{2}|\mu
_{e}|dvdx<\infty\right\}
\end{equation*}
and denote its inner product by $\left( \cdot,\cdot\right) _{\left\vert
\mu_{e}\right\vert }$. Let $\mathcal{P}$ be the projection operator of $%
L_{\left\vert \mu_{e}\right\vert }^{2}$ onto the kernel of $D$. We also
denote by $L_{P}^{p}$ $\left( H_{P}^{2}\right) $ the space of $P$-periodic $%
L_{x}^{p}$ $\left( H_{x}^{2}\right) $ functions for $p\ge1$.

Similarly to the two-species case, we define the following four operators,
each of which acts from $H_{P}^{2}$ to $L_{P}^{2}$, 
\begin{equation*}
\mathcal{A}_{1}^{0}h=-\partial_{x}^{2}h-\left( \int\mu_{e}dv\right)
h+\int\mu_{e}\ \mathcal{P}h\ dv,
\end{equation*}%
\begin{equation*}
\mathcal{A}_{2}^{0}h=-\partial_{x}^{2}h-\left( \int\hat{v}_{2}\mu
_{p}dv\right) h-\int\mu_{e}\hat{v}_{2}\mathcal{P}(\hat{v}_{2}h)\ dv,
\end{equation*}%
\begin{equation*}
\mathcal{B}^{0}h=\left( \int\mu_{p}dv\right) h+\int\mu_{e}\ \mathcal{P}(\hat{%
v}_{2}h)\ dv
\end{equation*}
and 
\begin{equation*}
\mathcal{L}^{0}=(\mathcal{B}^{0})^{\ast}(\mathcal{A}_{1}^{0})^{-1}\mathcal{B}%
^{0}+\mathcal{A}_{2}^{0}.
\end{equation*}
In these definitions one should keep in mind that $\mu\geq0$ is a function
of $x$ and $v$, that $\mu_{e}=\partial\mu/\partial e<0$ and that $\mu
_{p}=\partial\mu/\partial p$. It was shown in \cite{lw-linear} that $%
\mathcal{A}_{1}^{0}$ is invertible on the range of $\mathcal{B}^{0}$ so that 
$\mathcal{L}^{0}$ is well-defined. The following is the analogue of Theorem %
\ref{1.5d growing}.

%%%%%%%%%%%  Theorem 2.3  %%%%%%%%%%%%%

\begin{theorem}
\label{thm-1.5d-growing (1)} Assume (\ref{mu-assumption}) and (\ref%
{mu-assumption1}). \newline
(i) If $\mathcal{L}^{0}\ngeq0$, then there exists a real growing mode $%
[e^{\lambda t}f(x,v),e^{\lambda t}E(x),e^{\lambda t}B(x)]$ with $f,E,B\in
W^{1,1}$ and $\lambda>0$. \newline
(ii) The dimension of the space of growing modes equals the dimension of the
negative eigenspace of $\mathcal{L}^{0}$.
\end{theorem}

For the proof of this theorem we introduce the particle paths $%
(X(t;x,v),V(t;x,v))$, which are the characteristics of $D$. They are defined
as the solutions of%
\begin{equation}
\dot{X}=\hat{V}_{1},\quad\dot{V}_{1}=\partial_{x}\phi^{0}(X)-\hat{V}%
_{2}\partial_{x}\psi^{0}(X),\quad\dot{V}_{2}=\hat{V}_{1}\partial_{x}\psi
^{0}(X)  \label{particle-ode}
\end{equation}
with the initial conditions $X(0)=x,\ V(0)=v$. Using the particle paths, the
next three operators depending on a parameter $\lambda>0$ were already
introduced in \cite{lw-linear} 
\begin{equation*}
\mathcal{A}_{1}^{\lambda}h=-\partial_{x}^{2}h-\left( \int\mu_{e}dv\right)
h+\int\mu_{e}\int_{-\infty}^{0}\lambda e^{\lambda s}h(X(s))dsdv,
\end{equation*}%
\begin{equation*}
\mathcal{A}_{2}^{\lambda}h=-\partial_{x}^{2}h+\lambda^{2}h-\left( \int\hat {v%
}_{2}\mu_{p}dv\right) h-\int\hat{v}_{2}\mu_{e}\int_{-\infty}^{0}\lambda
e^{\lambda s}\hat{V}_{2}(s)h(X(s))dsdv,
\end{equation*}%
\begin{equation*}
\mathcal{B}^{\lambda}h=\left( \int\mu_{p}dv\right)
h+\int\mu_{e}\int_{-\infty}^{0}\lambda e^{\lambda s}\hat{V}%
_{2}(s)h(X(s))dsdv.
\end{equation*}
The following lemma in \cite{lw-linear} shows that $\mathcal{A}%
_{1}^{\lambda} $ is invertible on the range of $\mathcal{B}^{\lambda}$, so
that the operator 
\begin{equation*}
\mathcal{L}^{\lambda}=(\mathcal{B}^{\lambda})^{\ast}(\mathcal{A}%
_{1}^{\lambda })^{-1}\mathcal{B}^{\lambda}+\mathcal{A}_{2}^{\lambda}.
\end{equation*}
is also well-defined.

%%%%%%%%%%  Lemma 2.4  %%%%%%%%%%%%%%%

\begin{lemma}[\protect\cite{lw-linear}]
\label{1.5d operators} Assume $\lambda \geq 0$. \newline
(i) The operators $\mathcal{A}_{j}^{\lambda },\mathcal{L}^{\lambda }$ $%
(j=1,2)$ are self-adjoint on $L_{P}^{2}$ with the common domain $H_{P}^{2}$.
Their spectra are discrete. \newline
(ii) $\mathcal{A}_{1}^{\lambda }\geq 0$. \newline
(iii) The null-space $N(\mathcal{A}_{1}^{\lambda })$ consists of the
constant functions. The inverse $(\mathcal{A}_{1}^{\lambda })^{-1}$ is
bounded from $\{h\in L_{P}^{2}\ |\ \int_{0}^{P}hdx=0\}=N(\mathcal{A}%
_{1}^{\lambda })^{\perp }\supset R(\mathcal{B}^{\lambda })$ into $H_{P}^{2}$%
. \newline
\end{lemma}

We also introduce the following three functions that depend on $\lambda>0$. 
\begin{equation*}
b^{\lambda}\left( x\right) =\int\mu_{e}\int_{-\infty}^{0}\lambda e^{\lambda
s}\hat{V}_{1}(s)dsdv,
\end{equation*}%
\begin{equation*}
c^{\lambda}\left( x\right) =\int\hat{v}_{2}\mu_{e}\int_{-\infty}^{0}\lambda
e^{\lambda s}\hat{V}_{1}(s)dsdv,
\end{equation*}%
\begin{equation*}
d^{\lambda}=(\mathcal{B}^{\lambda})^{\ast}(\mathcal{A}_{1}^{%
\lambda})^{-1}b^{\lambda}-c^{\lambda}
\end{equation*}
and three constants 
\begin{equation*}
l^{\lambda}=\frac{1}{P}\int_{0}^{P}\int\hat{v}_{1}\mu_{e}\int_{-\infty}^{0}%
\lambda e^{\lambda s}\hat{V}_{1}(s)dsdvdx,
\end{equation*}
\begin{equation*}
m^{\lambda}=\frac{1}{P}\left( (\mathcal{A}_{1}^{\lambda})^{-1}b^{\lambda
},b^{\lambda}\right) ,\quad k^{\lambda}=P\left( \lambda^{2}-l^{\lambda
}-m^{\lambda}\right) .
\end{equation*}
Define $\mathcal{F}^{\lambda}$ to be the operator from $\mathbb{R }$ to $%
L_{P}^{2}$ by $\mathcal{F}^{\lambda}\left( b\right) =bd^{\lambda}$. Its
adjoint $\left( \mathcal{F}^{\lambda}\right) ^{\ast}$ mapping $L_{P}^{2}$ to 
$\mathbb{R }$ is defined by $\mathcal{F}^{\lambda}\left( \psi\right) =\left(
\psi,d^{\lambda}\right) $. We define the matrix operator $\mathcal{M}%
^{\lambda}$ from $H_{P}^{2}\times\mathbb{R }$ to $L_{P}^{2}\times\mathbb{R }$
by 
\begin{equation*}
\mathcal{M}^{\lambda}\left( 
\begin{array}{c}
\psi \\ 
b%
\end{array}
\right) =\left( 
\begin{array}{c}
\mathcal{L}^{\lambda}\psi+bd^{\lambda} \\ 
-bk^{\lambda}+\left( \psi,d^{\lambda}\right)%
\end{array}
\right) =\left( 
\begin{array}{cc}
\mathcal{L}^{\lambda} & \mathcal{F}^{\lambda} \\ 
\left( \mathcal{F}^{\lambda}\right) ^{\ast} & -k^{\lambda}%
\end{array}
\right) \left( 
\begin{array}{c}
\psi \\ 
b%
\end{array}
\right) .
\end{equation*}
By Lemma \ref{1.5d operators}, it is obvious that $\mathcal{M}^{\lambda}$ is
self-adjoint and has only discrete spectrum. The following lemma explains
the purpose of $\mathcal{M}^\lambda$.

%%%%%%%%%%%%%%%%  Lemma 2.5  %%%%%%%%%%%%%%%%%

\begin{lemma}
\label{lemma-1.5d formulation}If $\mathcal{M}^{\lambda }$ has a non-trivial
nullspace of even functions for some $\lambda >0$, then there exists a
purely growing mode in $W^{1,1}$ of (\ref{linearized-V}), (\ref{linearized-M}%
).
\end{lemma}

To clarify the ideas, below we present our original derivation of the matrix
operator $\mathcal{M}^{\lambda}$ from the equations satisfied by a growing
mode. The proof of Lemma \ref{lemma-1.5d formulation} is almost the reverse
process of this derivation, as in the proof of Lemma 2.5 of \cite{lw-linear}%
. So we skip it here.

To derive $\mathcal{M}^{\lambda }$, we start with a growing mode $%
[e^{\lambda t}f(x,v),e^{\lambda t}E(x),e^{\lambda t}B(x)]$. Since it was
shown in \cite{lw-linear} that a growing mode must be purely growing, we can
assume $\lambda >0$. Define the electromagnetic potentials $\phi ,\psi $ and
an number $b\in \mathbb{R}$ such that 
\begin{equation*}
B=\partial _{x}\psi ,\quad E_{2}=-\lambda \psi ,\quad E_{1}=-\partial
_{x}\phi -\lambda b.
\end{equation*}%
Then $[f(x,v),\phi ,\psi ,b]$ must satisfy 
\begin{equation}
\lambda f+Df=-\mu _{e}\hat{v}_{1}\partial _{x}\phi -\lambda b\mu _{e}\hat{v}%
_{1}-\mu _{p}\hat{v}_{1}\partial _{x}\psi -(\lambda \mu _{e}\hat{v}%
_{2}+\lambda \mu _{p})\psi  \label{linearized V-growing}
\end{equation}%
and 
\begin{equation}
\partial _{x}E_{1}=\rho ,\ \ \lambda E_{1}=-j_{1},\ \ \lambda E_{2}+\partial
_{x}B=-j_{2},\ \ \lambda B+\partial _{x}E_{2}=0  \label{linearized M-growing}
\end{equation}%
with $\rho =-\int fdv$ and $j_{i}=-\int \hat{v}_{i}fdv$. Integrating (\ref%
{linearized V-growing}) along the particle trajectory, after an integration
by parts we have 
\begin{equation}
f(x,v)=-\mu _{e}\phi (x)-\mu _{p}\psi (x)+\mu _{e}\int_{-\infty }^{0}\lambda
e^{\lambda s}\left[ \phi (X(s))-\hat{V}_{2}(s)\psi (X(s))-b\hat{V}_{1}(s)%
\right] ds.  \label{growing f-exp}
\end{equation}%
The first and third equations of (\ref{linearized M-growing}) are equivalent
to $-\partial _{x}^{2}\phi =\rho $ and $\left( -\partial _{x}^{2}+\lambda
^{2}\right) \psi =j_{2}$. After plugging (\ref{growing f-exp}) into them,
they become%
\begin{equation}
\mathcal{A}_{1}^{\lambda }\phi =\mathcal{B}^{\lambda }\psi +bb^{\lambda }
\label{equation-phi}
\end{equation}%
and 
\begin{equation}
\mathcal{A}_{2}^{\lambda }\psi =-(\mathcal{B}^{\lambda })^{\ast }\phi
+bc^{\lambda }.  \label{equation-pci}
\end{equation}%
The last equation in (\ref{linearized M-growing}) is automatic.

The second equation in (\ref{linearized M-growing}) is $\lambda E_{1}=-j_{1}$%
, from which we will now derive an equation for $b$. By the continuity
equation $\partial _{x}j_{1}+\lambda \rho =0$, we have $\partial
_{x}^{2}\phi =-\rho =\frac{1}{\lambda }\partial _{x}j_{1}$, which implies
that $\partial _{x}\phi =\frac{1}{\lambda }\left( j_{1}-\frac{1}{P}%
\int_{0}^{P}j_{1}dx\right) $. Thus $\lambda E_{1}=-j_{1}$ is equivalent to $%
\lambda ^{2}b=\frac{1}{P}\int_{0}^{P}j_{1}dx$. Plugging (\ref{growing f-exp}%
) into this result, we obtain 
\begin{align*}
\lambda ^{2}b& =\frac{1}{P}\int_{0}^{P}\int \hat{v}_{1}\mu _{e}\int_{-\infty
}^{0}\lambda e^{\lambda s} \left\{ -\phi (X(s)) + b \hat{V}_{1}(s) +\hat{V}%
_{2}(s)\psi (X(s)) \right\} dsdvdx \\
& =I+II+III.
\end{align*}
The first term is 
\begin{align*}
I& =-\frac{1}{P}\int_{-\infty }^{0}\lambda e^{\lambda s}\int_{0}^{P}\int \mu
_{e}\phi \left( x\right) \hat{V}_{1}(-s)dvdxds \\
& =\frac{1}{P}\int_{-\infty }^{0}\lambda e^{\lambda s}\int_{0}^{P}\int \mu
_{e}\phi \left( x\right) \hat{V}_{1}(s)dvdxds =\frac{1}{P}\left(
\phi,b^{\lambda }\right) ,
\end{align*}
where for the first equality we changed variables $\left( x,v\right)
\rightarrow \left( X(-s),\hat{V}(-s)\right) $ and for the second equality we
changed variable $v\rightarrow -v$ and used the trajectory property 
\begin{align*}
& \left( X(-s;x,-v_{1},v_{2}),-V_{1}\left( -s;x,-v_{1},v_{2}\right)
,V_{2}\left( -s;x,-v_{1},v_{2}\right) \right) \\
& =\left( X(s;x,v_{1},v_{2}),V_{1}\left( s;x,v_{1},v_{2}\right) ,V_{2}\left(
s;x,v_{1},v_{2}\right) \right) .
\end{align*}%
Similarly, $III=-\frac{1}{P}\left( \psi ,c^{\lambda }\right) $. By
definition, $II=bl^{\lambda }$. Thus the equation for $b$ is 
\begin{equation}
\left( \lambda ^{2}-l^{\lambda }\right) b=\frac{1}{P}\left[ \left( \phi
,b^{\lambda }\right) -\left( \psi ,c^{\lambda }\right) \right] .
\label{equation-b}
\end{equation}%
By (\ref{equation-phi}) we get 
\begin{equation*}
\phi =(\mathcal{A}_{1}^{\lambda })^{-1}\mathcal{B}^{\lambda }\psi +b(%
\mathcal{A}_{1}^{\lambda })^{-1}b^{\lambda }.
\end{equation*}%
Plugging this into (\ref{equation-pci}) and (\ref{equation-b}), we have the
pair of equations $\mathcal{L}^{\lambda }\psi +bd^{\lambda }=0$ and $%
-bk^{\lambda }+\left( \psi ,d^{\lambda }\right) =0$ by definition of $%
d^\lambda, k^\lambda$ and $\mathcal{L}^\lambda$. That is, the pair $\left(
\psi ,b\right) $ belongs to the kernel of the matrix operator $\mathcal{M}%
^{\lambda }$. We note that in the above formulation the equation $\lambda
E_{1}=-j_{1}$ is exactly taken care by the extra constant $b$.

Similar to the proof of Lemma 2.5 of \cite{lw-linear}, we can show that a
nontrivial kernel of $\mathcal{M}^{\lambda }$ indeed gives a growing mode.
Moreover, we also showed in \cite{lw-nonlinear} that for any growing mode, $%
f\in W^{1,1}$ and the linear instability implies nonlinear instability in
the macroscopic sense.

%%%%%%%%%%%%%%  Lemma 2.6  %%%%%%%%%%%%%%%%%%%%

\begin{lemma}
\label{continuation lemma}If $\mathcal{L}^{0}\not \geq 0$, then there exists 
$\lambda>0$ such that $\mathcal{M}^{\lambda}$ has a non-trivial nullspace.
\end{lemma}

\begin{proof}
Let $n^{\lambda }$ be the dimension of the eigenspace of $\mathcal{M}%
^{\lambda }$ corresponding to its negative eigenvalues. We first claim that
for sufficiently large $\lambda $, $n^{\lambda }\leq 1$. Indeed, it is shown
in \cite{lw-linear} that $\mathcal{L}^{\lambda }\geq \lambda ^{2}-C_{0}$ for
some constant $C_{0}$ independent of $\lambda $. It is also easy to show
that $\left\Vert d^{\lambda }\right\Vert _{L^{2}}\leq C_{1}$ for some
constant $C_{1}$ independent of $\lambda ,$ as in the proof of Lemma 2.4 of 
\cite{lw-linear}. So 
\begin{align*}
\left\langle \mathcal{M}^{\lambda }\left( 
\begin{array}{c}
\psi \\ 
b%
\end{array}%
\right) ,\left( 
\begin{array}{c}
\psi \\ 
b%
\end{array}%
\right) \right\rangle & =\left( \mathcal{L}^{\lambda }\psi ,\psi \right)
+2b\left( \psi ,d^{\lambda }\right) -k^{\lambda }b^{2} \\
& \geq \left( \lambda ^{2}-C_{0}\right) \left\Vert \psi \right\Vert
_{2}^{2}-2C_{1}\left\vert b\right\vert \left\Vert \psi \right\Vert
_{2}-\left\vert k^{\lambda }\right\vert b^{2} \\
& \geq -\left( C_{1}^{2}+\left\vert k^{\lambda }\right\vert \right) b^{2},
\end{align*}%
provided $\lambda ^{2}\geq C_{0}+1$. Since $b\in \mathbb{R}$, it follows
that $\mathcal{M}^{\lambda }$ has at most a one-dimensional negative
subspace. We now show that if $\lambda $ is small enough, then $n^{\lambda
}\geq 2$. It is shown in \cite{lw-linear} that $\mathcal{L}^{\lambda
}\rightarrow \mathcal{L}^{0}$ strongly when $\lambda \rightarrow 0$ and 
\begin{equation*}
\lim_{\lambda \searrow 0}\int_{-\infty }^{0}\lambda e^{\lambda
s}h(X(s),V(s))ds=\mathcal{P}h
\end{equation*}%
in the norm of $L_{\left\vert \mu _{e}\right\vert }^{2}$ for all $h\in
L_{\left\vert \mu _{e}\right\vert }^{2}$. As in the proof of Lemma 3.3 of 
\cite{lw-linear}, the projection operator $\mathcal{P}$ maps a function that
is odd or even in $v_{1}$ to another function with the same symmetry
property. So as $\lambda \rightarrow 0$, $b^{\lambda }\rightarrow \int \mu
_{e}\mathcal{P}\left( \hat{v}_{1}\right) dv=0$ and similarly $c^{\lambda
}\rightarrow 0$ in $L_{P}^{2}$ strongly. Thus $d^{\lambda }\rightarrow 0$
and $\mathcal{F}^{\lambda }\rightarrow 0$ in $L_{P}^{2}$ strongly as $%
\lambda \rightarrow 0$. So we have 
\begin{equation*}
\mathcal{M}^{\lambda }\left( 
\begin{array}{c}
\psi \\ 
b%
\end{array}%
\right) \rightarrow \mathcal{M}^{0}\left( 
\begin{array}{c}
\psi \\ 
b%
\end{array}%
\right) =\left( 
\begin{array}{cc}
\mathcal{L}^{0} & 0 \\ 
0 & -k^{0}%
\end{array}%
\right) \left( 
\begin{array}{c}
\psi \\ 
b%
\end{array}%
\right)
\end{equation*}%
strongly in $L_{P}^{2}\times \mathbb{R}$ as $\lambda \rightarrow 0$ for all $%
\psi \in H_{P}^{2}$ and $b\in \mathbb{R}$. Here 
\begin{equation*}
k^{0}=\int_{0}^{P}\int \left\vert \mu _{e}\right\vert \left( \mathcal{P}%
\left( \hat{v}_{1}\right) \right) ^{2}dvdx>0.
\end{equation*}%
Since $\mathcal{L}^{0}$ has at least one negative eigenvalue by assumption, $%
\mathcal{M}^{0}$ has at least two negative eigenvalues. Thus by (\cite{kato}%
, IV-3.5), $n^{\lambda }\geq 2$ if $\lambda $ is small enough.

For $\lambda>0$, it was shown in \cite{lw-linear} that $\mathcal{L}%
^{\lambda} $ is continuous in the operator norm. So $\mathcal{M}^{\lambda}$
is also continuous in the operator norm for $\lambda>0$. Thus if $\mathcal{M}%
^{\lambda}$ has no kernel for all $\lambda>0$, then $n^{\lambda}$ remains a
constant which is inconsistent with the behavior of $n^{\lambda}$ near zero
and infinity. So we conclude that for some $\lambda>0$, $\mathcal{M}%
^{\lambda}$ must have a nontrivial kernel. This completes the proof of the
Lemma.
\end{proof}

Theorem \ref{thm-1.5d-growing (1)} (i) on the existence of growing modes
follows immediately by combining Lemma \ref{lemma-1.5d formulation} and
Lemma \ref{continuation lemma}.

\bigskip For the proof of Theorem \ref{thm-1.5d-growing (1)}(\textit{ii}),
we need the following two lemmas. We consider real functions below, as all
growing modes should be by Theorem \ref{thm-1.5d-2s-sta}. The following
functionals were defined in \cite{lw-linear}. 
\begin{equation}
J(f,E_{1},\psi )=\iint \frac{1}{|\mu _{e}|}(f+\mu _{p}\psi )^{2}dvdx+\int %
\left[ E_{1}\right] ^{2}dx  \label{j-functional}
\end{equation}%
\begin{equation}
I(f,E_{1},\psi )=J(f,E_{1},\psi )-\iint \hat{v}_{2}\mu _{p}\psi
^{2}dvdx+\int [(\partial _{t}\psi )^{2}+(\partial _{x}\psi )^{2}]dx
\label{I-functional}
\end{equation}%
and we denote 
\begin{equation*}
J(f,E_{1},\psi ;\tilde{f},\tilde{E}_{1},\tilde{\psi})=\iint \frac{1}{|\mu
_{e}|}(f+\mu _{p}\psi )\left( \tilde{f}+\mu _{p}\tilde{\psi}\right)
dvdx+\int E_{1}\tilde{E}_{1}dx.
\end{equation*}

The next lemma follows immediately by polarization from Lemma 2.7 of \cite%
{lw-linear}.

%%%%%%%%%%%%%%%%%  Lemma 2.7  %%%%%%%%%%%%%%%%%

\begin{lemma}
\label{lemma-invariant 1.5} Consider two real solutions $\left( f^{i}\left(
t\right) ,E^{i}\left( t\right) ,B^{i}\left( t\right)
=\partial_{x}\psi^{i}\left( t\right) \right) ,i=1,2$ to the linearized
system (\ref{1.5D-one species}), with initial data $\left( f^{i}\left(
0\right) ,E^{i}\left( 0\right) ,B^{i}\left( 0\right) =\psi_{x}\left(
0\right) \right) \in$ $L^{1}$ in the constraint set 
\begin{equation*}
\mathcal{C}=\left\{ \iint f\left( 0\right) dvdx=0,\ \
\partial_{x}E_{1}\left( 0\right) =-\int f\left( 0\right) dv\ \text{ and }%
\int B\left( 0\right) dx=0\right\} ,
\end{equation*}
satisfying $J(f(0),E_{1}(0),\psi\left( 0\right) )<\infty$. Then the
functional 
\begin{align*}
& I(f^{1},E_{1}^{1},\psi^{1};f^{2},E_{1}^{2},\psi^{2})\left( t\right) \\
& =J(f^{1},E_{1}^{1},\psi^{1};f^{2},E_{1}^{2},\psi^{2})-\iint\hat{v}%
_{2}\mu_{p}\psi^{1}\psi^{2}dvdx+\int[\partial_{t}\psi^{1}\partial_{t}\psi
^{2}+\partial_{x}\psi^{1}\partial_{x}\psi^{2}]dx
\end{align*}
is independent of $t$. Furthermore, for all $g\in\ker D$, the functionals 
\begin{equation}
K_{g}(f^{i},\psi^{i})=\iint[f^{i}+(\hat{v}_{2}\mu_{e}+\mu_{p})\psi ^{i}]\ g\
dvdx  \label{K-functional}
\end{equation}
are also independent of $t$.
\end{lemma}

\begin{proof}[Proof of Theorem \protect\ref{thm-1.5d-growing (1)} (ii)]
Assume the linearized system (\ref{linearized-V}), (\ref{linearized-M}) has $%
l$ independent growing modes and the operator $\mathcal{L}^{0}$ has a $k-$%
dimensional negative eigenspace. By the proof of Lemma \ref{continuation
lemma}, as $\lambda $ increases from $0~$to $+\infty $, the negative
eigenvalues of $\mathcal{M}^{\lambda }$ must cross the imaginary axis at
least $n\left( \mathcal{M}^{0}\right) -1$ times, with $n\left( \mathcal{M}%
^{0}\right) =k+1$ being the number of negative eigenvalues of $\mathcal{M}%
^{0}$. Since we get a growing mode at each such crossing, there exist at
least $k$ growing modes. Thus $l\geq k$.

It remains to show that $l\leq k$. Suppose otherwise, $l>k$. Let $\left\{
\zeta _{1},\cdots ,\zeta _{k}\right\} \subset L_{P}^{2}$ span the negative
eigenspace of $\mathcal{L}^{0}$. Denote the $l$ linearly independent growing
modes by $e^{\lambda _{i}t}[f^{i}(x,v),E_{1}^{i}(x),\psi ^{i}(x)],i=1,\cdots
,l$, where $\psi ^{i}(x)$ is the magnetic potential $\partial _{x}\psi
^{i}=B^{i}$. By Theorem \ref{thm-1.5d-2s-sta} (ii), $\lambda _{i}$ are real
and positive and we only need to consider real functions below.

First we will prove that $\left\{ \psi ^{i}(x)\right\} _{i=1}^{l}$ are
linearly independent. Indeed suppose $\left( c_{1},\cdots ,c_{l}\right) \in 
\mathbb{R}^{l}$ such that $\psi ^{c}\left( x\right) =\sum_{i=1}^{l}c_{i}\psi
^{i}(x)=0$. We denote $f^{c}=\sum_{i=1}^{l}c_{i}f^{i}$ and $%
E_{1}^{c}=\sum_{i=1}^{l}c_{i}E_{1}^{i}$. Applying Lemma \ref{lemma-invariant
1.5} to any two growing modes $e^{\lambda
_{i}t}[f^{i}(x,v),E_{1}^{i}(x),\psi ^{i}(x)]$ and $e^{\lambda
_{j}t}[f^{j}(x,v),E_{1}^{j}(x),\psi ^{j}(x)]$ with $1\leq i,j\leq l$, we
have 
\begin{eqnarray*}
0 &=&I(f^{i},E_{1}^{i},\psi ^{i};f^{j},E_{1}^{j},\psi ^{j}) \\
&=&J(f^{i},E_{1}^{i},\psi ^{i};f^{j},E_{1}^{j},\psi ^{j})-\iint \hat{v}%
_{2}\mu _{p}\psi ^{i}\psi ^{j}dvdx+\int [\lambda _{i}\lambda _{j}\psi
^{i}\psi ^{j}+\psi _{x}^{i}\psi _{x}^{j}]dx.
\end{eqnarray*}
In particular, 
\begin{align}
0& =J\left( f^{c},E_{1}^{c},\psi ^{c}\right) -\iint \hat{v}_{2}\mu _{p}\left[
\psi ^{c}\right] ^{2}dvdx  \label{zero-I-combination} \\
& +\int \left[ \psi _{x}^{c}\right] ^{2}dx+\int \left( \sum_{i=1}^{l}\lambda
_{i}c_{i}\psi ^{i}\right) ^{2}dx.  \notag
\end{align}%
But $\psi ^{c}=0$ so that 
\begin{eqnarray}
0 &=&J(f^{c},E_{1}^{c},0)+\int \left( \sum_{i=1}^{l}\lambda _{i}c_{i}\psi
^{i}\right) ^{2}dx  \label{inequality-psi-combination} \\
&\geq &J(f^{c},E^{c},0)=\iint \frac{1}{|\mu _{e}|}\left[ f^{c}\right]
^{2}dvdx+\int \left[ E_{1}^{c}\right] ^{2}dx.  \notag
\end{eqnarray}
Thus we have $f^{c}=0,\ E_{1}^{c}=0$ and therefore $%
\sum_{i=1}^{l}c_{i}[f^{i}(x,v),E_{1}^{i}(x),\psi ^{i}(x)]=0$. It follows
that $c_{1}=\dots =c_{n}=0$ by the linear independence of $%
[f^{i}(x,v),E_{1}^{i}(x),\psi ^{i}(x)]_{i=1}^{l}$. This proves our claim
that $\left\{ \psi ^{i}(x)\right\} _{i=1}^{l}$ is linearly independent.

If $l>k$ , there exists a linear combination $\psi ^{d}\left( x\right)
=\sum_{i=1}^{l}d_{i}\psi ^{i}(x)$ with a nonzero vector $\left( d_{1},\cdots
,d_{l}\right) \in \mathbb{R}^{l},$ such that $\psi ^{d}\perp \zeta _{j}$ for
any $1\leq j\leq k$. Using the equation (\ref{zero-I-combination}) for $\psi
^{d}$, we have 
\begin{align}
0& =J\left( f^{d},E_{1}^{d},\psi ^{d}\right) -\iint \hat{v}_{2}\mu _{p}\left[
\psi ^{d}\right] ^{2}dvdx  \label{zero-I-combination-d} \\
& +\int \left[ \psi _{x}^{d}\right] ^{2}dx+\int \left( \sum_{i=1}^{l}\lambda
_{i}d_{i}\psi ^{i}\right) ^{2}dx.  \notag
\end{align}%
Now by Lemma \ref{lemma-invariant 1.5}, for all $g\in \ker D$ each
functional 
\begin{equation*}
K_{g}(f^{i},\psi ^{i})=\iint [f^{i}+(\hat{v}_{2}\mu _{e}+\mu _{p})\psi
^{i}]\ g\ dvdx
\end{equation*}%
vanishes, so that $K_{g}(f^{d},\psi ^{d})=0$. Thus by Lemma 2.8 of \cite%
{lw-linear}, we have 
\begin{equation*}
J\left( f^{d},E_{1}^{d},\psi ^{d}\right) \geq \iint \left\vert \mathcal{P}(%
\hat{v}_{2}\psi ^{d})\right\vert ^{2}|\mu _{e}|dvdx+((\mathcal{B}^{0})^{\ast
}(\mathcal{A}_{1}^{0})^{-1}\mathcal{B}^{0}\psi ^{d},\psi ^{d})
\end{equation*}%
and (\ref{zero-I-combination-d}) implies that%
\begin{align*}
0& \geq ((\mathcal{B}^{0})^{\ast }(\mathcal{A}_{1}^{0})^{-1}\mathcal{B}%
^{0}\psi ^{d},\psi ^{d}) \\
& \text{ \ \ }+\iint \left\{ |\mu _{e}||\mathcal{P}(\hat{v}_{2}\psi
^{d})|^{2}-\hat{v}_{2}\mu _{p}\left[ \psi ^{d}\right] ^{2}\right\} dvdx+\int %
\left[ \psi _{x}^{d}\right] ^{2}dx+\int \left( \sum_{i=1}^{l}\lambda
_{i}d_{i}\psi ^{i}\right) ^{2}dx \\
& =((\mathcal{B}^{0})^{\ast }(\mathcal{A}_{1}^{0})^{-1}\mathcal{B}^{0}\psi
^{d},\psi ^{d})+(\mathcal{A}_{2}^{0}\psi ^{d},\psi ^{d})+\int \left(
\sum_{i=1}^{l}\lambda _{i}d_{i}\psi ^{i}\right) ^{2}dx \\
& =\left( \mathcal{L}^{0}\psi ^{d},\psi ^{d}\right) +\int \left(
\sum_{i=1}^{l}\lambda _{i}d_{i}\psi ^{i}\right) ^{2}dx.
\end{align*}%
Since $\left( \mathcal{L}^{0}\psi ^{d},\psi ^{d}\right) \geq 0$, we deduce
that $\sum_{i=1}^{l}\lambda _{i}d_{i}\psi ^{i}=0$. So $\left\{ \psi
^{i}(x)\right\} _{i=1}^{l}$ is linearly dependent, which is a contradiction.
Therefore $l=k$. This completes the proof of Theorem \ref{thm-1.5d-growing
(1)}.
\end{proof}

%%%%%%%%%%%%%%%%%%%%%%%%%%%%%%%%%%%%%%%%%%%%%%%%%%
%%%%%%%%%%%%% Section 3 %%%%%%%%%%%%%%%%%%%%%%%%%%%%%%%

\section{Formulation of the 3D problem}

The 3D RVM for a non-neutral electron plasma with external fields is 
\begin{equation*}
\partial_{t}f+\hat{v}\cdot\nabla_{x}f-(\mathbf{E}+\mathbf{E}^{ext}+\hat {v}%
\times\left( \mathbf{B}+\mathbf{B}^{ext}\right) )\cdot\nabla_{v}f=0
\end{equation*}%
\begin{equation*}
\partial_{t}\mathbf{E}-\nabla\times\mathbf{B}=\int\hat{v}f\text{ }dv=-%
\mathbf{j}
\end{equation*}%
\begin{equation*}
\partial_{t}\mathbf{B}+\nabla\times\mathbf{E}=0
\end{equation*}%
\begin{equation*}
\nabla\cdot\mathbf{E}=-\int f\text{ }dv=\rho,\quad\nabla\cdot\mathbf{B}=0
\end{equation*}
where $x\in\mathbb{R}^{3},v\in\mathbb{R}^{3}$. We consider solutions of
finite energy. Thus they vanish in some averaged sense as $%
|x|\rightarrow\infty$.

We use the same notation as in \cite{lw-linear}. The cylindrical coordinates
in $\mathbb{R}^3$ are $(r,\theta,z)$ and the standard cylindrical basis is $%
\mathbf{e}_r,\mathbf{e}_\theta,\mathbf{e}_z$. The \textit{equilibrium}
distribution function is assumed to have the form $f^{0}=\mu\left(
e,p\right) ,$ with 
\begin{equation*}
e=\sqrt{1+\left\vert v\right\vert ^{2}}-\phi^{0}\left( r,z\right)
-\phi^{ext}\left( r,z\right) ,
\end{equation*}%
\begin{equation*}
p=r\left( v_{\theta}-A_{\theta}^{0}\left( r,z\right) -A_{\theta}^{ext}\left(
r,z\right) \right)
\end{equation*}
and the equilibrium fields are assumed to have the form 
\begin{equation*}
\mathbf{E}^{0}=-\partial_{r}\phi^{0}\mathbf{e}_{r}-\partial_{z}\phi^{0}%
\mathbf{e}_{z},\text{ \ }\mathbf{B}^{0}=-\partial_{z}A_{\theta}^{0}\mathbf{e}%
_{r}+\tfrac{1 }{r}\partial_{r}\left( rA_{\theta}^{0}\right) \mathbf{e}_{z},
\end{equation*}
with $\left( A_{\theta}^{0},\phi^{0}\right) $ satisfying the elliptic system
(\ref{eqn-equi-3d-ele}), (\ref{eqn-equi-3d-mag}). We assume $f^{0}$ has
compact support $S$ in $(x,v)$ space and $f^{0},E^{0},B^{0}$ are everywhere $%
C^1$. Such equilibria were constructed in the appendix of \cite{lw-linear}
for certain $\phi^{ext}$, $A_{\theta}^{ext}$ and $\mu$. We assume that 
\begin{equation*}
\mu_{e}<0 \quad \text { on the set } \{\mu>0\}.
\end{equation*}

For the perturbations $\mathbf{E}, \mathbf{B}$ of the electromagnetic
fields, we introduce scalar and vector potentials $\phi$ and $\mathbf{A}$
such that 
\begin{equation*}
\mathbf{E}=-\nabla\phi-\partial_{t}\mathbf{{A}\ \ \text{ }}\text{and } 
\mathbf{\ \ B=\nabla\times{A}}
\end{equation*}
and we impose the Coulomb gauge $\nabla\cdot\mathbf{A}=0$. We will consider
only \textit{axisymmetric} perturbations. In cylindrical coordinates we
write $\mathbf{A }=A_{r}\mathbf{e}_{r}+A_{\theta}\mathbf{e}_{\theta}+A_{z} 
\mathbf{e}_{z}$. We assume that $A_{r},$ $A_{\theta},$ $A_{z}$ and $\phi$ 
\textit{independent} of $\theta$. Some differentiation rules in cylindrical
coordinates are collected in the appendix. Then the corresponding fields are
given by 
\begin{equation*}
\mathbf{E}=\left( E_{r},E_{\theta},E_{z}\right) =\left( -\partial_{r}\phi-
\partial_t A_{r},-\partial_t A_\theta ,-\partial_{z}\phi -\partial_t
A_{z}\right) ,
\end{equation*}
\begin{equation*}
\mathbf{B}=\left( B_{r},B_{\theta},B_{z}\right) =\left( -\partial
_{z}A_{\theta},\ \partial_{z}A_{r}-\partial_{r}A_{z},\ \tfrac{1}{r}\partial
_{r}\left( rA_{\theta}\right) \right) .
\end{equation*}

Then the linearized Vlasov equation becomes 
\begin{equation}
\partial_{t}f+Df=-\mu_{e}D\phi-\mu_{e}\hat{v}\cdot\partial_{t}\mathbf{A }%
-r\mu_{p}\partial_{t}A_{\theta}-\mu_{p}D\left( rA_{\theta}\right) ,
\label{lin-3dvla}
\end{equation}
where 
\begin{equation*}
D=\hat{v}\cdot\nabla_{x}-\left( \mathbf{E}^{0}+\mathbf{E}^{ext}+\hat{v}%
\times\left( \mathbf{B}^{0}+\mathbf{B}^{ext}\right) \right) \cdot\nabla _{v}
\end{equation*}
(see the appendix). 
%%%% Note: we should explain in the Appendix how to get this equation.  %%%%%%
The Maxwell equations become the scalar equation 
\begin{equation}
\Delta\phi=-\rho=-\int fdv  \label{lin-maxwell-density}
\end{equation}
together with the vector equation 
\begin{equation}
\frac{\partial^{2}}{\partial t^{2}}\mathbf{A }+\frac{\partial}{\partial t}%
\nabla\phi-\Delta\mathbf{A }=\mathbf{j}=-\int\hat{v}fdv
\label{lin-maxwell-current}
\end{equation}

We are looking for a axisymmetric growing mode $\left[ e^{\lambda
t}f(x,v),e^{\lambda t}\mathbf{E}(x),e^{\lambda t}\mathbf{B}(x)\right] $,
which means we replace $\partial _{t}$ by $\lambda $ everywhere. Here $\text{%
Re} \lambda >0$ and $\left( \mathbf{E},\mathbf{B}\right) $ is independent of 
$\theta $. By Theorem \ref{3d theorem} of \cite{lw-linear}, $\lambda $ must
be real and so $\lambda >0$. Because of the Coulomb gauge condition, we have 
\begin{equation*}
0=\nabla \cdot \mathbf{A =}\frac{1}{r}\frac{\partial \left( rA_{r}\right) }{%
\partial r}+\frac{\partial A_{z}}{\partial z},
\end{equation*}%
so that we can introduce a \textit{super-potential }$\pi \left( r,z\right) $
such that 
\begin{equation*}
A_{r}=-\partial _{z}\pi \quad A_{z}=\frac{1}{r}\partial _{r}\left( r\pi
\right) =\partial _{r}\pi +\frac{1}{r}\pi .
\end{equation*}
Replacing $\partial_{t}$ by $\lambda$ and substituting $\hat v\cdot\mathbf{A 
}= \hat v_\theta A_\theta - \hat v_r(-\partial_z\pi) + \hat v_z
(\partial_r\pi+\frac1r\pi) $, we rewrite the Vlasov equation (\ref{lin-3dvla}%
) as 
\begin{align}
\left( \lambda+D\right) f & =-\mu_{e}D\phi-\left( \lambda+D\right) \left(
r\mu_{p}A_{\theta}\right) -\mu_{e}\lambda\hat{v}_{\theta}A_{\theta }
\label{formula-f-opeator} \\
& -\mu_{e}\lambda\left[ -\hat{v}_{r}\partial_{z}+\hat{v}_{z}\left(
\partial_{r}+\tfrac{1}{r}\right) \right] \pi.  \notag
\end{align}

We can explicitly invert the operator $(\lambda+D)$ by introducing the
particle paths $(X(t;x,v),V(t;x,v))$, which are the characteristics of $D$.
They are defined as the solutions of the ODE 
\begin{equation}
{\dot{X}=\hat{V},}\text{ }\quad{\dot{V}=-}\left( \mathbf{E}^{0}+\mathbf{E}%
^{ext}\right) \left( X\right) -\hat{V}\times\left( \mathbf{B}^{0}+\mathbf{B}%
^{ext}\right) \left( X\right)  \label{trajectory}
\end{equation}
with the initial conditions $X(0)=x,\ V(0)=v$. Integrating (\ref%
{formula-f-opeator}) along the path from $t=-\infty$ to $t=0$, we get 
\begin{align}
f\left( x,v\right) & =-\mu_{e}\phi+\mu_{e}\int_{-\infty}^{0}\lambda
e^{\lambda s}\phi\left( X(s)\right) ds-\mu_{p}rA_{\theta}
\label{formula-f-integral} \\
& -\mu_{e}\int_{-\infty}^{0}\lambda e^{\lambda s}{\hat{V}}_{\theta}\left(
s\right) A_{\theta}\left( X(s)\right) ds  \notag \\
& -\mu_{e}\int_{-\infty}^{0}\lambda e^{\lambda s}\left\{ -{\hat{V}}%
_{r}\left( s\right) \partial_{z}\pi\left( X(s)\right) +{\hat{V}}_{z}\left(
s\right) \left( \partial_{r}+\tfrac{1}{r}\right) \pi\left( X(s)\right)
\right\} ds.  \notag
\end{align}

%%%%%%%%%% DEFINITIONS %%%%%%%%%%%%%%%%%%%%%
Now it is convenient to introduce the following operators depending on a
positive parameter $\lambda $. These operators will be used throughout the
rest of the paper. For $k=k\left( x,v\right) $ define 
\begin{equation*}
\left( \mathcal{Q}^{\lambda }k\right) \left( x,v\right) =\int_{-\infty
}^{0}\lambda e^{\lambda s}k\left( X(s;x,v),V(s;x,v)\right) ds
\end{equation*}%
and 
\begin{equation*}
Gk=-\hat{v}_{r}\partial _{z}k+\hat{v}_{z}\left( \partial _{r}+\tfrac{1}{r}%
\right) k\ ,\quad G^{\ast }k=\hat{v}_{r}\partial _{z}k-\hat{v}_{z}\partial
_{r}k.
\end{equation*}%
For $h=h\left( r,z\right) $, define all of the following operators. 
\begin{equation*}
\mathcal{A}_{1}^{\lambda }h=-\Delta h-\left( \int \mu _{e}dv\right) h+\int
\mu _{e}\mathcal{Q}^{\lambda }h\ dv
\end{equation*}%
\begin{equation*}
\mathcal{A}_{2}^{\lambda }h=\left( -\Delta +\frac{1}{r^{2}}+\lambda
^{2}\right) h-r\left( \int \hat{v}_{\theta }\mu _{p}dv\right) h-\int \hat{v}%
_{\theta }\mu _{e}\mathcal{Q}^{\lambda }\left( \hat{v}_{\theta }h\right) dv
\end{equation*}%
\begin{equation*}
\mathcal{B}^{\lambda }h=-\left( \int \hat{v}_{\theta }\mu _{e}dv\text{ }%
\right) h+\int \mu _{e}\mathcal{Q}^{\lambda }\left( \hat{v}_{\theta
}h\right) dv
\end{equation*}%
\begin{equation*}
\mathcal{L}^{\lambda }=\left( \mathcal{B}^{\lambda }\right) ^{\ast }\left( 
\mathcal{A}_{1}^{\lambda }\right) ^{-1}\mathcal{B}^{\lambda }+A_{2}^{\lambda
}\ 
\end{equation*}%
\begin{equation*}
\mathcal{C}^{\lambda }h=\int \hat{v}_{\theta }\mu _{e}\mathcal{Q}^{\lambda
}\left( Gh\right) dv,\quad \left( \mathcal{C}^{\lambda }\right) ^{\ast
}h=\int G^{\ast }\left( \mu _{e}\mathcal{Q}^{\lambda }\left( \hat{v}_{\theta
}h\right) \right) dv
\end{equation*}%
\begin{equation*}
\mathcal{D}^{\lambda }h=\int \mu _{e}\mathcal{Q}^{\lambda }\left( Gh\right)
dv,\quad \left( \mathcal{D}^{\lambda }\right) ^{\ast }h=-\int G^{\ast
}\left( \mu _{e}\mathcal{Q}^{\lambda }\left( h\right) \right) dv
\end{equation*}%
\begin{equation*}
\mathcal{E}^{\lambda }h=\int G^{\ast }\left( \mu _{e}\mathcal{Q}^{\lambda
}\left( Gh\right) \right)
\end{equation*}%
\begin{equation*}
\mathcal{F}^{\lambda }=\left( \mathcal{D}^{\lambda }\right) ^{\ast }\left( 
\mathcal{A}_{1}^{\lambda }\right) ^{-1}\mathcal{B}^{\lambda }-\left( 
\mathcal{C}^{\lambda }\right) ^{\ast }.
\end{equation*}%
\begin{equation*}
\mathcal{G}^{\lambda }=\mathcal{E}^{\lambda }+(\mathcal{D}^{\lambda })^{\ast
}(\mathcal{A}_{!}^{\lambda })^{-1}\mathcal{D}^{\lambda }
\end{equation*}%
\begin{equation*}
\mathcal{A}_{3}^{\lambda }=\left( -\Delta +\tfrac{1}{r^{2}}\right) \left(
-\Delta +\tfrac{1}{r^{2}}+\lambda ^{2}\right) -\mathcal{E}^{\lambda }
\end{equation*}%
\begin{equation*}
\mathcal{A}_{4}^{\lambda }=\mathcal{A}_{3}^{\lambda }-\left( \mathcal{D}%
^{\lambda }\right) ^{\ast }\left( \mathcal{A}_{1}^{\lambda }\right) ^{-1}%
\mathcal{D}^{\lambda }=\left( -\Delta +\tfrac{1}{r^{2}}\right) \left(
-\Delta +\tfrac{1}{r^{2}}+\lambda ^{2}\right) -\mathcal{G}^{\lambda }
\end{equation*}%
Here these operators are defined formally. In the next section, they will be
defined carefully and key properties will be derived.

Using these operators, we can rewrite the formula (\ref{formula-f-integral})
as 
\begin{equation}
f=-\mu _{e}\phi +\mu _{e}\mathcal{Q}^{\lambda }\phi -\mu _{p}rA_{\theta
}-\mu _{e}\mathcal{Q}^{\lambda }\left( \hat{v}_{\theta }A_{\theta }\right)
-\mu _{e}\mathcal{Q}^{\lambda }\left( G\pi \right) .
\label{formula-f-i-operator}
\end{equation}%
Moreover, substituting (\ref{formula-f-i-operator}) into the Poisson
equation $-\Delta \phi =\int fdv$, we obtain%
\begin{align*}
-\Delta \phi & =-\left( \int \mu _{e}dv\right) \phi +\int \mu _{e}\mathcal{Q}%
^{\lambda }\phi dv-\left( \int \mu _{p}dv\right) rA_{\theta } \\
& -\int \mu _{e}\mathcal{Q}^{\lambda }\left( \hat{v}_{\theta }A_{\theta
}\right) dv-\int \mu _{e}\mathcal{Q}^{\lambda }\left( G\pi \right) dv.
\end{align*}%
Since $r\int \mu _{p}dv=-\int \hat{v}_{\theta }\mu _{e}dv$, this result can
be written as 
\begin{equation}
\mathcal{A}_{1}^{\lambda }\phi =\mathcal{B}^{\lambda }A_{\theta }+\mathcal{D}%
^{\lambda }\pi .  \label{equation-phi-3d}
\end{equation}%
With $\partial _{t}$ replaced by $\lambda $, the Maxwell equation (\ref%
{lin-maxwell-current}) becomes%
\begin{equation}
\lambda ^{2}\mathbf{A}+\lambda \nabla \phi -\Delta \mathbf{A}=\mathbf{{j}.}
\label{equation-current-lambda}
\end{equation}%
Taking the $\theta -$component of (\ref{equation-current-lambda}) and
substituting (\ref{formula-f-opeator}), 
\begin{align*}
\left( \lambda ^{2}-\Delta \right) A_{\theta }& =-\int \hat{v}_{\theta }fdv
\\
& =\left( \int \hat{v}_{\theta }\mu _{e}dv\right) \phi -\int \hat{v}_{\theta
}\mu _{e}\mathcal{Q}^{\lambda }\phi dv+\left( \int \hat{v}_{\theta }\mu
_{p}dv\right) rA_{\theta } \\
& \ \ \ \ +\int \hat{v}_{\theta }\mu _{e}\mathcal{Q}^{\lambda }\left( \hat{v}%
_{\theta }A_{\theta }\right) dv+\int \hat{v}_{\theta }\mu _{e}\mathcal{Q}%
^{\lambda }\left( G\pi \right) dv.
\end{align*}%
That is,%
\begin{equation}
\mathcal{A}_{2}^{\lambda }A_{\theta }=-\left( \mathcal{B}^{\lambda }\right)
^{\ast }\phi +\mathcal{C}^{\lambda }\pi .  \label{equation-atheta-3d}
\end{equation}

%%%%%%%%%%%%% Lemma 3.1 %%%%%%%%%%%%%

\begin{lemma}
\label{lemma-3.1}%
\begin{equation}
\mathcal{A}_{3}^{\lambda}\pi=\left( \mathcal{D}^{\lambda}\right)
^{\ast}\phi-\left( \mathcal{C}^{\lambda}\right) ^{\ast}A_{\theta}.
\label{equation-pi-3d}
\end{equation}
\end{lemma}

\begin{proof}
First we claim that 
\begin{equation}
\left( -\Delta +\tfrac{1}{r^{2}}\right) \left( -\Delta +\tfrac{1}{r^{2}}%
+\lambda ^{2}\right) \pi =\partial _{z}j_{r}-\partial _{r}j_{z}.
\label{equation-pi-current}
\end{equation}%
Indeed, let $\mathbf{K}=j_{r}\mathbf{e}_{r}+j_{z}\mathbf{e}_{z}$ and $%
\mathbf{I}=\left( -\Delta \right) ^{-1}\mathbf{K}$ so that $\mathbf{e}%
_{\theta }\cdot \mathbf{I}=0$. By the continuity equation $\partial _{t}\rho
+\nabla \cdot \mathbf{j}=0$, for a growing mode we have 
\begin{equation*}
\nabla \cdot \mathbf{K}=\left( \partial _{r}+\tfrac{1}{r}\right)
j_{r}+\partial _{z}j_{z}=\nabla \cdot \mathbf{j}=-\lambda \rho =\lambda
\Delta \phi .
\end{equation*}%
Thus the vector identity 
\begin{equation*}
\nabla \times \left( \nabla \times \mathbf{K}\right) =-\Delta \mathbf{K}%
+\nabla \left( \nabla \cdot \mathbf{K}\right)
\end{equation*}%
takes the form%
\begin{equation*}
-\nabla \times \left( \nabla \times \Delta \mathbf{I}\right) =-\Delta 
\mathbf{K}+\lambda \nabla \Delta \phi
\end{equation*}%
or%
\begin{equation*}
\nabla \times \left( \nabla \times \mathbf{I}\right) =\vec{K}-\lambda \nabla
\phi .
\end{equation*}%
Now the $r$ and $z$ components of the Maxwell equation (\ref%
{equation-current-lambda}) can be written as 
\begin{equation*}
\left( \lambda ^{2}-\Delta \right) \left( A_{r}\mathbf{e}_{r}+A_{z}\mathbf{e}%
_{z}\right) =\mathbf{K}-\lambda \nabla \phi .
\end{equation*}%
Furthermore, 
\begin{equation*}
A_{r}\mathbf{e}_{r}+A_{z}\mathbf{e}_{z}=-(\partial _{z}\pi )\mathbf{e}%
_{r}-\left( \left( \partial _{r}+\tfrac{1}{r}\right) \pi \right) \mathbf{e}%
_{z}=\nabla \times \left( \pi \mathbf{e}_{\theta }\right) .
\end{equation*}%
Combining the last three equations, we have 
\begin{equation*}
\nabla \times \left( \lambda ^{2}-\Delta \right) \left( \pi \mathbf{e}%
_{\theta }\right) =\nabla \times \left( \nabla \times \mathbf{I}\right) ,
\end{equation*}%
which is satisfied if 
\begin{equation*}
\left( \lambda ^{2}-\Delta \right) \left( \pi \mathbf{e}_{\theta }\right)
=\nabla \times \mathbf{I}=\left( \partial _{z}I_{r}-\partial
_{r}I_{z}\right) \mathbf{e}_{\theta }.
\end{equation*}%
Noting that $\Delta \mathbf{e}_{\theta }=-\tfrac{1}{r^{2}}\mathbf{e}_{\theta
},$ we deduce 
\begin{equation*}
\left( \lambda ^{2}-\Delta +\tfrac{1}{r^{2}}\right) \pi =\partial
_{z}I_{r}-\partial _{r}I_{z}.
\end{equation*}%
Applying $-\Delta +\frac{1}{r^{2}}$ to this result yields 
\begin{align*}
\left( -\Delta +\tfrac{1}{r^{2}}\right) \left( -\Delta +\tfrac{1}{r^{2}}%
+\lambda ^{2}\right) \pi & =\partial _{z}\left( -\Delta +\tfrac{1}{r^{2}}%
\right) I_{r}-\partial _{r}\left( -\Delta \right) I_{z} \\
& =\partial _{z}j_{r}-\partial _{r}j_{z}
\end{align*}%
since $\left[ \partial _{r},-\Delta \right] =\frac{1}{r^{2}}\partial _{r}$.
This proves the claim.

Upon substituting (\ref{formula-f-i-operator}) into $j_r=\int\hat v_r f\,dv$%
, the first and third terms vanish because they are odd in $v_r$. The same
reasoning is valid for $j_z=\int\hat v_z f\,dv$. Therefore 
\begin{align}
\partial_{z}j_{r}-\partial_{r}j_{z} & =-\partial_{z}\int\hat{v}%
_{r}fdv+\partial_{r}\int\hat{v}_{z}fdv  \label{formula-curl-current} \\
& =-\partial_{z}\int\hat{v}_{r}\mu_{e}\mathcal{Q}^{\lambda}\phi\
dv+\partial_{r}\int\hat{v}_{z}\mu_{e}\mathcal{Q}^{\lambda}\phi\ dv  \notag \\
& -\partial_{z}\int\hat{v}_{r}\mu_{e}\mathcal{Q}^{\lambda}\left( \hat {v}%
_{\theta}A_{\theta}\right) dv+\partial_{r}\int\hat{v}_{z}\mu _{e}\mathcal{Q}%
^{\lambda}\left( \hat{v}_{\theta}A_{\theta}\right) dv  \notag \\
& +\partial_{z}\int\hat{v}_{r}\mu_{e}\mathcal{Q}^{\lambda}\left( G\pi\right)
dv-\partial_{r}\int\hat{v}_{z}\mu_{e}\mathcal{Q}^{\lambda}\left( G\pi\right)
dv.  \notag
\end{align}
The last four terms in (\ref{formula-curl-current}) equal 
\begin{align*}
& -\int G^{\ast}\left[ \mu_{e}\mathcal{Q}^{\lambda}\left( \hat{v}_{\theta
}A_{\theta}\right) \right] dv+\int G^{\ast}\left[ \mu_{e}\mathcal{Q}%
^{\lambda}\left( G\pi\right) \right] dv. \\
& =-\left( \mathcal{C}^{\lambda}\right) ^{\ast}A_{\theta}+\mathcal{E}%
^{\lambda}\pi
\end{align*}
In (\ref{formula-curl-current}) call the first two terms $T\left(\phi\right) 
$. Then 
\begin{align*}
\left( T\left( \phi\right) ,\psi\right) _{L^{2}\left( \mathbb{R }%
^{3}\right)} & =2\pi\iint T\left( \phi\right) \psi\,rdrdz \\
& =\left\langle G^{\ast}\left[ \mu_{e}\mathcal{Q}^{\lambda}\phi\right]
,\psi\right\rangle _{L^{2}\left( \mathbb{R }^{6}\right) }=\left\langle 
\mathcal{Q}^{\lambda}\phi,\mu_{e}G\psi\right\rangle \\
& =\left\langle \phi,\mu_{e}\mathcal{Q}^{\lambda}G\psi\right\rangle
\end{align*}
by Lemma 4.1(\textit{d}) below since $\phi$ is independent of $v$. The last
expression equals 
\begin{equation*}
\left\langle \phi,\mathcal{Q}^{\lambda}\left[ \mu_{e}G\psi\right]
\right\rangle =\left( \phi,\mathcal{D}^{\lambda}\psi\right) _{L^{2}\left( 
\mathbb{R }^{3}\right) }=\left( \left( \mathcal{D}^{\lambda}\right) ^{\ast
}\phi,\psi\right) _{L^{2}\left( \mathbb{R }^{3}\right) }.
\end{equation*}
So $T\left(\phi\right) =\left( \mathcal{D}^{\lambda}\right) ^{\ast}\phi$.
Thus by (\ref{equation-pi-current}) and (\ref{formula-curl-current}), 
\begin{equation*}
\left( -\Delta+\tfrac{1}{r^{2}}\right) \left( -\Delta+\tfrac{1}{r^{2}}%
+\lambda^{2}\right) \pi=\left( \mathcal{D}^{\lambda}\right)
^{\ast}\phi-\left( \mathcal{C}^{\lambda}\right) ^{\ast}A_{\theta}+\mathcal{E}%
^{\lambda}\pi
\end{equation*}
Hence 
\begin{equation*}
\mathcal{A}_{3}^{\lambda}\pi=\left( \mathcal{D}^{\lambda}\right)
^{\ast}\phi-\left( \mathcal{C}^{\lambda}\right) ^{\ast}A_{\theta}.
\end{equation*}
\end{proof}

We now have three equations (\ref{equation-phi-3d}), (\ref%
{equation-atheta-3d}) and (\ref{equation-pi-3d}) that link the unknowns $%
\phi,A_{\theta}$ and $\pi$. Using (\ref{equation-phi-3d}) to eliminate $\phi$%
, we obtain 
\begin{equation*}
\mathcal{A}_{2}^{\lambda}A_{\theta}=-\left( \mathcal{B}^{\lambda}\right)
^{\ast}\left( \mathcal{A}_{1}^{\lambda}\right) ^{-1}[\mathcal{B}^{\lambda
}A_{\theta}+\mathcal{D}^{\lambda}\pi]+\mathcal{C}^{\lambda}\pi,
\end{equation*}

\begin{equation*}
\mathcal{A}_{3}^{\lambda}\pi=\left( \mathcal{D}^{\lambda}\right) ^{\ast
}\left( \mathcal{A}_{1}^{\lambda}\right) ^{-1}[\mathcal{B}^{\lambda
}A_{\theta}+\mathcal{D}^{\lambda}\pi]-\left( \mathcal{C}^{\lambda}\right)
^{\ast}A_{\theta}.
\end{equation*}
That is 
\begin{equation}
\mathcal{L}^{\lambda}A_{\theta}=-\left( \mathcal{F}^{\lambda}\right) ^{\ast
}\pi\text{, \ }  \label{equation-atheta-reduced}
\end{equation}
and 
\begin{equation}
\mathcal{A}_{4}^{\lambda}\pi=\mathcal{F}^{\lambda}A_{\theta}.
\label{equation-pi-reduced}
\end{equation}
These are the basic reduced equations of which we want to find a non-zero
solution. Motivated by (\ref{equation-atheta-reduced}) and (\ref%
{equation-pi-reduced}), we define the matrix operator 
\begin{equation}
\mathcal{M}^{\lambda}=\left( 
\begin{array}{cc}
\mathcal{L}^{\lambda} & \left( \mathcal{F}^{\lambda}\right) ^{\ast} \\ 
\mathcal{F}^{\lambda} & -\mathcal{A}_{4}^{\lambda}%
\end{array}
\right)  \label{matrix-operator}
\end{equation}
of which we want to find a non-trivial nullspace.

%%%%%%%%%%%%%%  Section 4  %%%%%%%%%%%%%%%%%
%%%%%%%%%%%%%%  Section 4  %%%%%%%%%%%%%%%%%

\section{The Operators}

Let the space ${L}_{S}^{2}$ consist of the cylindrically symmetric functions
(functions of $r$ and $z$ only) in ${L}^{2}\left( \mathbb{R}^{3}\right) $.
For any positive integer $k$, let %\begin{align*}
\begin{equation*}
H^{k\dagger }=\left\{ \psi \in L_{S}^{2}\left( \mathbb{R}^{3}\right) \text{ }%
\Big|\text{ }e^{i\theta }\psi \in H^{k}\left( \mathbb{R}^{3}\right) \right\}
\end{equation*}%
and $\left\Vert \psi \right\Vert _{H^{k\dagger }}=\left\Vert e^{i\theta
}\psi \right\Vert _{\mathbf{H}^{k}\left( \mathbb{R}^{3}\right) }^{2}$.
Furthermore, we define $V^{k\dag }$ to be the closure of the cylindrically
symmetric functions in $C_{c}^{\infty }(\mathbb{R}^{3})$ with respect to the 
$\dot{H}^{k}$ semi-norm 
\begin{equation*}
\Vert \psi \Vert _{V^{k\dag }}^{2}=\sum_{|\alpha |=k}\Vert \partial ^{\alpha
}\left( e^{i\theta }\psi \right) \Vert _{L^{2}}^{2}\ 
\end{equation*}%
We denote ${H}^{-k\dagger }=\left( {H}^{k\dagger }\right) ^{\ast }$ and $%
V^{-k\dagger }=\left( V^{k\dagger }\right) ^{\ast }$. It follows easily that 
$\psi \left( r,z\right) \in {H}^{1\dagger }$ is equivalent to $\psi ,\psi
_{r},\psi _{z},\psi /r\in {L}^{2}\left( \mathbb{R}^{3}\right) $.
Furthermore, $\psi \left( r,z\right) \in {H}^{2\dagger }$ is equivalent to $%
\psi ,\psi _{rr},\psi _{zz},\left( \psi /r\right) _{r}\in {L}^{2}\left( 
\mathbb{R}^{3}\right) $, and such a function also satisfies $\psi _{r},\psi
_{z},\psi /r$ $\in {L}^{2}\left( \mathbb{R}^{3}\right) $. We also define the
space $W^{2\dagger }=V^{2\dagger }\cap V^{1\dagger }$ with the norm 
\begin{equation*}
\Vert \psi \Vert _{W^{2\dagger }}=\Vert \Delta \left( e^{i\theta }\psi
\right) \Vert _{L^{2}}+\Vert \nabla \left( e^{i\theta }\psi \right) \Vert
_{L^{2}}
\end{equation*}%
and $W^{-2\dagger }=\left( W^{2\dagger }\right) ^{\ast }$. We also denote $%
V^{k}$ to be the closure of the functions in $C_{c}^{\infty }(\mathbb{R}%
^{3}) $ with respect to the norm 
\begin{equation*}
\Vert \psi \Vert _{V^{k}}^{2}=\sum_{|\alpha |=k}\Vert \partial ^{\alpha
}\psi \Vert _{L^{2}}^{2}
\end{equation*}%
and $V^{-k}=\left( V^{k}\right) ^{\ast }$.

As noted by F. H. Lin (see \cite{lw-linear}), for any function $\psi\left(
r,z\right) $ we have 
\begin{equation}
-\Delta\left( \psi e^{i\theta}\right) = e^{i\theta} \left(
-\partial_{zz}\psi -\partial_{rr}\psi-\frac{1}{r}\partial_{r}\psi+\frac{1}{%
r^{2}}\psi\right) .  \label{singular-remove}
\end{equation}
We can apply the usual elliptic regularity theorem to the operator $%
-\partial_{zz}-\partial_{rr}-\frac{1}{r}\partial_{r}+\frac{1}{r^{2}}$ and
the singular factor $1/r^{2}$ is artificial, introduced merely by the change
of coordinates. The daggered spaces are designed to take account of this
singular factor.

We denote by $\left\vert {\ \ }\right\vert _{2}$ the norm in $%
L_{S}^{2}\left( \mathbb{R }^{3}\right) $, by $\left( \ ,\right) $ the inner
product in $L_{S}^{2}\left( \mathbb{R }^{3}\right) $, by $\left\langle \
,\right\rangle $ the pairing of dual spaces, and by $\left\langle \text{ }%
,\right\rangle _{\left\vert \mu_{e}\right\vert }$ the inner product in $%
L_{\left\vert \mu _{e}\right\vert }^{2}\left( \mathbb{R }^{6}\right) $ where 
$\left\vert \mu_{e}\left( x,v\right) \right\vert $ is the weight with $%
\left\Vert {\ \ }\right\Vert _{\left\vert \mu_{e}\right\vert }$ the
corresponding norm. We defined the operator $\mathcal{Q}^{\lambda}$ in the
previous section.

%%%%%%%%% Lemma 4.1 %%%%%%%%

\begin{lemma}
\label{lemma-property-Qlambda}(Properties of $\mathcal{Q}^{\lambda}$) Let $%
0<\lambda<\infty.$

(a) $\mathcal{Q}^{\lambda}:L_{\left\vert \mu_{e}\right\vert }^{2}\left( 
\mathbb{R }^{6}\right) \rightarrow L_{\left\vert \mu_{e}\right\vert
}^{2}\left( \mathbb{R }^{6}\right) $ with operator norm $=1$.

(b) For all $m\in L_{\left\vert \mu_{e}\right\vert }^{2}\left( \mathbb{R }%
^{6}\right) $, $\left\Vert \mathcal{Q}^{\lambda}m-\mathcal{P}m\right\Vert
_{\left\vert \mu_{e}\right\vert }\rightarrow0$ as $\lambda\rightarrow0$,
where $\mathcal{P}$ is defined in the introduction.

(c) If $\sigma>0$, then $\left\Vert \mathcal{Q}^{\lambda}-\mathcal{Q}%
^{\sigma }\right\Vert =O\left( \left\vert \lambda-\sigma\right\vert \right) $
as $\lambda\rightarrow\sigma$, where $\left\Vert {\ }\right\Vert $ denotes
the operator norm from $L_{\left\vert \mu_{e}\right\vert }^{2}$ to $%
L_{\left\vert \mu_{e}\right\vert }^{2}$.

(d) For $v=v_{r}\mathbf{e}_{r}+v_{\theta}\mathbf{e}_{\theta}+v_{z}\mathbf{e}%
_{z}$, denote $\tilde{v}=-v_{r}\mathbf{e}_{r}+v_{\theta}\mathbf{e}%
_{\theta}-v_{z}\mathbf{e}_{z}$ and $\tilde{n}\left( x,v\right) =n\left( x,%
\tilde{v}\right) $. Then $\left\langle \mathcal{Q}^{\lambda}m,n\right\rangle
_{\left\vert \mu _{e}\right\vert }=\left\langle m,\mathcal{Q}^{\lambda}%
\tilde{n}\right\rangle _{\left\vert \mu_{e}\right\vert }$, for any $m,n\in
L_{\left\vert \mu _{e}\right\vert }^{2}\left( \mathbb{R }^{6}\right) $.

(e) For all $m\in L_{\left\vert \mu_{e}\right\vert }^{2}\left( \mathbb{R }%
^{6}\right) $, $\left\Vert \mathcal{Q}^{\lambda}m-m\right\Vert _{\left\vert
\mu_{e}\right\vert }\rightarrow0$ as $\lambda\rightarrow+\infty$.
\end{lemma}

\begin{proof}
To prove (a), 
\begin{align*}
\left\langle \mathcal{Q}^{\lambda}m,n\right\rangle _{\left\vert \mu
_{e}\right\vert } & =\int_{-\infty}^{0}\lambda e^{\lambda s}\iint\left( m%
\sqrt{\left\vert \mu_{e}\right\vert }\right) \left( X\left( s\right)
,V\left( s\right) \right) \cdot \left( n\sqrt{\left\vert \mu_{e}\right\vert }%
\right) \left( x,v\right) dvdxds \\
& \leq\left\Vert m\right\Vert _{\left\vert \mu_{e}\right\vert }\left\Vert
n\right\Vert _{\left\vert \mu_{e}\right\vert }.
\end{align*}
Moreover, $\mathcal{Q}^{\lambda} 1 = 1$.

Assertion (b) was proven in Lemma 2.6 of \cite{lw-linear}. As for (c), we
estimate 
\begin{align*}
\left\Vert \mathcal{Q}^{\lambda}m-\mathcal{Q}^{\sigma}m\right\Vert
_{\left\vert \mu_{e}\right\vert } & \leq\int_{-\infty}^{0}\left\vert \lambda
e^{\lambda s}-\sigma e^{\sigma s}\right\vert \left\Vert m\left( X\left(
s\right) ,V\left( s\right) \right) \right\Vert _{\left\vert \mu
_{e}\right\vert }ds \\
& =\int_{-\infty}^{0}\left\vert \lambda e^{\lambda s}-\sigma e^{\sigma
s}\right\vert ds\left\Vert m\right\Vert _{\left\vert \mu_{e}\right\vert } \\
\, & \leq C\left\vert \ln\lambda-\ln\sigma\right\vert \left\Vert
m\right\Vert _{\left\vert \mu_{e}\right\vert }.
\end{align*}

To prove (d), note that the characteristic ODE is invariant under the
transformation $s\rightarrow-s,$\ $r\rightarrow+r,$\ $z\rightarrow +z,\
v_{r}\rightarrow-v_{r},\ v_{\theta}\rightarrow+v_{\theta},\
v_{z}\rightarrow-v_{e}.$ Thus 
\begin{equation*}
n(X(-s;x,v),V(-s;x,v))=\tilde{n}\left( X(s;x,v),V(s;x,v)\right) .
\end{equation*}
Now%
\begin{equation*}
\left\langle \mathcal{Q}^{\lambda}m,n\right\rangle _{\left\vert \mu
_{e}\right\vert }=\int_{-\infty}^{0}\lambda e^{\lambda s}\iint\left\vert
\mu_{e}\right\vert m\left( X(s),V(s)\right) n\left( x,v\right) dvdxds.
\end{equation*}
We change variables $\left( X(s),V(s)\right) \rightarrow\left( x,v\right) $
and $\left( x,v\right) \rightarrow\left( X(-s),V(-s)\right) $ with Jacobian $%
=1$ to obtain 
\begin{align*}
\left\langle \mathcal{Q}^{\lambda}m,n\right\rangle _{\left\vert \mu
_{e}\right\vert } & =\int_{-\infty}^{0}\lambda e^{\lambda s}\int
\int\left\vert \mu_{e}\right\vert m\left( x,v\right) \tilde{n}\left(
X(-s),V(-s)\right) dvdxds \\
& =\left\langle m,\mathcal{Q}^{\lambda}\tilde{n}\right\rangle _{\left\vert
\mu_{e}\right\vert }.
\end{align*}

Although assertion (e) was essentially proven in Lemma 2.6 of \cite%
{lw-linear}, we outline the proof here. Letting $M$ denote the spectral
measure of the self-adjoint operator $-iD$ in the space $L^2_{|\mu_e|}$, we
have 
\begin{equation*}
\mathcal{Q}^\lambda m - m = \int_\mathbb{R }\left( \frac{\lambda}{%
\lambda+i\alpha} - 1 \right) dM(\alpha) m.
\end{equation*}
Thus 
\begin{equation*}
\| \mathcal{Q}^\lambda m - m \|_{|\mu_e|}^2 \leq \int_\mathbb{R }\left| 
\frac{\lambda}{\lambda+i\alpha} - 1 \right|^2 d\|M(\alpha)m\|_{|\mu_e|}^2\to
0
\end{equation*}
as $\lambda \to +\infty$.
\end{proof}

\begin{remark}
Since $\int_{-\infty }^{0}\lambda e^{\lambda s}ds=1$, the function%
\begin{equation*}
\left( \mathcal{Q}^{\lambda }m\right) \left( x,v\right) =\int_{-\infty
}^{0}\lambda e^{\lambda s}m\left( X(t;x,v),V(t;x,v)\right) ds
\end{equation*}%
is a weighted time average of the observable $m$ along the particle
trajectory. Lemma \ref{lemma-property-Qlambda} (b) tells us that as $\lambda
\rightarrow 0$, the limit of this weighted time average equals the phase
space average. This is the same as the ergodic theorem for the usual time
average, that is 
\begin{equation*}
\lim_{T\rightarrow \infty }\frac{1}{T}\int_{0}^{T}m\left(
X(t;x,v),V(t;x,v)\right) ds=\lim_{\lambda \rightarrow 0+}\mathcal{Q}%
^{\lambda }m=\mathcal{P}m.
\end{equation*}%
In particular, if the particle motion is ergodic in the set $S_{e,p}$
determined by the two invariants $e$ and $p$, and if $d\sigma _{e,p}$
denotes the induced measure on $S_{e,p}$, then 
\begin{equation*}
\mathcal{P}m=\frac{1}{\sigma _{e,p}\left( S_{e,p}\right) }%
\int_{S_{e,p}}m\left( x\right) d\sigma _{e,p}\left( x\right) .
\end{equation*}%
For non-ergodic particles, we do not have such an explicit expression, but $%
\mathcal{P}m$ still equals the phase space average of $m$ on the set traced
by the particle.
\end{remark}

%%%%%%%%%%%%%%% Lemma 4.2 %%%%%%%%%%%%

\begin{lemma}
\label{lemma-operator-property}Let $0<\lambda<\infty$.

(a) $\mathcal{B}^{\lambda}$ maps $L^{2}\rightarrow L^{2}$ with operator
bound independent of $\lambda$.

(b) $\mathcal{A}_{1}^{\lambda},\ \mathcal{A}_{2}^{\lambda}$ and $\mathcal{L}
^{\lambda}$ are self-adjoint on $L^{2}$ with domains $H^{2},\ H^{2\dagger}$
and $H^{2\dagger}$ respectively.

(c) The essential spectrum of $\mathcal{A}_{1}^{\lambda}$ is $[0,\infty)$,
while that of $\mathcal{A}_{2}^{\lambda}$ and $\mathcal{L}^{\lambda}$ is $%
[\lambda^{2},\infty).$

(d) $\left\langle \mathcal{A}_{1}^{\lambda}h,h\right\rangle >0$ for all $%
0\neq h\in H^{2}$.

(e) $\left( \mathcal{A}_{1}^{\lambda }\right) ^{-1}$ maps $V^{-1}$ into $%
V^{1}$ with operator bound $\leq 1$.

(f) For all $h\in L^{2},\ \left( \mathcal{L}^{\lambda}-\mathcal{L}
^{0}\right) h\rightarrow0$ strongly in $L^{2}$ as $\lambda\rightarrow0.$

(g) If $\sigma>0$, then as $\lambda\rightarrow\sigma$, the operator norm
from $L^{2}$ to $L^{2}$ of $\mathcal{A}_{2}^{\lambda}-\mathcal{A}%
_{2}^{\sigma}$ tends to zero. The same is true of $\mathcal{B}^{\lambda},\ 
\mathcal{A} _{1}^{\lambda},\ \left( \mathcal{A}_{1}^{\lambda}\right) ^{-1}$
and $\mathcal{L}^{\lambda}$.
\end{lemma}

\begin{proof}
Assertions (a), (b), (c), (d) and (f) were proven in Lemma 3.1 of \cite%
{lw-linear}. As for (e), let us define $\mathcal{A=A}_{1}^{\lambda }$ for
brevity. Let $\phi \in V^{1}$. Then 
\begin{align*}
\left\langle \mathcal{A}\phi ,\phi \right\rangle & =\left\vert \nabla \phi
\right\vert _{2}^{2}+\int \int \left\vert \mu _{e}\right\vert dv\phi
^{2}dx-\left\langle \mathcal{Q}^{\lambda }\phi ,\phi \right\rangle
_{\left\vert \mu _{e}\right\vert } \\
& \geq \left\vert \nabla \phi \right\vert _{L^{2}}^{2}=\left\Vert \phi
\right\Vert _{V^{1}}^{2}
\end{align*}%
by Lemma \ref{lemma-property-Qlambda}(a). Denoting $h=\mathcal{A}\phi $, we
therefore have 
\begin{equation*}
\left\Vert \mathcal{A}^{-1}h\right\Vert _{V^{1}}^{2}=\left\vert \nabla \phi
\right\vert _{2}^{2}\leq \left\langle \mathcal{A}\phi ,\phi \right\rangle
=\left\langle h,\mathcal{A}^{-1}h\right\rangle \leq \left\Vert h\right\Vert
_{V^{-1}}\left\Vert \mathcal{A}^{-1}h\right\Vert _{V^{1}}.
\end{equation*}%
Thus $\left\Vert \mathcal{A}^{-1}h\right\Vert _{V^{1}}\leq \left\Vert
h\right\Vert _{V^{-1}}$. Finally, Assertion (g) follows directly from Lemma %
\ref{lemma-property-Qlambda}(c).
\end{proof}

%%%%%%%%%%%%% Remark 1 %%%%%%%%%%

\begin{remark}
The supports are under control in the following sense. Recall that we assume 
$f^{0}\left( x,v\right) =\mu\left( e,p\right) $ has compact support $\subset
S\subset\mathbb{R }_{x}^{3}\times\mathbb{R }_{v}^{3}$. We may assume $%
S=S_{x}\times S_{v}$, both balls in $\mathbb{R }^{3}$. Let $\chi=\chi\left(
r,z\right) $ be a smooth cut-off function for the spatial support of $f^{0}$
in $S_{x}$; that is, $\chi=1$ on the spatial support of $f^0$ and has
compact support inside $S_x$. Let $M_\chi$ be the operator of multiplication
by $\chi$. Then 
\begin{equation*}
\mathcal{B}^{\lambda}=\mathcal{B}^{\lambda}M_\chi =M_\chi \mathcal{B}%
^{\lambda }=M_\chi \mathcal{B}^{\lambda}M_\chi
\end{equation*}
and the same is true for all the operators $\mathcal{C}^{\lambda },\ 
\mathcal{D}^{\lambda},\ \mathcal{E}^{\lambda},\ \mathcal{F}^{\lambda
},\left( \ \mathcal{C}^{\lambda}\right) ^{\ast},\ \left( \mathcal{D}%
^{\lambda}\right) ^{\ast},\ \left( \mathcal{F}^{\lambda}\right) ^{\ast}$.
Indeed, 
\begin{equation*}
\mu_{e}\left( x,v\right) =\mu_{e}\left( X(s;x,v),V(s;x,v)\right)
\end{equation*}
because of the invariance of $e$ and $p$ under the flow. So for example 
\begin{align*}
\left( \mathcal{B}^{\lambda}h\right) \left( x\right) & =-h\int\hat {v}%
_{\theta}\mu_{e}dv+\int\mu_{e}\mathcal{Q}^{\lambda}\left( \hat{v}_{\theta
}h\right) dv \\
& =-h\int\hat{v}_{\theta}\mu_{e}dv+\int\mathcal{Q}^{\lambda}\left( \mu _{e}%
\hat{v}_{\theta}h\right) dv = (M_\chi\mathcal{B}^\lambda M_\chi h)(x).
\end{align*}
\end{remark}

Below, for any function space $Y$, we denote by $Y_{c} = \{h\in Y\ | \
supp(h) \subset S_x \}$. Then $V_c^k = \dot{H}_{c}^{k}=H_{c}^{k}$ and $%
V_c^{k\dagger} = \dot{H}_{c}^{k\dagger}=H_{c}^{k\dagger}$. By mollification, 
$H_c^k$ is dense in $V^k$. Furthermore, $(H^{-k})_c \subset V^{-k}$. The
multiplication operator $M_\chi$ maps $V^1$ into $H^1$.

%%%%%%%%% Lemma 4.3 %%%%%%%%%%%%

\begin{lemma}
\label{operator bound}For any $\lambda>0,$

\begin{equation*}
\mathcal{C}^{\lambda},\ \mathcal{D}^{\lambda},\ \left( \mathcal{F}^{\lambda
}\right) ^{\ast}:H_{loc}^{1\dagger}\rightarrow L_{c}^{2}
\end{equation*}%
\begin{equation*}
\left( \ \mathcal{C}^{\lambda}\right) ^{\ast},\ \left( \mathcal{D}%
^{\lambda}\right) ^{\ast},\ \mathcal{F}^{\lambda}:L_{loc}^{2}\rightarrow
H_{c}^{-1\dagger}
\end{equation*}
\begin{equation*}
\mathcal{E}^{\lambda}:H_{loc}^{1\dagger}\rightarrow H_{c}^{-1\dagger}.
\end{equation*}
All these operator bounds are independent of $\lambda$. Furthermore, all
these operators are continuous functions of $\lambda\,$in the operator
norms. As $\lambda\rightarrow0+$, all these operators converge to $0$
strongly (but not in operator norm).
\end{lemma}

\begin{proof}
By the preceding remark, the images all have support in the fixed set $S_{x}$
and the operators act on functions $h$ depending only on $\chi h$. Now
\thinspace%
\begin{align*}
\left\langle \mathcal{C}^{\lambda}h,k\right\rangle & =\iint\hat {v}%
_{\theta}\mu_{e}\mathcal{Q}^{\lambda}\left( G\chi h\right) \chi k\ dvdx \\
& =-\left\langle \mathcal{Q}^{\lambda}\left( G\chi h\right) ,\hat {v}%
_{\theta}\chi k\right\rangle _{\left\vert \mu_{e}\right\vert }
\end{align*}
so that 
\begin{equation*}
\left\vert \left\langle \mathcal{C}^{\lambda}h,k\right\rangle \right\vert
\leq C\left\Vert \chi h\right\Vert _{H^{1\dagger}}\left\Vert \chi
k\right\Vert _{L^{2}}
\end{equation*}
with $C$ independent of $\lambda$. The same proof works for all of the
operators (in their appropriate spaces), except $\mathcal{F}^{\lambda}$ and $%
\left( \mathcal{F}^{\lambda}\right) ^{\ast}$. For $\mathcal{F}^{\lambda
}=\left( \mathcal{D}^{\lambda}\right) ^{\ast}\left( \mathcal{A}%
_{1}^{\lambda}\right) ^{-1}\mathcal{B}^{\lambda}-\left( \mathcal{C}%
^{\lambda}\right) ^{\ast}$, it follows from Lemma \ref%
{lemma-operator-property}(a) that the operator $\left( \mathcal{D}%
^{\lambda}\right) ^{\ast}\left( \mathcal{A}_{1}^{\lambda}\right) ^{-1}%
\mathcal{B}^{\lambda}=\left( \mathcal{D}^{\lambda}\right) ^{\ast}M_\chi
\left( \mathcal{A}_{1}^{\lambda}\right) ^{-1}\mathcal{B}^{\lambda}$ maps 
\begin{equation*}
L_{loc}^{2}\rightarrow L_{c}^{2}\subset H_c^{-1}\subset V^{-1} \rightarrow
V^{1}\rightarrow H_{c}^{1}\subset L^2 \rightarrow H_{c}^{-1\dagger}.
\end{equation*}
Similarly for $\left( \mathcal{F}^{\lambda}\right) ^{\ast}.$

The continuity follows directly from Lemma \ref{lemma-property-Qlambda}(c).
Now let us consider the behavior as $\lambda\to 0$. For any function $%
\phi\in $ $H_{c}^{1\dagger}$, by Lemma \ref{lemma-property-Qlambda}(b) we
have 
\begin{equation*}
\mathcal{C}^{\lambda}\phi\rightarrow\int\hat{v}_{\theta}\mu_{e}\mathcal{P}%
\left( G\phi\right) dv
\end{equation*}
strongly in $L^{2}$ as $\lambda\rightarrow0+$. Clearly $G\phi$ is odd in $%
(v_r,v_z)$. By Lemma 3.3 in \cite{lw-linear}, it follows that $\mathcal{P}%
\left( Gh\right) $ is also odd in$\left( v_{r},v_{z}\right)$. But $\hat{v}%
_{\theta}\mu_{e}$ is even, so that the integral $\int\hat{v} _{\theta}\mu_{e}%
\mathcal{P}\left( Gh\right) dv$ vanishes. Therefore $\mathcal{C}%
^{\lambda}\rightarrow0$ strongly as $\lambda \rightarrow0+$. The proof is
the same for the other operators.
\end{proof}

We study the mapping properties of the operator $\mathcal{A}_{4}^{\lambda}$
in the following lemma.

%%%%%%%%%%%% Lemma 4.4 %%%%%%%%%%%%%%

\begin{lemma}
\label{lemma-operator a5} There exists $\lambda _{1}>0$ such that for any $%
0<\lambda <\lambda _{1}$, the operator $\mathcal{A}_{4}^{\lambda }$ maps $%
W^{2\dagger }$ in a one-to-one manner onto $W^{-2\dagger }$. Therefore it
has a bounded inverse from $W^{-2\dagger }$ onto $W^{2\dagger }$.
Furthermore, $\left( \mathcal{A}_{4}^{\lambda }\right) ^{-1}$, if restricted
to $V^{-2\dagger }$, maps $V^{-2\dagger }$ into $V^{2\dagger }$ with
operator bound independent of $\lambda $.
\end{lemma}

\begin{proof}
It is convenient to introduce yet another operator $\mathcal{A}_5^\lambda$
so that 
\begin{equation*}
\mathcal{A}_{4}^{\lambda}=\mathcal{U}^{\lambda}\mathcal{A}_{5}^{\lambda }%
\mathcal{U}^{\lambda}
\end{equation*}
where $\mathcal{U}^{\lambda}=\left( -\Delta+\frac{1}{r^{2}}+\lambda
^{2}\right) ^{\frac{1}{2}}$. Then 
\begin{equation*}
\mathcal{A}_{5}^{\lambda} = -\Delta+\tfrac{1}{r^{2}}-\left( \mathcal{U}%
^{\lambda}\right) ^{-1} \mathcal{G}^\lambda \left( \mathcal{U}%
^{\lambda}\right) ^{-1},
\end{equation*}
where $\mathcal{G}^\lambda = \mathcal{E}^{\lambda}+\left( \mathcal{D}
^{\lambda}\right) ^{\ast}\left( \mathcal{A}_{1}^{\lambda}\right) ^{-1}%
\mathcal{D}^{\lambda}$. We remark that the operator $\mathcal{E}^\lambda
\leq 0$; however, this fact is not useful because the other operator $\left( 
\mathcal{D}^{\lambda}\right) ^{\ast}\left( \mathcal{A}_{1}^{\lambda}\right)
^{-1}\mathcal{D}^{\lambda} \geq 0$ so that the two signs are in conflict.

By (\ref{singular-remove}), for $\phi \in L_{S}^{2}$ we have 
\begin{equation*}
e^{i\theta }\left( \mathcal{U}^{\lambda }\right) ^{-1}\phi =\left( -\Delta
+\lambda ^{2}\right) ^{-\frac{1}{2}}\left( e^{i\theta }\phi \right)
\end{equation*}%
so that $\left( \mathcal{U}^{\lambda }\right) ^{-1}:L_{S}^{2}\rightarrow
H^{1\dagger }$ and $H^{-1\dagger }\rightarrow L_{S}^{2}$. We consider the
two terms in $\mathcal{G}^{\lambda }$ separately. The operator $\left( 
\mathcal{U}^{\lambda }\right) ^{-1}\mathcal{E}^{\lambda }\left( \mathcal{U}%
^{\lambda }\right) ^{-1}=\left( \mathcal{U}^{\lambda }\right) ^{-1}M_{\chi }%
\mathcal{E}^{\lambda }M_{\chi }\left( \mathcal{U}^{\lambda }\right) ^{-1}$
maps 
\begin{equation*}
H^{2\dagger }\rightarrow H^{3\dagger }\rightarrow H^{1\dagger }\rightarrow
H_{C}^{-1\dagger }\rightarrow L^{2}.
\end{equation*}%
Since the mapping $M_{\chi }:$ $H^{3\dagger }\rightarrow H^{1\dagger }$ is
compact, the operator $\left( \mathcal{U}^{\lambda }\right) ^{-1}\mathcal{E}%
^{\lambda }\left( \mathcal{U}^{\lambda }\right) ^{-1}$ is relatively compact
with respect to $-\Delta +\frac{1}{r^{2}}$. Similarly, the operator 
\begin{equation*}
\left( \mathcal{U}^{\lambda }\right) ^{-1}\left( \mathcal{D}^{\lambda
}\right) ^{\ast }\left( \mathcal{A}_{1}^{\lambda }\right) ^{-1}\mathcal{D}%
^{\lambda }\left( \mathcal{U}^{\lambda }\right) ^{-1}=\left( \mathcal{U}%
^{\lambda }\right) ^{-1}\left( \mathcal{D}^{\lambda }\right) ^{\ast }M_{\chi
}\left( \mathcal{A}_{1}^{\lambda }\right) ^{-1}\mathcal{D}^{\lambda }M_{\chi
}\left( \mathcal{U}^{\lambda }\right) ^{-1}
\end{equation*}%
maps 
\begin{equation*}
H^{2\dagger }\rightarrow H^{3\dagger }\rightarrow H_{c}^{1\dagger
}\rightarrow L_{c}^{2}\subset H_{c}^{-1}\subset V^{-1}\rightarrow
V^{1}\rightarrow H_{c}^{1}\subset L_{c}^{2}\rightarrow H_{c}^{-1\dagger
}\rightarrow L^{2}
\end{equation*}%
and it is relatively compact with respect to $-\Delta +\frac{1}{r^{2}}$.
Therefore by the Kato-Rellich and Weyl theorems, $\mathcal{A}_{5}^{\lambda }$
is self-adjoint on $L_{S}^{2}$ with domain $H^{2\dagger }$ and its essential
spectrum equals $[0,+\infty )$.

We split $\mathcal{A}_5^\lambda$ into two parts as 
\begin{equation*}
\mathcal{A}_5^\lambda = \tfrac12\left( -\Delta + \tfrac 1{r^2} \right) + 
\mathcal{A}_6^\lambda, \qquad \mathcal{A}_6^\lambda = \tfrac12\left( -\Delta
+ \tfrac 1{r^2} \right) - \left( \mathcal{U}^{\lambda}\right) ^{-1} \mathcal{%
G}^{\lambda}\left( \mathcal{U}^{\lambda}\right) ^{-1}
\end{equation*}
and claim that 
\begin{equation*}
\mathcal{A}_6^\lambda \ge 0
\end{equation*}
for sufficiently small $\lambda$. To prove the claim, first note that $%
\mathcal{A}_6^\lambda$ too is self-adjoint on $L_S^2$ with domain $%
H^{2\dagger}$ and its essential spectrum equals $[0,\infty)$. So we merely
need to show that the point spectrum of $\mathcal{A}_6^\lambda$ is also
contained in $[0,\infty)$ for sufficiently small $\lambda$. We prove this by
contradiction. If it were not true, then there would be sequences $\lambda_n
\searrow 0,\ \kappa_n>0$ and $0\ne u_n \in H^{2\dagger}$ such that $\mathcal{%
A}_6^{\lambda_n} u_n = -\kappa_n^2 u_n$. Let $h_n = e^{i\theta} \mathcal{U}%
^{\lambda_n} u_n$. Then $0\ne h_n \in H^3$ and 
\begin{equation*}
\tfrac12 (-\Delta)(-\Delta+\lambda_n^2) h_n = e^{i\theta} \mathcal{G}%
^{\lambda_n} e^{-i\theta} h_n - \kappa_n^2(-\Delta+\lambda_n^2) h_n.
\end{equation*}
Because of the support properties of $\mathcal{G}^\lambda$, we can insert
the cut-off function $\chi$ freely both before and after the exponentials.
So if $\chi h_n=0$, then $\frac12 (-\Delta+\kappa_n^2)(-\Delta+\lambda_n^2)
h_n = 0$, whence $h_n=0$. Therefore $\chi h_n\ne0$. We normalize $\| \chi
h_n \|_{V^1} = 1$.

By Lemma \ref{operator bound}, $e^{i\theta} \mathcal{G}^{\lambda_n}
e^{-i\theta}$ is bounded from $H^1$ to $H^{-1}$ uniformly in $\lambda$.
Hence 
\begin{equation*}
(-\tfrac12 \Delta + \kappa_n^2)(-\Delta+\lambda_n^2) h_n = \chi e^{i\theta} 
\mathcal{G}^{\lambda_n} e^{-i\theta}\chi h_n
\end{equation*}
is a bounded sequence in $H^{-1}$. Multiplying this equation by $h_n$, we
get $\| h_n \|_{V^2}^2 \le C\|\chi h_n \|_{H^1} \le C^{\prime }\|h_n
\|_{V^2} $. Thus $h_n$ is bounded in $V^2$.

Taking a subsequence, we therefore have $h_n \rightharpoonup h$ weakly in $%
V^2$. Since $\chi$ has compact support, it follows that $\chi h_n \to \chi h$
strongly in $V^1$ and that $\|\chi h\|_{V^1} = 1$. Now for any $\ell\in H^1$%
, we have 
\begin{equation*}
\left|\langle e^{i\theta} \mathcal{G}^{\lambda_n} e^{-i\theta} h_n , \ell
\rangle\right| = \left|\langle \chi h_n, e^{i\theta} \mathcal{G}^{\lambda_n}
e^{-i\theta} \chi \ell \rangle\right| \leq \| \mathcal{G}^{\lambda_n}
e^{-i\theta} \chi\ell \|_{H^{-1\dagger}}
\end{equation*}
since $\chi h_n$ is bounded in $V^1$. By Lemma \ref{operator bound}, the
right side tends to zero as $n\to\infty$. Thus $e^{i\theta} \mathcal{G}%
^{\lambda_n} e^{-i\theta} h_n \rightharpoonup 0$ weakly in $H^{-1}$.

Letting $n\rightarrow \infty ,\lambda _{n}\rightarrow 0,\kappa
_{n}\rightarrow \kappa _{0}$, the limit satisfies $(-\frac{1}{2}\Delta
+\kappa _{0}^{2})(-\Delta )h=0$, where $h\in V^{2}$. Since $\Delta h\in
L^{2} $, we deduce $\Delta h=0$. We do not know that $h$ or $\nabla h$
belong to $L^{2}$, but we can use Hardy's inequality (valid for functions in 
$V^{2}$) to estimate 
\begin{equation*}
|\nabla h(x_{0})|\leq \frac{C}{R^{3}}\int_{\{|x-x_{0}|<R\}}|\nabla h|dx\leq 
\frac{C^{\prime }}{R^{3}}\left( \int \frac{|\nabla h|^{2}}{\left\vert
x-x_{0}\right\vert ^{2}}dx\right) ^{\frac{1}{2}}(R^{5})^{\frac{1}{2}}=O(R^{-%
\frac{1}{2}})
\end{equation*}%
for every point $x_{0}$. Therefore $h$ is a constant. Since $h\in
V^{2},\,h\equiv 0$. This contradicts $\Vert \chi h\Vert _{V^{1}}=1$, which
proves the claim.

The claim we have just proven means that $\langle \mathcal{A}_{6}^{\lambda
}u,u\rangle \geq 0$ for all $u$ in the domain $H^{2\dagger }$ of the
operator. Thus 
\begin{equation*}
\langle \mathcal{A}_{5}^{\lambda }u,u\rangle \geq \frac{1}{2}\int (|\nabla
u|^{2}+\frac{1}{r^{2}}u^{2})dx.
\end{equation*}
The right side is the squared norm of $u$ in $V^{1\dag }$. The left side
defines a bilinear form $a(u,u)$ that extends continuously to $V^{1\dag
}\times V^{1\dag }$. So by the Lax-Milgram lemma, the operator $\mathcal{A}%
_{5}^{\lambda }:V^{1\dagger }\rightarrow V^{-1\dag }$ is one-to-one onto.

But $\mathcal{A}_{4}^{\lambda }=\mathcal{U}^{\lambda }\mathcal{A}%
_{5}^{\lambda }\mathcal{U}^{\lambda }$. Since for fixed $\lambda >0$, the
operator $\mathcal{U}^{\lambda }$ is an isomorphism: $W^{2\dagger
}\rightarrow V^{1\dagger }$ and also $V^{-1\dagger }\rightarrow W^{-2\dagger
}$, we deduce that $\mathcal{A}_{4}^{\lambda }$ maps $W^{2\dagger }$ to $%
W^{-2\dagger }$ in a one-to-one onto fashion. It is also clear that $\Vert
h\Vert _{V^{2\dagger }}\leq C\Vert \mathcal{U}^{\lambda }h\Vert
_{V^{1\dagger }}$ so that $\mathcal{(}U^{\lambda })^{-1}:V^{1\dagger
}\rightarrow V^{2\dagger }$ with a bound independent of $\lambda $ and $%
\mathcal{(}U^{\lambda })^{-1}:V^{-2\dagger }\rightarrow V^{-1\dagger }$ with
a bound independent of $\lambda $. Therefore 
\begin{equation*}
\left( \mathcal{A}_{4}^{\lambda }\right) ^{-1}:V^{-2\dagger }\rightarrow
V^{-1\dagger }\rightarrow V^{+1\dagger }\rightarrow V^{+2\dagger }
\end{equation*}%
with a bound independent of $\lambda $.
\end{proof}

%%%%%%%%%%%% Lemma 4.5 %%%%%%%%%

\begin{lemma}
\label{lemma-a4-inverse-bound} If $S$ is a ball in $\mathbb{R}^3$, there
exist constants $C>0$ and $\lambda_{2}\in\left(0,\lambda_{1}\right)$ such
that 
\begin{equation*}
\langle \mathcal{A}_4^\lambda u,u \rangle \ge C \|u\|_{V^{1\dagger}}^2
\end{equation*}
for all $u\in V^{2\dagger}$ with support in $S$ and all $\lambda\in
(0,\lambda_2]$.
\end{lemma}

\begin{proof}
We argue by contradiction in a similar way to the preceding proof. If the
lemma were false, then there would be sequences $\lambda_n\to 0$ and $u_n\in
V^{2\dagger}$ with supports in $S$ such that $\|u_n\|_{V^{1\dagger}}=1$ but $%
\langle \mathcal{A}_4^{\lambda_n} u_n,u_n \rangle \to 0$. By definition of $%
\mathcal{A}_4^\lambda$, 
\begin{equation*}
\left\langle \left( -\Delta + \tfrac1{r^2}\right) \left( -\Delta +
\tfrac1{r^2} + \lambda_n^2 \right) u_n -\mathcal{G}^{\lambda_n}u_n, u_n
\right\rangle \to 0.
\end{equation*}
Letting $h_n=e^{i\theta}u_n$, we have 
\begin{equation*}
\left\langle \left( -\Delta\right) \left( -\Delta + \lambda_n^2 \right) h_n,
h_n\right\rangle - \left\langle e^{i\theta}\mathcal{G}^{\lambda_n}
e^{-i\theta} h_n, h_n \right\rangle \to 0 .
\end{equation*}
Thus 
\begin{equation*}
\|\Delta h_n\|_{L^2}^2 + \lambda_n^2\|\nabla h_n\|_{L^2}^2 \le \|\mathcal{G}%
^{\lambda_n}\|_{H^{1\dagger}\mapsto H^{-1\dagger}} \|h_n\|_{H^1}^2 + 1.
\end{equation*}
Because the right side is bounded, we therefore have a bound for $\Delta h_n$
so that $h_n$ is bounded in $V^2$. Taking a subsequence, we have $%
h_n\rightharpoonup h_0$ weakly in $V^2$ and consequently $u_n\rightharpoonup
e^{-i\theta}h_0$ weakly in $V^{2\dagger}$. Because of the uniformly bounded
support, we can replace $V^{2\dagger}$ by $H^{2\dagger}$ and use the compact
embedding to deduce that $u_n\to e^{-i\theta}h_0$ strongly in $H^{1\dagger}$%
. Therefore $1 = \|e^{-i\theta} h_0\|_{V^{1\dagger}}= \|h_0\|_{V^1} $. By
the strong convergence of $u_n$ in $H^{1\dagger}$, and the strong
convergence of $\mathcal{G}^{\lambda_n}$ as $\lambda_n\to0$ from Lemma \ref%
{operator bound}, we have $\left\langle \mathcal{G}^{\lambda_n}u_n, u_n
\right\rangle \to 0 $. Therefore 
\begin{equation*}
\left\langle \left( -\Delta\right) \left( -\Delta + \lambda_n^2 \right) h_n,
h_n\right\rangle \to0 .
\end{equation*}
So $h_n$ tends to zero strongly in $V^2$ and so also in $V^1$ (due to the
bounded support), which contradicts $\|h_0\|_{V^1}=1$.
\end{proof}

It follows immediately from either of the two preceding lemmas that $M_\chi (%
\mathcal{A}_4^\lambda)^{-1} M_\chi$ maps $H^{-1\dagger}$ into $H^{1\dagger}$
with a bound independent of $\lambda$.

%%%%%%%%%%%%% Section 5 %%%%%%%%%%%%%%%%%%

\section{Behavior for small $\protect\lambda$}

\begin{lemma}
\label{lemma-operator-N-lambda} There exists $\lambda_3>0$ such that for any 
$\lambda\in(0,\lambda_3]$ the operator 
\begin{equation*}
\mathcal{N}^{\lambda}=\mathcal{L}^{\lambda}+\left( \mathcal{F}^{\lambda
}\right) ^{\ast}\left( \mathcal{A}_{4}^{\lambda}\right) ^{-1}\mathcal{F}%
^{\lambda}
\end{equation*}
is self-adjoint on $L_S^2$ with domain $H^{2\dagger}$ and has essential
spectrum $[\lambda^{2},\infty)$. Moreover, if $\mathcal{L}^{0}$ has a
negative eigenvalue, then $\mathcal{N}^{\lambda}$ also has a negative
eigenvalue.
\end{lemma}

\begin{proof}
The bound $\lambda_{2}$ is given in Lemma \ref{lemma-a4-inverse-bound}. By
the proof of Lemma 3.1 in \cite{lw-linear}, the operator $\mathcal{L}%
^{\lambda}$ is relatively compact with respect to $-\Delta
+1/r^{2}+\lambda^{2}$. By Lemmas \ref{operator bound} and \ref%
{lemma-a4-inverse-bound}, the operator $\left( \mathcal{F}^{\lambda}\right)
^{\ast} \left( \mathcal{A}_{4}^{\lambda}\right) ^{-1} \mathcal{F}^{\lambda}
=M_\chi \cdot \left( \mathcal{F}^{\lambda}\right) ^{\ast} \cdot \{ M_\chi
\left( \mathcal{A}_{4}^{\lambda}\right) ^{-1} M_\chi \} \cdot \mathcal{F}%
^{\lambda}\cdot M_\chi $ maps 
\begin{equation*}
H^{2\dagger}\rightarrow L_{c}^{2}\rightarrow H^{-1\dagger}\rightarrow
H^{1\dagger}\rightarrow L^{2}\rightarrow L_c^2,
\end{equation*}
which implies that it is relatively compact with respect to $-\Delta
+1/r^{2}+\lambda^{2}$. So the self-adjoint and the essential spectrum
properties follow from the Kato-Rellich and Weyl theorems.

Assume now that $\mathcal{L}^{0}$ has a negative eigenvalue $k^{0}<0$ and
let $\zeta ^{0}\in H^{2\dagger }$ be a normalized eigenvector. Write 
\begin{align*}
\left\langle \mathcal{N}^{\lambda }\zeta ^{0},\zeta ^{0}\right\rangle
-k^{0}& =\left\langle \mathcal{N}^{\lambda }\zeta ^{0},\zeta
^{0}\right\rangle -\left\langle \mathcal{L}^{0}\zeta ^{0},\zeta
^{0}\right\rangle \\
& =\left\langle \left( \mathcal{L}^{\lambda }-\mathcal{L}^{0}\right) \zeta
^{0},\zeta ^{0}\right\rangle +\left\langle \left( \mathcal{F}^{\lambda
}\right) ^{\ast }\left( \mathcal{A}_{4}^{\lambda }\right) ^{-1}\mathcal{F}%
^{\lambda }\zeta ^{0},\zeta ^{0}\right\rangle .
\end{align*}%
By Lemma \ref{lemma-operator-property}(\textit{f}), the first term on the
right is less than $\left\Vert \left( \mathcal{L}^{\lambda }-\mathcal{L}%
^{0}\right) \zeta ^{0}\right\Vert _{L^{2}}\rightarrow 0$, as $\lambda
\searrow 0$. By Lemma \ref{lemma-a4-inverse-bound}, the second term is
bounded by 
\begin{align*}
\left\vert \left\langle \left( \mathcal{A}_{4}^{\lambda }\right)
^{-1}M_{\chi }\mathcal{F}^{\lambda }\zeta ^{0},M_{\chi }\mathcal{F}^{\lambda
}\zeta ^{0}\right\rangle \right\vert & \leq \left\Vert M_{\chi }\left( 
\mathcal{A}_{4}^{\lambda }\right) ^{-1}M_{\chi }\right\Vert _{H^{-1\dagger
}\mapsto H^{1\dagger }}\left\Vert \mathcal{F}^{\lambda }\zeta
^{0}\right\Vert _{H^{-1\dagger }}^{2} \\
& \leq C\left\Vert \mathcal{F}^{\lambda }\zeta ^{0}\right\Vert
_{H^{-1\dagger }}^{2}\rightarrow 0\quad \text{ as }\lambda \searrow 0
\end{align*}%
because $C$ is independent of $\lambda $, and using Lemma \ref{operator
bound}. Thus $\left\langle \mathcal{N}^{\lambda }\zeta ^{0},\zeta
^{0}\right\rangle \rightarrow k^{0}<0$ as $\lambda \searrow 0$. So if $%
\lambda _{3}$ is small enough and $0<\lambda \leq \lambda _{3}$, then $%
\mathcal{N}^{\lambda }$ has a negative eigenvalue.
\end{proof}

%%%%%%%%% Projection %%%%%%%%%%%%
Now we perform a finite-dimensional truncation of the matrix operator (\ref%
{matrix-operator}). Let $\{\sigma_{1},\sigma_{2},\cdots\}$ be a sequence of
functions in $H_c^{2\dagger}$, for which the finite linear combinations are
dense in $V^{2\dagger}$. Orthogonalize them so that they form an orthonormal
set in $L_S^2$. As before, $\langle\ ,\ \rangle$ denotes the usual $L^2$
pairing and we will denote the standard inner product in $\mathbb{R}^n$ by a
dot. Let $n$ be a positive integer. Define the projection operator $%
P_{n}:V^{-2\dagger}\rightarrow\mathbb{R}^{n}$ and its $L^2$-adjoint $%
P_{n}^{\ast}:\mathbb{R}^{n}\rightarrow V^{2\dagger}$ by 
\begin{equation*}
P_{n}h = \left\{ \left\langle h,\sigma _{j}\right\rangle \right\}
_{j=1}^{n}\ , \qquad P_{n}^{\ast} b = \sum_{j=1}^{n}b^{j}\sigma_{j}\ ,
\end{equation*}
where $h\in V^{-2\dagger}$ and $b=\left( b^{1},\cdots,b^{n}\right)\in\mathbb{%
R}^{n}. $ Then $P_{n}P_{n}^{\ast}b=b$ for any $b\in\mathbb{R}^{n}$, and $%
P_{n}^{\ast}P_{n}h=\sum_{j=1}^{n}\left\langle h,\sigma_{j}\right\rangle
\sigma_{j}$ for any $h\in V^{-2\dagger}$. Define the ``approximate matrix
operator" 
\begin{equation*}
\mathcal{M}_{n}^{\lambda}=\left( 
\begin{array}{cc}
\mathcal{L}^{\lambda} & \left( \mathcal{F}^{\lambda}\right)
^{\ast}P_{n}^{\ast} \\ 
P_{n}\mathcal{F}^{\lambda} & -P_{n}\mathcal{A}_{4}^{\lambda}P_{n}^{\ast}%
\end{array}
\right)
\end{equation*}
which takes $V^{2\dagger}\times\mathbb{R}^{n}$ into $L_{S}^{2}\times \mathbb{%
R}^{n}$.

%%%%%%%%%% Lemma 5.2 %%%%%%%%%%

\begin{lemma}
\label{n-bound} Let $0<\lambda\leq\lambda_{3}$. For any $\eta\in L_{loc}^{2}$%
, define $d_{n}=(P_{n}\mathcal{A}_{4}^{\lambda}P_{n}^{\ast})^{-1}P_{n}%
\mathcal{F}^{\lambda}\eta$. Then 
\begin{equation*}
\sup_{n}\Vert P_{n}^{\ast}d_{n}\Vert_{V^{2\dagger}}<\infty.
\end{equation*}
\end{lemma}

\begin{proof}
Because $\lambda $ is fixed, for brevity we denote $\mathcal{A}=\mathcal{A}%
_{4}^{\lambda }\ $and $\mathcal{F}=\mathcal{F}^{\lambda }$. Note that $%
\alpha =P_{n}\mathcal{A}P_{n}^{\ast }$ is the $n\times n$ symmetric
positive-definite matrix with entries $\alpha _{jk}=\left\langle \mathcal{A}%
_{4}\sigma _{k},\sigma _{j}\right\rangle $. Let $c\left( n\right)
=\left\Vert \chi P_{n}^{\ast }d_{n}\right\Vert _{H^{1\dagger }}$. We will
show that $c(n)$ is bounded. Suppose on the contrary that $c(n)\rightarrow
\infty $. Let $u_{n}=P_{n}^{\ast }d_{n}/c(n)$ so that $\Vert \chi u_{n}\Vert
_{H^{1\dagger }}=1$. Then $P_{n}\mathcal{A}u_{n}=P_{n}\mathcal{A}P_{n}^{\ast
}d_{n}/c(n)=P_{n}\mathcal{F}\eta /c(n)$ so that 
\begin{eqnarray*}
\langle \mathcal{A}u_{n},u_{n}\rangle &=&\langle \mathcal{A}u_{n},\frac{1}{%
c(n)}P_{n}^{\ast }d_{n}\rangle =\frac{1}{c(n)}P_{n}\mathcal{A}u_{n}\cdot
d_{n} \\
&=&\frac{1}{c^{2}(n)}P_{n}\mathcal{F}\eta \cdot d_{n}=\frac{1}{c(n)}\langle 
\mathcal{F}\eta ,u_{n}\rangle .
\end{eqnarray*}%
Thus 
\begin{equation*}
\left\langle \left( -\Delta +\frac{1}{r^{2}}\right) \left( -\Delta +\frac{1}{%
r^{2}}+\lambda ^{2}\right) u_{n},u_{n}\right\rangle =\langle \mathcal{G}%
^{\lambda }u_{n},u_{n}\rangle +\frac{1}{c(n)}\langle \mathcal{F}\eta ,\chi
u_{n}\rangle
\end{equation*}%
so that, as in the proof of Lemma \ref{lemma-operator a5}, 
\begin{equation*}
\Vert u_{n}\Vert _{W^{2\dagger }}^{2}\leq C\Vert \chi u_{n}\Vert
_{H^{1\dagger }}^{2}+\frac{1}{c(n)}\Vert \chi u_{n}\Vert _{H^{1\dagger
}}\leq C+1.
\end{equation*}%
We take a subsequence so that $u_{n}\rightharpoonup u_{0}$ weakly in $%
W^{2\dagger }$. Then $\chi u_{n}\rightarrow \chi u_{0}$ strongly in $%
H^{1\dagger }$, so that $\Vert \chi u_{0}\Vert _{H^{1\dagger }}=1$. Fix an
integer $m\geq 1$ and let $n\geq m$. Then $P_{n}^{\ast }\delta _{m}=\sigma
_{m}$, where $(\delta _{m})_{j}=1$ for $j=m$ and is otherwise 0. Then 
\begin{equation*}
\langle \mathcal{A}u_{n},\sigma _{m}\rangle =P\mathcal{A}u_{n}\cdot \delta
_{m}=\frac{1}{c(n)}P_{n}\mathcal{F}\eta \cdot \delta _{m}=\frac{1}{c(n)}%
\langle \mathcal{F}\eta ,\sigma _{m}\rangle \rightarrow 0
\end{equation*}%
as $n\rightarrow \infty $ since $\langle \mathcal{F}\eta ,\sigma _{m}\rangle 
$ is independent of $n$. Thus $\langle \mathcal{A}u_{0},\sigma _{m}\rangle
=0 $ for all $m$, so that $\mathcal{A}u_{0}=0$. So $u_{0}=0$, which
contradicts $\Vert \chi u_{0}\Vert _{H^{1\dagger }}=1$. Thus $c(n)$ is
indeed bounded.

Now substituting $u_{n} = P_{n}^{*}d_{n} / c(n) $ into the inequality above,
we get 
\begin{equation*}
\|\frac1{c(n)} P_{n}^{*} d_{n}\|_{W^{2\dagger}}^{2} =
\|u_{n}\|_{W^{2\dagger}}^{2} \le C\|\chi u_{n}\|_{H^{1\dagger}}^{2} + \frac
C{c(n)}\| \chi u_{n}\|_{H^{1\dagger}}
\end{equation*}
Multiplying by $c^{2}(n)$, we find 
\begin{equation*}
\| P_{n}^{*} d_{n}\|_{W^{2\dagger}}^{2} \le C\|\chi
P_{n}^{*}d_{n}\|_{H^{1\dagger}}^{2} + C\| \chi
P_{n}^{*}d_{n}\|_{H^{1\dagger}} = Cc^{2}(n) + Cc(n) \le C^{\prime}.
\end{equation*}
Therefore $P_{n}^{*}d_{n}$ is bounded in $W^{2\dagger}$, hence in $%
V^{2\dagger}$.
\end{proof}

%%%%%%%%%% Lemma 5.3 %%%%%%%%%%%

\begin{lemma}
\label{lemma-small-lambda}

Fix $0<\lambda\le \lambda_3$. There exists a positive integer $%
N=N(\lambda_3) $ such that for $n\geq N$, the matrix operator 
\begin{equation*}
\mathcal{M}_{n}^{\lambda}=\left( 
\begin{array}{cc}
\mathcal{L}^{\lambda} & \left( \mathcal{F}^{\lambda}\right)
^{\ast}P_{n}^{\ast} \\ 
P_{n}\mathcal{F}^{\lambda} & -P_{n}\mathcal{A}_{4}^{\lambda}P_{n}^{\ast}%
\end{array}
\right)
\end{equation*}
is self-adjoint on $L_{S}^{2}\times\mathbb{R}^{n}$ with domain $%
H^{2\dagger}\times\mathbb{R}^n$, has essential spectrum $[\lambda^2,\infty)$
and has at least $n+1$ negative eigenvalues.
\end{lemma}

\begin{proof}
We recall that $\mathcal{L}^{\lambda}$ is self-adjoint with essential
spectrum $[\lambda^{2},\infty)$. However, the symmetric operator 
\begin{equation*}
\left( 
\begin{array}{cc}
0 & \left( \mathcal{F}^{\lambda}\right) ^{\ast}P_{n}^{\ast} \\ 
P_{n}\mathcal{F}^{\lambda} & -P_{n}\mathcal{A}_{4}^{\lambda}P_{n}^{\ast}%
\end{array}
\right)
\end{equation*}
has finite-dimensional range and so it is compact. The theorems of
Kato-Rellich and Weyl's apply here directly to prove the first two
assertions of the lemma. It remains to consider the negative spectrum.

The last assertion is equivalent to saying that there is an $\left(
n+1\right) $-dimensional subspace $S\subset H^{2\dagger}\times\mathbb{R}^{n}$
such that $\left\langle \mathcal{M}_{n}^{\lambda}z,z\right\rangle <0$ for
all $z\in S \setminus \{0\} $. For simplicity, we temporarily drop the
superscript $\lambda$ as it is fixed in this proof. As above, let $\alpha$
be the $n\times n$ symmetric positive matrix with entries $%
\alpha_{jk}=\left\langle \mathcal{A}_{4}\sigma_{k},\sigma_{j}\right\rangle $%
. Let 
\begin{equation*}
\mathcal{J}_n=\left( 
\begin{array}{cc}
I & 0 \\ 
\alpha^{-1}P_{n}\mathcal{F} & I%
\end{array}
\right) .
\end{equation*}
Then 
\begin{equation*}
\mathcal{J}_{n}^{\ast}\mathcal{M}_n\mathcal{J}_n = \left( 
\begin{array}{cc}
\mathcal{L}+\left( \mathcal{F}\right) ^{\ast}P_{n}^{\ast}\mathcal{\alpha }%
^{-1}P_{n}\mathcal{F} & 0 \\ 
0 & -\alpha%
\end{array}
\right)
\end{equation*}
has the same number of negative eigenvalues as $\mathcal{M}_{n}$. But $%
-\alpha$ has exactly $n$ negative eigenvalues, so it suffices to prove that 
\begin{equation*}
\mathcal{N}_{n}=\mathcal{L}+\left( \mathcal{F}\right) ^{\ast}P_{n}^{\ast }%
\mathcal{\alpha}^{-1}P_{n}\mathcal{F}
\end{equation*}
has a negative eigenvalue when $n$ is large.

By Lemma \ref{lemma-operator-N-lambda}, the untruncated operator $\mathcal{%
N=L}+\left( \mathcal{F}\right) ^{\ast }\left( \mathcal{A}_{4}\right) ^{-1}%
\mathcal{F}$ has a negative eigenvalue. Let $\eta =\eta ^{\lambda }$ be an
eigenvector of $\mathcal{N}=\mathcal{N}^{\lambda }$ as in Lemma \ref{n-bound}
with eigenvalue $\mu <0$ and $\Vert \eta \Vert _{L^{2}}=1$. Let $\xi =%
\mathcal{A}_{4}^{-1}\mathcal{F}\eta $. Since $\eta \in L^{2}$, we have $%
\mathcal{F}\eta \in H^{-1\dagger }$ and $\xi \in V^{2\dagger }$. Recall that 
$d_{n}=\alpha ^{-1}P_{n}\mathcal{F}\eta $. By these definitions, we have 
\begin{eqnarray*}
&&\langle \mathcal{N}_{n}\eta ,\eta \rangle -\langle \mathcal{N}\eta ,\eta
\rangle \\
&=&\langle (\mathcal{F}^{\ast }P_{n}^{\ast }\alpha ^{-1}P_{n}\mathcal{F}-%
\mathcal{F}^{\ast }\mathcal{A}_{4}^{-1}\mathcal{F})\eta ,\eta \rangle
=\langle (P_{n}^{\ast }\alpha ^{-1}P_{n}\mathcal{F}-\mathcal{A}_{4}^{-1}%
\mathcal{F})\eta ,\mathcal{F}\eta \rangle \\
&=&\langle P_{n}^{\ast }d_{n}-\xi ,\mathcal{A}_{4}\xi \rangle =\langle 
\mathcal{A}_{4}(P_{n}^{\ast }d_{n}-\xi ),\xi \rangle .
\end{eqnarray*}%
Choose a sequence $\xi _{n}$ such that $\Vert \xi _{n}-\xi \Vert
_{V^{2\dagger }}\rightarrow 0$ and such that each $\xi _{n}$ is a linear
combination of $\{\sigma _{1},\dots ,\sigma _{n}\}$. Then $\xi _{n}$ belongs
to the range of $P_{n}^{\ast }$. Because $P_{n}\mathcal{A}_{4}(P_{n}^{\ast
}d_{n}-\xi )=\alpha d_{n}-P_{n}\mathcal{F}\eta =0$, it follows that 
\begin{eqnarray*}
|\langle \mathcal{N}_{n}\eta ,\eta \rangle -\langle \mathcal{N}\eta ,\eta
\rangle | &=&|\langle \mathcal{A}_{4}(P_{n}^{\ast }d_{n}-\xi ),\xi -\xi
_{n}\rangle | \\
&\leq &\Vert \mathcal{A}_{4}(P_{n}^{\ast }d_{n}-\xi )\Vert _{V^{-2\dagger
}}\Vert \xi -\xi _{n}\Vert _{V^{2\dagger }} \\
&\leq &C\Vert P_{n}^{\ast }d_{n}-\xi \Vert _{V^{2\dagger }}\Vert \xi -\xi
_{n}\Vert _{V^{2\dagger }} \\
&\leq &C^{\prime }\Vert \xi -\xi _{n}\Vert _{V^{2\dagger }}\rightarrow 0
\end{eqnarray*}%
as $n\rightarrow \infty $. Since $\langle \mathcal{N}\eta ,\eta \rangle <0$,
it follows that $\langle \mathcal{N}_{n}\eta ,\eta \rangle <0$ for
sufficiently large $n$, so that $\mathcal{N}_{n}$ must have a negative
eigenvalue.
\end{proof}

%%%%%%%%% Section 6 %%%%%%%%%%%%%%%%%%%

\section{Approximate growing mode}

Now we consider the behavior for large $\lambda$.

\begin{lemma}
\label{lemma-large-lambda}There exists $\lambda_{5}>0$ such that if $%
\lambda\geq\lambda_{5}$, then for each $n$ the operator $\mathcal{M}
_{n}^{\lambda}$ has at most $n$ negative eigenvalues.
\end{lemma}

\begin{proof}
For $h\in H^{2\dagger},$ 
\begin{align*}
\left( \mathcal{L}^{\lambda}h,h\right) & \geq\left( \mathcal{A}
_{2}^{\lambda}h,h\right) =\left( \left( -\Delta+\frac{1}{r^{2}}+\lambda
^{2}\right) h,h\right) -\iint r\hat{v}_{\theta}\mu_{p}dv\text{ } \left\vert
h\right\vert ^{2}dx \\
& -\iint\hat{v}_{\theta}\mu_{e}\mathcal{Q}^{\lambda}\left( \hat {v}
_{\theta}h\right) dvhdx.
\end{align*}
The last term is bounded by 
\begin{equation*}
\left\vert \left\langle \mathcal{Q}^{\lambda}\left( \hat{v}_{\theta}h\right)
,\hat{v}_{\theta}h\right\rangle _{\left\vert \mu_{e}\right\vert }\right\vert
\leq \left\Vert \hat{v}_{\theta}h\right\Vert _{\left\vert\mu_{e}\right\vert
}^{2} \leq \left( \sup_{x} \int\left\vert\mu_{e}\right\vert dv\right)
\left\vert h\right\vert _{2}^2
\end{equation*}
by Lemma \ref{lemma-property-Qlambda} (a). Thus $\left( \mathcal{L}%
^{\lambda}h,h\right) \geq\left( \lambda^{2}-C_{0}\right) \left\vert
h\right\vert _{2}^{2}$, where $C_{0}=\sup_{x} \left( \int\left( \left\vert
r\mu _{p}\right\vert +\left\vert \mu_{e}\right\vert \right) dv\right) $. Now
for any $\left( h,b\right) \in H^{2\dagger}\times\mathbb{R}^{n}$, 
\begin{align*}
\left\langle \mathcal{M}_{n}^{\lambda}\left( 
\begin{array}{c}
h \\ 
b%
\end{array}
\right) ,\left( 
\begin{array}{c}
h \\ 
b%
\end{array}
\right) \right\rangle & =\left( \mathcal{L}^{\lambda}h,h\right)
+2\left\langle \mathcal{F}^{\lambda}h,P_{n}^{\ast}b\right\rangle
-\left\langle \mathcal{A}_{4}^{\lambda}P_{n}^{\ast}b,P_{n}^{\ast}b\right%
\rangle \\
& \geq \left( \lambda^{2}-C_{0}\right) \left\vert h\right\vert
_{2}^{2}-\left\Vert \mathcal{F}^{\lambda}h\right\Vert
_{H^{-1\dagger}}\left\Vert \chi P_{n}^{\ast}b\right\Vert
_{H^{1\dagger}}-C\left( \lambda\right) \left\Vert P_{n}^{\ast}b\right\Vert
_{ V^{2\dagger}}^{2} \\
& \geq \left( \lambda^{2}-C_{0}\right) \left\vert h\right\vert
_{2}^{2}-2C_{1}\left\vert h\right\vert _{2}\left\Vert
P_{n}^{\ast}b\right\Vert _{V^{2\dagger}}-C\left( \lambda\right) \left\Vert
P_{n}^{\ast }b\right\Vert _{V^{2\dagger}}^{2} \\
& \geq -\left( C_{1}^{2}+C\left( \lambda\right) ^{2}\right) \left\Vert
P_{n}^{\ast}b\right\Vert _{V^{2\dagger}}^{2}
\end{align*}
provided $\lambda\geq\lambda_{5}=\sqrt{C_{0}+1}.$ Since $b\in\mathbb{R}^{n},$
it follows that $\mathcal{M}_{n}^{\lambda}$ has at most $n$ negative
eigenvalues.
\end{proof}

%%%%%%%%%% Lemma 6.2 %%%%%%%%%%
Now we are ready to exhibit an approximate growing mode.

\begin{lemma}
\label{lemma-approximate-n}For each positive integer $n\geq N(\lambda _{3}),$
there exists $\lambda _{n}\in \left[ \lambda _{3},\lambda _{5}\right] $ such
that $\mathcal{M}_{n}^{\lambda _{n}}$ has a nontrivial kernel. Here $\lambda
_{4},\lambda _{5}$ are in Lemmas \ref{lemma-small-lambda} and \ref%
{lemma-large-lambda}.
\end{lemma}

\begin{proof}
We emphasize that $\lambda_3$ and $\lambda_5$ do not depend on $n$. We use
continuation in $\lambda$. First, $\mathcal{M}_{n}^{\lambda}$ is a
continuous family of operators of $\lambda$ in the sense that if $\sigma>0$,
then there exists $C,\delta>0$ such that 
\begin{equation*}
\left\Vert \mathcal{M}_{n}^{\lambda}-\mathcal{M}_{n}^{\sigma}\right\Vert
\leq C\left\vert \lambda-\sigma\right\vert
\end{equation*}
for $\left\vert \lambda-\sigma\right\vert <\delta$, $\lambda,\sigma\in\left(
0,\infty\right) ,$ where $\left\Vert {}\right\Vert $ denotes the operator
norm from $L_{S}^{2}\times\mathbb{R}^{n}$ to $L_{S}^{2}\times\mathbb{R}^{n}$%
. This continuity property follows immediately from Lemma \ref%
{lemma-operator-property}.

By Lemma \ref{lemma-small-lambda}, $\mathcal{M}_{n}^{\lambda _{4}}$ has at
least $\left( n+1\right) $ negative eigenvalues. By Lemma \ref%
{lemma-large-lambda}, $\mathcal{M}_{n}^{\lambda _{5}}$ has at most $n$
negative eigenvalues. By (\cite{kato}, IV-3.5), the eigenvalues of $\mathcal{%
M}_{n}^{\lambda }$ within the interval $\left[ \lambda _{4},\lambda _{5}%
\right] $ are continuous functions of $\lambda .$ In particular, the
dimension of the corresponding eigenspace is a constant. hence at least one
eigenvalue must cross from negative to positive. So there exists some $%
\lambda _{n}\in \left[ \lambda _{4},\lambda _{5}\right] $ such that $%
\mathcal{M}_{n}^{\lambda _{n}}$ has a nontrivial kernel.
\end{proof}

%%%%%%%%%%%% Section 7 %%%%%%%%%%%%%%

\section{Limit as $n\rightarrow+\infty$}

\begin{lemma}
There exist $\lambda_{0},\ h_{0},\ k_{0}\ $such that $0<\lambda_{0}<\infty,\
h_{0}\in H^{2\dagger},\ k_{0}\in H^{2\dagger}$ and 
\begin{equation}
\mathcal{L}^{\lambda_{0}}h_{0}+\left( \mathcal{F}^{\lambda_{0}}\right)
^{\ast}k_{0}=0,  \label{eqn-h0}
\end{equation}%
\begin{equation}
\mathcal{F}^{\lambda_{0}}h_{0}-\mathcal{A}_{4}^{\lambda_{0}}k_{0}=0
\label{eqn-k0}
\end{equation}
with $\left( h_{0},k_{0}\right) \neq\left( 0,0\right) $.
\end{lemma}

\begin{proof}
By Lemma \ref{lemma-approximate-n}, for each $n\geq N(\lambda _{3})$ there
exists $\lambda _{n}\in \left[ \lambda _{3},\lambda _{5}\right] $ \ and a
nonzero solution $\left( h_{n},b_{n}\right) \in $ $H^{2\dagger }\times 
\mathbb{R}^{n}$ such that 
\begin{equation}
\mathcal{L}^{\lambda _{n}}h_{n}+\left( \mathcal{F}^{\lambda _{n}}\right)
^{\ast }P_{n}^{\ast }b_{n}=0,  \label{eqn-approximate-h}
\end{equation}%
\begin{equation}
P_{n}\mathcal{F}^{\lambda _{n}}h_{n}-P_{n}\mathcal{A}_{4}^{\lambda
_{n}}P_{n}^{\ast }b_{n}=0.  \label{eqn-approximate-b}
\end{equation}%
We normalize $h_{n},b_{n}$ such that 
\begin{equation*}
\left\Vert h_{n}\right\Vert _{L^{2}}+\left\Vert \chi P_{n}^{\ast
}b_{n}\right\Vert _{H^{1\dagger }}=1
\end{equation*}%
by Lemma \ref{lemma-a4-inverse-bound}. We claim that $h_{n}$ is bounded in $%
H^{2\dagger }$. In deed, $\left\Vert \chi P_{n}^{\ast }b_{n}\right\Vert
_{H^{1\dagger }}\leq 1$, so that $\left( \mathcal{F}^{\lambda _{n}}\right)
^{\ast }P_{n}^{\ast }b_{n}$ is bounded in $L^{2}$, and $\mathcal{L}^{\lambda
_{n}}h_{n}$ is bounded in $L^{2}$. Since $\left\Vert h_{n}\right\Vert
_{L^{2}}\leq 1$, $\left( \mathcal{B}^{\lambda _{n}}\right) ^{\ast }\left( 
\mathcal{A}_{1}^{\lambda _{n}}\right) ^{-1}\mathcal{B}^{\lambda _{n}}h_{n}$
is also bounded in $L^{2},$ and so are $\mathcal{A}_{2}^{\lambda _{n}}h_{n}$
and $\left( -\Delta +\frac{1}{r^{2}}+\lambda _{n}^{2}\right) h_{n}$.
Therefore $h_{n}$ is bounded in $H^{2\dagger }$. By (\ref{eqn-approximate-b}%
) we have 
\begin{equation*}
\left\langle \mathcal{A}_{4}^{\lambda _{n}}P_{n}^{\ast }b_{n},P_{n}^{\ast
}b_{n}\right\rangle =\left\langle \mathcal{F}^{\lambda
_{n}}h_{n},P_{n}^{\ast }b_{n}\right\rangle .
\end{equation*}%
The right side of this equation is bounded. So $\langle (-\Delta +\frac{1}{%
r^{2}})(-\Delta +\frac{1}{r^{2}}+\lambda _{n}^{2})P_{n}^{\ast
}b_{n},P_{n}^{\ast }b_{n}\rangle $ is also bounded. Therefore $P_{n}^{\ast
}b_{n}$ is bounded in $V^{1\dagger }\cap V^{2\dagger }$. Now we take
subsequences such that $\lambda _{n}\rightarrow \lambda _{0}\in \left[
\lambda _{4},\lambda _{5}\right] $, $h_{n}\rightarrow h_{0}$ weakly in $%
H^{2\dagger }$, $P_{n}^{\ast }b_{n}\rightarrow k_{0}$ weakly in $V^{2\dagger
}$. We look at each term for (\ref{eqn-approximate-h}), (\ref%
{eqn-approximate-b}) separately. First, for any $l\in H^{2\dagger },$%
\begin{align*}
& \left\vert \left( \mathcal{L}^{\lambda _{n}}h_{n}-\mathcal{L}^{\lambda
_{0}}h_{0},l\right) \right\vert \\
& \leq \left\vert \left( \mathcal{L}^{\lambda _{0}}\left( h_{n}-h_{0}\right)
,l\right) \right\vert +\left\vert \left( \left( \mathcal{L}^{\lambda _{n}}-%
\mathcal{L}^{\lambda _{0}}\right) h_{n},l\right) \right\vert \\
& \leq \left\vert \left( \left( h_{n}-h_{0}\right) ,\mathcal{L}^{\lambda
_{0}}l\right) \right\vert +\left\Vert \mathcal{L}^{\lambda _{n}}-\mathcal{L}%
^{\lambda _{0}}\right\Vert _{L^{2}\rightarrow L^{2}}\left\Vert
h_{n}\right\Vert _{L^{2}}\left\Vert l\right\Vert _{L^{2}}\rightarrow 0
\end{align*}%
as $n\rightarrow \infty $, by Lemma \ref{lemma-operator-property}. Thus $%
\mathcal{L}^{\lambda _{n}}h_{n}\rightarrow \mathcal{L}^{\lambda _{0}}h_{0}$
weakly in $H^{-2\dagger }$.

Secondly, for any $l\in L_{S}^{2}$, 
\begin{align*}
& \left\vert \left\langle \left( \mathcal{F}^{\lambda_{n}}\right) ^{\ast
}P_{n}^{\ast}b_{n}-\left( \mathcal{F}^{\lambda_{0}}\right)
^{\ast}k_{0},l\right\rangle \right\vert \\
& \leq\left\vert \left\langle \left( \left( \mathcal{F}^{\lambda_{n}}\right)
^{\ast}-\left( \mathcal{F}^{\lambda_{0}}\right) ^{\ast}\right)
k_{0},l\right\rangle \right\vert +\left\vert \left\langle \left( \mathcal{F}%
^{\lambda_{0}}\right) ^{\ast}\left( P_{n}^{\ast}b_{n}-k_{0}\right)
,l\right\rangle \right\vert \\
& \leq\left\Vert \mathcal{F}^{\lambda_{n}}-\mathcal{F}^{\lambda_{0}}\right%
\Vert _{L^{2}\rightarrow H_{C}^{-1\dagger}}\left\Vert \chi k_{0}\right\Vert
_{H^{1\dagger}}\left\Vert l\right\Vert _{L^{2}}+\left\vert \left\langle
\left( P_{n}^{\ast}b_{n}-k_{0}\right) ,\mathcal{F}^{\lambda
_{0}}l\right\rangle \right\vert \rightarrow0
\end{align*}
as $n\rightarrow\infty$ by Lemma \ref{operator bound}. Thus $\left( \mathcal{%
F}^{\lambda_{n}}\right) ^{\ast}P_{n}^{\ast}b_{n}\rightarrow\left( \mathcal{F}%
^{\lambda_{0}}\right) ^{\ast}k_{0}$ weakly in $L_{S}^{2}$.

Thirdly, for any $g\in H^{2\dagger}$, let $g_{n}=\sum_{j=1}^{n} c_n^j
\sigma_j \to g$ strongly in $H^{2\dagger}$ as $n\to\infty$. Let $%
\gamma_{n}=\left\{ c_n^j \right\} _{j=1}^{n}\in\mathbb{R}^{n}$. Then $%
P_{n}^{\ast} \gamma_{n}=g_{n}\ $and $\left\langle P_{n}\mathcal{F}%
^{\lambda_{n}}h_{n},\gamma_{n}\right\rangle =\left\langle \mathcal{F}%
^{\lambda_{n}}h_{n},g_{n}\right\rangle .$ Hence again using Lemma \ref%
{operator bound}, 
\begin{align*}
& \left\vert \left\langle P_{n}\mathcal{F}^{\lambda_{n}}h_{n},\gamma
_{n}\right\rangle -\left\langle \mathcal{F}^{\lambda_{0}}h_{0},g\right%
\rangle \right\vert \\
& \leq \left\vert \left\langle \left( \mathcal{F}^{\lambda_{n}}-\mathcal{F}%
^{\lambda_{0}}\right) h_{n},g_{n}\right\rangle \right\vert +\left\vert
\left\langle \mathcal{F}^{\lambda_{0}}h_{n},g_{n}-g\right\rangle \right\vert
+\left\vert \left\langle \mathcal{F}^{\lambda_{0}}\left( h_{n}-h_{0}\right)
,g\right\rangle \right\vert \\
& \leq \left\Vert \mathcal{F}^{\lambda_{n}}-\mathcal{F}^{\lambda_{0}}\right%
\Vert _{L^{2}\rightarrow H^{-1\dagger}}\left\Vert h_{n}\right\Vert
_{L^{2}}\left\Vert \chi g_{n}\right\Vert _{H^{1\dagger}}+\left\Vert \mathcal{%
F}^{\lambda_{0}}h_{n}\right\Vert _{H^{-1\dagger}}\left\Vert \chi\left(
g_{n}-g\right) \right\Vert _{H^{1\dagger}} \\
& +\left\vert \left\langle \left( h_{n}-h_{0}\right) ,\left( \mathcal{F}%
^{\lambda_{0}}\right) ^{\ast}g\right\rangle \right\vert \\
& \leq C_{1}\left\Vert \mathcal{F}^{\lambda_{n}}-\mathcal{F}%
^{\lambda_{0}}\right\Vert _{L^{2}\rightarrow H^{-1\dagger}}\left\Vert
h_{n}\right\Vert _{L^{2}}\left\Vert g\right\Vert _{{H}^{2\dagger}}+C_{2}%
\left\Vert h_{n}\right\Vert _{L^{2}}\left\Vert g_{n}-g\right\Vert _{{H}%
^{2\dagger}} \\
& + \left\vert \left\langle \left( h_{n}-h_{0}\right) ,\left( \mathcal{F}%
^{\lambda_{0}}\right) ^{\ast}g\right\rangle \right\vert \\
& \rightarrow0
\end{align*}
as $n\rightarrow0$. Thus $\left\langle P_{n}\mathcal{F}^{\lambda_{n}}h_{n},%
\gamma_{n}\right\rangle \rightarrow\left\langle \mathcal{F}^{\lambda
_{0}}h_{0},g\right\rangle $ for all $g\in H^{2\dagger}$.

Fourthly, using the same $g\in H^{2\dagger }$ as above,%
\begin{align*}
& \left\vert \left\langle P_{n}\mathcal{A}_{4}^{\lambda _{n}}P_{n}^{\ast
}b_{n},\gamma _{n}\right\rangle -\left\langle \mathcal{A}_{4}^{\lambda
_{0}}k_{0},g\right\rangle \right\vert \\
& =\left\vert \left\langle \mathcal{A}_{4}^{\lambda _{n}}P_{n}^{\ast
}b_{n},g_{n}\right\rangle -\left\langle \mathcal{A}_{4}^{\lambda
_{0}}k_{0},g\right\rangle \right\vert \\
& \leq \left\vert \left\langle \left( \mathcal{A}_{4}^{\lambda _{n}}-%
\mathcal{A}_{4}^{\lambda _{0}}\right) P_{n}^{\ast }b_{n},g\right\rangle
\right\vert +\left\vert \left\langle \mathcal{A}_{4}^{\lambda
_{0}}P_{n}^{\ast }b_{n},g_{n}-g\right\rangle \right\vert +\left\vert
\left\langle \mathcal{A}_{4}^{\lambda _{0}}\left( P_{n}^{\ast
}b_{n}-k_{0}\right) ,g\right\rangle \right\vert \\
& =I+II+III.
\end{align*}%
The first term on the right is estimated as 
\begin{align*}
I& \leq \left( \lambda _{n}^{2}-\lambda _{0}^{2}\right) \left\Vert
P_{n}^{\ast }b_{n}\right\Vert _{V^{1\dagger }}\left\Vert g\right\Vert
_{V^{1\dagger }} \\
& +\left\Vert \mathcal{G}^{\lambda _{n}}-\mathcal{G}^{\lambda
_{0}}\right\Vert _{H^{1\dagger }\rightarrow H^{-1\dagger }}\left\Vert \chi
P_{n}^{\ast }b_{n}\right\Vert _{H^{1\dagger }}\left\Vert \chi g\right\Vert
_{H^{1\dagger }} \\
& \rightarrow 0\,\text{, as }n\rightarrow \infty ,
\end{align*}%
where 
\begin{equation*}
\mathcal{G}^{\lambda }=\mathcal{E}^{\lambda }+\left( \mathcal{D}^{\lambda
}\right) ^{\ast }\left( \mathcal{A}_{1}^{\lambda }\right) ^{-1}\mathcal{D}%
^{\lambda }:H^{1\dagger }\rightarrow H^{-1\dagger }.
\end{equation*}%
By Lemma \ref{lemma-operator a5}$\left\Vert \mathcal{A}_{4}^{\lambda
_{0}}\right\Vert _{V^{2\dagger }\rightarrow V^{-2\dagger }}\leq C$, so 
\begin{equation*}
II\leq C\left\Vert P_{n}^{\ast }b_{n}\right\Vert _{V^{2\dagger }}\left\Vert
g_{n}-g\right\Vert _{V^{2\dagger }}\rightarrow 0\text{, as }n\rightarrow
\infty \text{. }
\end{equation*}%
As for the third term, $\mathcal{A}_{4}^{\lambda _{0}}g\in H^{-2\dagger }$
so that 
\begin{equation*}
III=\left\vert \left\langle \left( P_{n}^{\ast }b_{n}-k_{0}\right) ,\mathcal{%
A}_{4}^{\lambda _{0}}g\right\rangle \right\vert \rightarrow 0\text{, as }%
n\rightarrow \infty \text{.}
\end{equation*}%
So $\left\langle P_{n}\mathcal{A}_{4}^{\lambda _{n}}P_{n}^{\ast
}b_{n},\gamma _{n}\right\rangle \rightarrow \left\langle \mathcal{A}%
_{4}^{\lambda _{0}}k_{0},g\right\rangle $ for all $g\in H^{2\dagger }$. Thus
all four terms in (\ref{eqn-approximate-h}), (\ref{eqn-approximate-b})
converge and the limits satisfy (\ref{eqn-h0}) and (\ref{eqn-k0}).

It remains to show that $\left( h_{0},k_{0}\right) \neq\left( 0,0\right) $.
Let us write (\ref{eqn-approximate-h}) explicitly, using the definition of $%
\mathcal{L}^{\lambda_{n}}$, as 
\begin{equation*}
\left( -\Delta+\lambda_{n}^{2}\right) \left( e^{i\theta}h_{n}\right)
=e^{i\theta}\left( -\Delta+\tfrac{1}{r^{2}}+\lambda_{n}^{2}\right)
h_{n}=f_{n},
\end{equation*}
where 
\begin{align*}
f_{n} & =-e^{i\theta}\left( \mathcal{F}^{\lambda_{n}}\right)
^{\ast}P_{n}^{\ast}b_{n} + e^{i\theta}\int r\hat{v}_{\theta}\mu_{p}dv\,h_{n}
\\
& + e^{i\theta}\int\hat{v}_{\theta}\mu_{e}\mathcal{Q}^{\lambda}\left( \hat {v%
}_{\theta}h\right) dv-e^{i\theta}\left( \mathcal{B}^{\lambda_{n}}\right)
^{\ast}\left( \mathcal{A}_{1}^{\lambda_{n}}\right) ^{-1}\mathcal{B}%
^{\lambda_{n}}h_{n}.
\end{align*}
By Lemmas \ref{lemma-operator-property} and \ref{operator bound}, $f_{n}$ is
bounded in $L^{2}\left( \mathbb{R}^{3}\right) $ and has support in the fixed
bounded set $S_{x}\subset\mathbb{R}^{3}$. Therefore the inversion of the
operator $\left( -\Delta+\lambda_{n}^{2}\right) $ with $\lambda_{n}\geq%
\lambda_{3}>0$ implies that $h_{n}$ decays exponentially as $\left\vert
x\right\vert \rightarrow\infty\,,$ uniformly in $n$. Thus $\left\{
h_{n}\right\} $ is compact in $L^{2}$, so that $h_{n}\rightarrow h_{0}$
strongly in $L^{2}$. Since $\left\Vert P_{n}^{\ast}b_{n}\right\Vert
_{V^{2\dagger}}$ is uniformly bounded, $\chi
P_{n}^{\ast}b_{n}\rightarrow\chi k_{0}$ strongly in $H^{1\dagger}$.
Therefore, we have $\left\Vert h_{0}\right\Vert _{L^{2}}+\left\Vert \chi
k_{0}\right\Vert _{H^{1\dagger}}=1$ and so $\left( h_{0},k_{0}\right)
\neq\left( 0,0\right) $.
\end{proof}

%%%%%%%%%%%% Section 8 %%%%%%%%%%%%%

\section{Growing mode}

Changing notation, $A_{\theta}=h_{0},\ \pi=k_{0},\ $ and replacing $%
\lambda_{0}$ by $\lambda$, we have from (\ref{eqn-h0}) and (\ref{eqn-k0})
the pair of equations 
\begin{equation}
\mathcal{L}^{\lambda}A_{\theta}=-\left( \mathcal{F}^{\lambda}\right) ^{\ast
}\pi,\text{ \ }\mathcal{A}_{4}^{\lambda}\pi=\mathcal{F}^{\lambda}A_{\theta}
\label{eqn-growing mode}
\end{equation}
where $\left( A_{\theta},\pi\right) \neq\left( 0,0\right) ,\ A_{\theta}\in
H^{2\dagger},\ \pi\in H^{2\dagger},\ \lambda\in\left( 0,+\infty\right) $. We
must define $f,\phi$ and $\mathbf{A }$ so that (\ref{lin-3dvla}), (\ref%
{lin-maxwell-density}) and (\ref{lin-maxwell-current}) are satisfied by $%
e^{\lambda t}\left( f,\phi,\mathbf{A }\right) $. Indeed, motivated by
Section 3, we define

\begin{equation*}
A_{r}=-\partial _{z}\pi ,\ A_{z}=\tfrac{1}{r}\partial _{r}\left( r\pi
\right) ,\ \mathbf{A}=\left( A_{r},A_{\theta },A_{z}\right) ,
\end{equation*}%
\begin{equation}
\phi =\left( \mathcal{A}_{1}^{\lambda }\right) ^{-1}\left( \mathcal{B}%
^{\lambda }A_{\theta }+\mathcal{D}^{\lambda }\pi \right) ,
\label{eqn-electric}
\end{equation}%
\begin{equation*}
\mathbf{E}=-\nabla \phi -\lambda \mathbf{A}\ ,\quad \mathbf{\ \ B=\nabla
\times {A}},
\end{equation*}%
and 
\begin{equation}
f\left( x,v\right) =-\mu _{e}\phi +\mu _{e}\mathcal{Q}^{\lambda }\phi -\mu
_{p}rA_{\theta }-\mu _{e}\mathcal{Q}^{\lambda }\left( \hat{v}_{\theta
}A_{\theta }\right) -\mu _{e}\mathcal{Q}^{\lambda }\left( G\pi \right) .
\label{f-formula}
\end{equation}%
It follows directly that $\nabla \cdot {\mathbf{A}}=0,\ {\mathbf{A}}\in
H^{1},\ \phi \in V^{1},\ \mathbf{E}\in L^{2},\ \mathbf{B}\in L^{2}$, $%
A_{\theta }\in L^{\infty }$ and by Lemmas \ref{lemma-operator-property}, \ref%
{operator bound} and \ref{lemma-property-Qlambda}, $f\in L^{2}(\mathbb{R}%
^{3}\times \mathbb{R}^{3})$.

%%%%%%%%%%% Lemma 8.1 %%%%%%%%%%%

\begin{lemma}
The Poisson equation $-\Delta \phi =\rho $ is satisfied. Moreover, $\phi \in 
$ $H^{2}\left( \mathbb{R}^{3}\right) $ and $f\in L^{\infty }(\mathbb{R}%
^{3}\times \mathbb{R}^{3})$.
\end{lemma}

\begin{proof}
By (\ref{eqn-electric}), we have $\mathcal{A}_{1}^{\lambda }\phi =\mathcal{B}%
^{\lambda }A_{\theta }+\mathcal{D}^{\lambda }\pi ,\ $which is written
explicitly as%
\begin{align*}
-\Delta \phi & =\left( \int \mu _{e}dv\right) \phi -\int \mu _{e}\mathcal{Q}%
^{\lambda }\phi dv-\left( \int \hat{v}_{\theta }\mu _{e}dv\right) A_{\theta }
\\
& +\int \mu _{e}\mathcal{Q}^{\lambda }\left( \hat{v}_{\theta }A_{\theta
}\right) dv+\int \mu _{e}\mathcal{Q}^{\lambda }\left( G\pi \right) dv.
\end{align*}
On the other hand, by (\ref{f-formula}) and $\int \left( r\mu _{p}+\hat{v}%
_{\theta }\mu _{e}\right) dv=0,$ we get exactly the same expression for $%
\rho =-\int fdv$. Now integrating (\ref{f-formula}) in $v$ and $x$, we find
that the first and second terms cancel, the third and fourth terms cancel,
and the fifth term $\iint \mu_e\, G\pi\, dvdx$ vanishes by the oddness of
the integrand in $(v_r,v_z)$. Thus $\int \rho dx=-\int \int fdxdv=0$.
Furthermore, $\rho $ has compact support. So by the proof of Lemma 3.2 of 
\cite{lw-linear}, $\phi \in L^{2}$. Since $\rho \in L^{2}$, by elliptic
regularity we have $\phi \in H^{2}\subset L^\infty.$

Moreover, 
\begin{eqnarray*}
\left( -\Delta +\tfrac{1}{r^{2}}\right) \left( -\Delta +\tfrac{1}{r^{2}}%
+\lambda ^{2}\right) \pi &=&\mathcal{A}_{4}^{\lambda }\pi +\mathcal{G}%
^{\lambda }\pi \\
&=&\mathcal{F}^{\lambda }A_{\theta }+\mathcal{E}^{\lambda }\pi +(\mathcal{D}%
^{\lambda })^{\ast }(\mathcal{A}_{1}^{\lambda })^{-1}\mathcal{D}^{\lambda
}\pi \in H^{-1\dagger }
\end{eqnarray*}%
so that $\pi \in V^{3\dagger }$ and $G\pi \in L^{\infty }$. Therefore from (%
\ref{f-formula}), $f\in L^{\infty }(\mathbb{R}^{3}\times \mathbb{R}^{3})$.
\end{proof}

%%%%%%%%%%% Lemma 8.2 %%%%%%%%%%

\begin{lemma}
The function $f$ defined by (\ref{f-formula}) satisfies (\ref{lin-3dvla}).
\end{lemma}

\begin{proof}
We have 
\begin{equation*}
f=-\mu _{e}\phi +\mu _{e}\mathcal{Q}^{\lambda }\phi -\mu _{p}rA_{\theta
}-\mu _{e}\mathcal{Q}^{\lambda }\left( \hat{v}\cdot {\mathbf{A}}\right) .
\end{equation*}%
To show that $f$ is a weak solution of (\ref{lin-3dvla}), we take any $g\in
C_{c}^{1}\left( \mathbb{R}^{3}\times \mathbb{R}^{3}\right) ,$ and compute 
\begin{align*}
& \iint_{\mathbb{R}^{3}\times \mathbb{R}^{3}}\left( Dg\right) fdxdv \\
& =\iint_{\mathbb{R}^{3}\times \mathbb{R}^{3}}\left( Dg\right) \left( -\mu
_{e}\phi \right) dxdv+\iint_{\mathbb{R}^{3}\times \mathbb{R}^{3}}\left(
Dg\right) \mu _{e}\mathcal{Q}^{\lambda }\phi dxdv \\
& +\iint_{\mathbb{R}^{3}\times \mathbb{R}^{3}}\left( Dg\right) \left( -\mu
_{p}rA_{\theta }\right) dxdv-\iint_{\mathbb{R}^{3}\times \mathbb{R}%
^{3}}\left( Dg\right) \mu _{e}\mathcal{Q}^{\lambda }\left( \hat{v}\cdot 
\mathbf{A}\right) dxdv \\
& =I+II+III+IV.
\end{align*}%
Since $D$ is skew-adjoint, the first term is 
\begin{equation*}
I=\iint_{\mathbb{R}^{3}\times \mathbb{R}^{3}}gD\left( \mu _{e}\phi \right)
dxdv=\iint_{\mathbb{R}^{3}\times \mathbb{R}^{3}}\mu _{e}gD\phi dxdv.
\end{equation*}%
Similarly, 
\begin{equation*}
III=\iint_{\mathbb{R}^{3}\times \mathbb{R}^{3}}\mu _{p}gD\left( rA_{\theta
}\right) dxdv.
\end{equation*}
By definition of $\mathcal{Q}$, 
\begin{align*}
II& =\int_{-\infty }^{0}\lambda e^{\lambda s}\iint_{\mathbb{R}^{3}\times 
\mathbb{R}^{3}}\mu _{e}\ Dg(x,v)\ \phi \left( X(s;x,v)\right) dxdvds \\
& =\int_{-\infty }^{0}\lambda e^{\lambda s}\iint_{\mathbb{R}^{3}\times 
\mathbb{R}^{3}}\mu _{e}\left( Dg\right) \left( X(-s),V(-s)\right) \phi
\left( x\right) dxdvds \\
& =\iint_{\mathbb{R}^{3}\times \mathbb{R}^{3}}\mu _{e}\int_{-\infty
}^{0}\lambda e^{\lambda s}\left( -\frac{d}{ds}g\left( X(-s),V(-s)\right)
\right) ds\ \phi \left( x\right) dxdv \\
& =\iint_{\mathbb{R}^{3}\times \mathbb{R}^{3}}\mu _{e}\left\{ -\lambda
g\left( x,v\right) +\int_{-\infty }^{0}\lambda ^{2}e^{\lambda s}g\left(
X(-s),V(-s)\right) ds\right\} \phi \left( x\right) dxdv \\
& =\iint_{\mathbb{R}^{3}\times \mathbb{R}^{3}}\left\{ -\mu _{e}\lambda \phi
\left( x\right) +\mu _{e}\int_{-\infty }^{0}\lambda ^{2}e^{\lambda s}\phi
\left( X(s),V(s)\right) ds\right\} g\left( x,v\right) dxdv \\
& =\lambda \iint_{\mathbb{R}^{3}\times \mathbb{R}^{3}}\left\{ -\mu _{e}\phi
+\mu _{e}\mathcal{Q}^{\lambda }\phi \right\} g\ dxdv.
\end{align*}
The preceding calculations are valid since $\phi$ belongs to $H^{2}$ and
thus is continuous. Similarly, 
\begin{equation*}
IV=-\lambda \iint_{\mathbb{R}^{3}\times \mathbb{R}^{3}}\left\{ -\mu _{e}\hat{%
v}\cdot {A}+\mu _{e}\mathcal{Q}^{\lambda }\left( \hat{v}\cdot \mathbf{A}%
\right) \right\} g\ dxdv.
\end{equation*}
So we have 
\begin{align*}
& \iint_{\mathbb{R}^{3}\times \mathbb{R}^{3}}\left( Dg\right) fdxdv \\
& =\iint_{\mathbb{R}^{3}\times \mathbb{R}^{3}}\lambda \left\{ -\mu _{e}\phi
+\mu _{e}\mathcal{Q}^{\lambda }\phi +\mu _{e}\mathcal{Q}^{\lambda }\left( 
\hat{v}\cdot {\mathbf{A}}\right) \right\} g\ dxdv \\
& +\iint_{\mathbb{R}^{3}\times \mathbb{R}^{3}}\left\{ \mu _{e}D\phi +\mu
_{p}D\left( rA_{\theta }\right) +\lambda \mu _{e}\hat{v}\cdot \mathbf{A}%
\right\} g\ dxdv \\
& =\iint_{\mathbb{R}^{3}\times \mathbb{R}^{3}}\left\{ \lambda \left( f+\mu
_{p}rA_{\theta }\right) +\mu _{e}D\phi +\mu _{p}D\left( rA_{\theta }\right)
+\lambda \mu _{e}\hat{v}\cdot \mathbf{A}\right\} g\ dxdv.
\end{align*}%
So $f$ satisfies weakly the equation%
\begin{equation*}
\left( \lambda +D\right) f=-\mu _{e}D\phi -\mu _{p}D\left( rA_{\theta
}\right) -\lambda \mu _{p}rA_{\theta }-\lambda \mu _{e}\hat{v}\cdot \mathbf{A%
}
\end{equation*}%
which is exactly (\ref{lin-3dvla}).
\end{proof}

%%%%%%%%%%%% Lemma 8.3 %%%%%%%%%%%%%%

\begin{lemma}
Denoting $\rho=-\int fdv$ and $\mathbf{j}=-\int\hat{v}f$ $dv,$ we have the
continuity equation $\lambda\rho+\nabla\cdot\mathbf{j}=0$.
\end{lemma}

\begin{proof}
By the last lemma, $f$ satisfies (\ref{lin-3dvla}) weakly, which can be
written as 
\begin{align}
& \lambda f+\nabla_{x}\cdot\left( \hat{v}f\right) -\nabla_{v}\cdot\left\{
\left( \mathbf{E}^{0}+\mathbf{E}^{ext}+\hat{v}\times\left( \mathbf{B}^{0}+%
\mathbf{B}^{ext}\right) \right) f\right\}  \label{eqn-f-divergence} \\
& =-\nabla_{v}\cdot\left\{ \left( \mathbf{E}+\hat{v}\times\mathbf{B}\right)
f^{0}\right\} .  \notag
\end{align}
The last equality follows in the same way that (\ref{lin-3dvla}) was
derived. Let $\varsigma\left( v\right) \in C_{c}^{1}\left( \mathbb{R}%
^{3}\right) $ to be a cut-off function for the $v$-support of $\mu(e,p)$.
Taking any $h\left( x\right) \in C_{c}^{1}\left( \mathbb{R}^{3}\right) $ and
using $\varsigma\left( v\right) h\left( x\right) $ as a test function for (%
\ref{eqn-f-divergence}), all the terms coming from $v$-divergences vanish
and we have 
\begin{equation*}
\int\lambda\rho\left( x\right) h\left( x\right) dx-\int\mathbf{j}\cdot\nabla
hdx=0.
\end{equation*}
So $\lambda\rho+\nabla\cdot\mathbf{j}=0$ weakly$.$
\end{proof}

%%%%%%%%%%%% Lemma 8.4 %%%%%%%%%%%%%%

\begin{lemma}
The Maxwell equation (\ref{lin-maxwell-current}) is satisfied.
\end{lemma}

\begin{proof}
By (\ref{f-formula}), we have%
\begin{align*}
\mathbf{j}& =-\int\hat{v}fdv=\left( \int\hat{v}\mu_{e}dv\right) \phi -\int%
\hat{v}\mu_{e}\mathcal{Q}^{\lambda}\phi\ dv \\
& +\left( \int\hat{v}\mu_{p}dv\right) rA_{\theta}+\int\hat{v}\mu _{e}%
\mathcal{Q}^{\lambda}\left( \hat{v}_{\theta}A_{\theta}\right) dv+\int\hat{v}%
\mu_{e}\mathcal{Q}^{\lambda}\left( G\pi\right) dv.
\end{align*}
Its $\theta$-component can be written as 
\begin{equation*}
j_{\theta}=-\left( \mathcal{B}^{\lambda}\right) ^{\ast}\phi -\mathcal{A}%
_{2}^{\lambda}A_{\theta}+\left( -\Delta+\frac{1}{r^{2}}+\lambda^{2}\right)
A_{\theta}+\mathcal{C}^{\lambda}\pi.
\end{equation*}
By the definition of $\phi$, 
\begin{equation*}
-\left( \mathcal{B}^{\lambda}\right) ^{\ast}\phi=-\left( \mathcal{B}%
^{\lambda}\right) ^{\ast}\left( \mathcal{A}_{1}^{\lambda}\right) ^{-1}%
\mathcal{B}^{\lambda}A_{\theta}-\left( \mathcal{B}^{\lambda}\right)
^{\ast}\left( \mathcal{A}_{1}^{\lambda}\right) ^{-1}\mathcal{D}^{\lambda}\pi.
\end{equation*}
By the definition of $\mathcal{L}^{\lambda},$ 
\begin{equation*}
-\mathcal{A}_{2}^{\lambda}A_{\theta}=-\mathcal{L}^{\lambda}A_{\theta}+\left( 
\mathcal{B}^{\lambda}\right) ^{\ast}\left( \mathcal{A}_{1}^{\lambda}\right)
^{-1}\mathcal{B}^{\lambda}A_{\theta}.
\end{equation*}
By (\ref{eqn-growing mode}), 
\begin{equation*}
-\mathcal{L}^{\lambda}A_{\theta}=\left( \mathcal{F}^{\lambda}\right) ^{\ast
}\pi=\left( \mathcal{B}^{\lambda}\right) ^{\ast}\left( \mathcal{A}%
_{1}^{\lambda}\right) ^{-1}\mathcal{D}^{\lambda}\pi-\mathcal{C}^{\lambda}\pi.
\end{equation*}
Adding the last three equations, we obtain%
\begin{equation*}
-\left( \mathcal{B}^{\lambda}\right) ^{\ast}\phi-\mathcal{A}%
_{2}^{\lambda}A_{\theta}+\mathcal{C}^{\lambda}\pi=0,
\end{equation*}
so that 
\begin{equation*}
j_{\theta}=\left( -\Delta+\tfrac{1}{r^{2}}+\lambda^{2}\right) A_{\theta}
\end{equation*}
and 
\begin{equation*}
j_{\theta}\mathbf{e}_{\theta}=\left( \lambda^{2}-\Delta\right) \left(
A_{\theta}\mathbf{e}_{\theta}\right) .
\end{equation*}
Because $\nabla\phi$ has no $\theta$-component, this result is the $\theta $%
-component of the Maxwell equation (\ref{lin-maxwell-current}).

It remains to derive the $r$ and $z$ components of (\ref{lin-maxwell-current}%
). By (\ref{eqn-growing mode}), (\ref{eqn-electric}) and (\ref{f-formula}),
it follows exactly as in the proof of Lemma \ref{lemma-3.1} that 
\begin{align*}
\left( -\Delta+\tfrac{1}{r^{2}}\right) \left( -\Delta+\tfrac{1}{r^{2}}%
+\lambda^{2}\right) \pi & =\left( \mathcal{D}^{\lambda}\right) ^{\ast
}\phi-\left( \mathcal{C}^{\lambda}\right) ^{\ast}A_{\theta}+\mathcal{E}%
^{\lambda}\pi \\
& =\partial_{z}j_{r}-\partial_{r}j_{z}.
\end{align*}
As in that proof, we introduce $\mathbf{K }=j_{r}\mathbf{e}_{r}+j_{z}\mathbf{%
e}_{z}$ and $\mathbf{I }=\left( -\Delta\right) ^{-1}\mathbf{K }$. Then 
\begin{equation*}
\left( -\Delta+\tfrac{1}{r^{2}}\right) \left( -\Delta+\tfrac{1}{r^{2}}%
+\lambda^{2}\right) \pi=\left( -\Delta+\tfrac{1}{r^{2}}\right) \left(
\partial_{z}I_{r}-\partial_{r}I_{z}\right)
\end{equation*}
so that 
\begin{equation*}
\left( -\Delta+\tfrac{1}{r^{2}}+\lambda^{2}\right)
\pi=\partial_{z}I_{r}-\partial_{r}I_{z}.
\end{equation*}
This result can be rewritten as 
\begin{equation*}
\left( \lambda^{2}-\Delta\right) \left( \pi\mathbf{e}_{\theta}\right)
=\nabla\times\mathbf{I }.
\end{equation*}
Taking the curl of both sides, 
\begin{equation*}
\left( \lambda^{2}-\Delta\right) \left( A_{r}\mathbf{e}_{r}+A_{z}\mathbf{e}%
_{z}\right) =-\Delta\mathbf{I }+\nabla\left( \nabla\cdot\mathbf{I }\right) .
\end{equation*}
But 
\begin{equation*}
\nabla\cdot\mathbf{I }=\nabla\cdot\left( -\Delta\right) ^{-1}\mathbf{K }%
=\left( -\Delta\right) ^{-1}\nabla\cdot\vec{j}=\lambda\Delta^{-1}\rho=-%
\lambda \phi,
\end{equation*}
so that%
\begin{equation*}
\left( \lambda^{2}-\Delta\right) \left( A_{r}\mathbf{e}_{r}+A_{z}\mathbf{e}%
_{z}\right) =\mathbf{K }-\lambda\nabla\phi.
\end{equation*}
In components, this means 
\begin{equation*}
\left( \lambda^{2}-\Delta\right) A_{r}=j_{r}-\lambda\partial_{r}\phi\ ,\
\quad \left( \lambda^{2}-\Delta\right) A_{z}=j_{z}-\lambda\partial _{z}\phi\
,
\end{equation*}
which are precisely the $r$ and $z$ components of (\ref{lin-maxwell-current}%
).
\end{proof}

This completes the proof of Theorem \ref{3d growing} (i). To prove Theorem %
\ref{3d growing} (ii) on the number of growing modes, we first note that for
each $n-$truncated problem, it follows from the continuation argument that
the number of approximate growing modes is bounded below by the dimension of
the negative eigenspace of $\mathcal{L}^{0}$. Since we have the uniform
control of the converging process as $n\rightarrow \infty \,,$ the lower
bound for the number of exact growing modes follows. The proof of the upper
bound is the same as in the $1\frac{1}{2}$D case and we omit it. \qed

\begin{remark}
In this $3$D case we do not have much regularity of $f$ and the growing mode
is only shown to satisfy the linearized equation weakly. This is mainly due
to the complicated behavior of the $3D$ particle trajectories. To see this
difficulty more clearly, we formally differentiate $f$ given by (\ref%
{f-formula}) and look at a typical term 
\begin{equation*}
\int_{-\infty }^{0}\int \int \mu _{e}\lambda e^{\lambda s}\bigtriangledown
_{x}\phi \left( X(s;x,v)\right) \frac{\partial X(s;x,v)}{\partial v}dxdvds.
\end{equation*}
If the stretching factor $\frac{\partial X(s;x,v)}{\partial v}$ grows like $%
e^{a\left\vert s\right\vert }$ with $a>\lambda $, the integral diverges and
we lose the differentiability of $f$. In the $1\frac{1}{2}$D case it is
possible to prove (see \cite{lw-nonlinear}) some regularity of $f$ by
estimating an averaged Liapunov exponent for the quantity $\int \int
\left\vert \frac{\partial X(s;x,v)}{\partial v}\right\vert dxdv$. This idea
was first introduced in the $1$D Vlasov-Poisson in \cite{lin-cpam} and it
works for integrable trajectories. However, the $3$D trajectory in general
is non-integrable so that the idea fails. For this reason we have had to
study the operators $\left( \ \mathcal{C}^{\lambda }\right) ^{\ast },\
\left( \mathcal{D}^{\lambda }\right) ^{\ast },\ \mathcal{E}^{\lambda },\ 
\mathcal{F}^{\lambda }$ and $\mathcal{A}_{4}^{\lambda }$ with ranges in
negative Sobolev spaces. We note as well that the non-integrability of
trajectories is the main reason for the difficulty of passing from linear to
nonlinear instability.
\end{remark}

%%%%%%%%%%%%%%%% Section 9 %%%%%%%%%%%%%%%%%%%

\section{Non-monotone Equilibria}

In case $\mu _{e}$ changes sign, it does not seem possible to extend the
methods of \cite{lw-linear} to get linear stability. However, we can still
get sufficient conditions for linear \textit{instability} by extending the
matrix formulation of this paper. If $\mu _{e}$ changes sign, we will
reformulate the growing mode problem as a $3\times 3$ matrix operator $%
\mathcal{M}^{\lambda }$ depending on a positive parameter $\lambda >0$ and
then look for the change of the signature of $\mathcal{M}^{\lambda }$ as $%
\lambda$ goes from $0$ to $+\infty $.

In the discussion below, we illustrate this idea only for a simple case,
namely a purely magnetic equilibrium of $1\frac{1}{2}D$ RVM system with two
species. Assume now that 
\begin{equation}
\mu ^{+}(e,p)=\mu ^{-}(e,-p).  \label{parity}
\end{equation}%
Then an purely magnetic equilibrium is obtained with electric potential $%
\phi ^{0}\equiv 0$ and magnetic potential $\psi ^{0}$ satisfying the ODE 
\begin{equation*}
\partial _{x}^{2}\psi ^{0}=2\int \hat{v}_{2}\mu ^{-}(\langle v\rangle
,v_{2}-\psi ^{0}(x))dv.
\end{equation*}%
We use the same notation as in \cite{lw-linear} and \cite{lw-nonlinear}.
Define 
\begin{equation*}
\mathcal{A}_{1}^{0}h=-\partial _{x}^{2}h-\left( \int 2\mu _{e}^{-}dv\right)
h+\int 2\mu _{e}^{-}\ \mathcal{P}^{-}h\ dv,
\end{equation*}%
\begin{equation*}
\mathcal{A}_{2}^{0}h=-\partial _{x}^{2}h-\left( 2\int \hat{v}_{2}\mu
_{p}^{-}dv\right) h-\int 2\mu _{e}^{-}\hat{v}_{2}\mathcal{P}^{-}(\hat{v}%
_{2}h)\ dv,
\end{equation*}%
\begin{equation*}
k^{0}=\int_{0}^{P}\int \mu _{e}^{-}\left( \mathcal{P}^{-}\left( \hat{v}%
_{1}\right) \right) ^{2}dvdx
\end{equation*}%
where $\mathcal{P}^{-}\ $is the projection operator of $L_{\left\vert \mu
_{e}^{-}\right\vert }^{2}$ onto $\ker D^{-}$ and $D^{-}=\hat{v}_{1}\partial
_{x}-\hat{v}_{2}B^{0}\partial _{v_{1}}+\hat{v}_{1}B^{0}\partial _{v_{2}}$.
Denote by $n\left( \mathcal{A}_{1}^{0}\right) $ and $n\left( \mathcal{A}%
_{2}^{0}\right) $ the number of negative eigenvalues of $\mathcal{A}_{1}^{0}$
and $\mathcal{A}_{2}^{0}$.

\begin{theorem}
\label{theorem-pure-extension}Consider a periodic purely magnetic
equilibrium as above. Assume $\ker \mathcal{A}_{1}^{0} =\left\{ 0\right\} $.
Then the equilibrium is spectrally unstable if either

(i) $l^{0}<0$ and $n\left( \mathcal{A}_{1}^{0}\right) $ $\neq n\left( 
\mathcal{A}_{2}^{0}\right) .$ \newline
or

(ii) $l^{0}>0$ and $n\left( \mathcal{A}_{1}^{0}\right) +1\neq n\left( 
\mathcal{A}_{2}^{0}\right) .$
\end{theorem}

\begin{proof}
(sketched) As we are merely sketching the extension of our results to this
case, let us take just one species and use the notation in Section 2.
Finding a growing mode $e^{\lambda t}\left( f,E_{1},E_{2},B\right) $ with $%
\lambda>0$ is equivalent to solving (\ref{equation-phi}), ( \ref%
{equation-pci}) and (\ref{equation-b}) for $\left( \phi,\psi,b\right) $
where $\left( \phi,\psi\right) $ is the electromagnetic potential and $b\in 
\mathbb{R}^{1}$. We define the rank-one operators $\mathcal{C}^{\lambda},\ 
\mathcal{D} ^{\lambda}:\mathbb{R}^{1}\rightarrow L_{P}^{2}$ by $\mathcal{C}
^{\lambda}\left( b\right) =bb^{\lambda}$ and $\mathcal{D}^{\lambda}\left(
b\right) =bc^{\lambda}.$ Then $\left( \phi,\psi,b\right) $ satisfies the
matrix equation 
\begin{equation*}
\left( 
\begin{array}{ccc}
-\mathcal{A}_{1}^{\lambda} & \mathcal{B}^{\lambda} & C^{\lambda} \\ 
(\mathcal{B}^{\lambda})^{\ast} & \mathcal{A}_{2}^{\lambda} & -\mathcal{D}
^{\lambda} \\ 
\left( C^{\lambda}\right) ^{\ast} & -\left( \mathcal{D}^{\lambda}\right)
^{\ast} & -P\left( \lambda^{2}-l^{\lambda}\right)%
\end{array}
\right) \left( 
\begin{array}{c}
\phi \\ 
\psi \\ 
b%
\end{array}
\right) =\mathcal{M}^{\lambda}\left( 
\begin{array}{c}
\phi \\ 
\psi \\ 
b%
\end{array}
\right) =0.
\end{equation*}
This $3\times 3$ matrix $\mathcal{M}^\lambda$ is different from the $2\times
2$ one of the previous sections. Notice that $\mathcal{M}^{\lambda}$ is
formally self-adjoint.

Let us look at the asymptotic behavior of $\mathcal{M}^{\lambda }$. As $%
\lambda \rightarrow +\infty $, we can show that the off-diagonal terms $%
\mathcal{B}^{\lambda },\ \mathcal{C}^{\lambda },\ \mathcal{D}^{\lambda
}\rightarrow 0$ and $\mathcal{A}_{1}^{\lambda }\rightarrow -\frac{d}{dx^{2}}%
>0$, by noticing that 
\begin{equation*}
\lim_{\lambda \rightarrow \infty }\int_{-\infty }^{0}\lambda e^{\lambda
s}h(X(s))ds\rightarrow h\left( x\right)
\end{equation*}%
strongly in $L_{P}^{2}$, which is the analogue of Lemma \ref%
{lemma-property-Qlambda}(\textit{e}). We also have $\mathcal{A}_{2}^{\lambda
}>0$ for large $\lambda $. As $\lambda \searrow 0$, we have $\mathcal{C}%
^{\lambda },\ \mathcal{D}^{\lambda }\rightarrow 0$ as shown in the proof of
Lemma \ref{continuation lemma}. Moreover, it was shown in Lemmas 4.2 and 3.1
of \cite{lw-linear} that for a purely magnetic equilibrium, $\mathcal{B}%
^{\lambda }\rightarrow 0$ strongly as $\lambda \searrow 0$. So $\mathcal{M}%
^{\lambda }$ tends to a diagonal operator as $\lambda $ tends to $0$ and the
same as $\lambda$ tends to $\infty $. Now $\mathcal{A}_{1}^{\lambda },\ 
\mathcal{A}_{2}^{\lambda }$ and $l^{\lambda }$ tend to $\mathcal{A}%
_{1}^{0},\ \mathcal{A}_{2}^{0}$ and $l^{0}\ $as $\lambda \searrow 0$.

We want to show that $\mathcal{M}^{\lambda} $ has a different signature for
small and large $\lambda $. For then a continuation argument should ensure
the existence of a nontrivial kernel for some $\mathcal{M}^{\lambda }$.
However since $\mathcal{M}^{\lambda }$ is not bounded either from below or
from above, in order to make the argument rigorous we must truncate as in
the $3D$ case. We truncate the $\phi $-component (but not the other
components) to an $n$-dimensional subspace which does not spoil the negative
space of $\mathcal{A}_1^0$; that is, we project onto the lowest $n$ modes of 
$\mathcal{A}_1^0$. We denote the resulting truncated matrix operator by $%
\mathcal{M}_{n}^{\lambda }$. Then for large $\lambda$, say $\lambda \ge
\Lambda$, $\mathcal{M}_{n}^{\lambda }$ has $n+0+1$ negative eigenvalues. In
case $l^0<0$, $\mathcal{M}_{n}^0$ has $(n-n(\mathcal{A}_1^0))+n(\mathcal{A}%
_2^0) +1$ negative eigenvalues. In case $l^0>0$, $\mathcal{M}_{n}^0$ has $%
(n-n(\mathcal{A}_1^0))+n(\mathcal{A}_2^0) +0$ negative eigenvalues.
Therefore $\mathcal{M}_{n}^0$ and $\mathcal{M}_{n}^\Lambda$ have a different
number of negative eigenvalues in both cases (i) and (ii). By continuation, $%
\mathcal{\ M}_{n}^{\lambda }$ has a nontrivial kernel for some $\lambda>0$.
Then we let $n$ go to $+\infty $ to obtain a nontrivial kernel for $\mathcal{%
M}^{\lambda }$. As the details are somewhat similar to the 3D cylindrical
case, we omit them.
\end{proof}

For purely magnetic equilibria, in case $\mu_{e}<0$, we have $\mathcal{A}%
_{1}^{0}>0$ and $l^{0}<0$. In this case, it was shown in \cite{lw-linear}
that $n\left( \mathcal{A}_{2}^{0}\right) \neq0$ is the sharp condition for
linear instability. So Theorem \ref{theorem-pure-extension} is a
generalization of that instability result to the case of a general purely
magnetic equilibrium with nonmonotone $\mu$. Moreover, it was shown in \cite%
{lw-nonlinear} that these linear instability results imply nonlinear
instability in the macroscopic sense.

For the $3D$ case with $\mu _{e}$ of general sign, one can also use the same
idea. The equations (\ref{equation-phi-3d}), (\ref{equation-atheta-3d}) and (%
\ref{equation-pi-3d}) for $\left( \phi ,A_{\theta },\pi \right) $ can be
rewritten as 
\begin{equation*}
\left( 
\begin{array}{ccc}
-\mathcal{A}_{1}^{\lambda } & \mathcal{B}^{\lambda } & -\mathcal{D}^{\lambda
} \\ 
(\mathcal{B}^{\lambda })^{\ast } & \mathcal{A}_{2}^{\lambda } & C^{\lambda }
\\ 
-\left( \mathcal{D}^{\lambda }\right) ^{\ast } & \left( \mathcal{C}^{\lambda
}\right) ^{\ast } & \mathcal{A}_{3}^{\lambda }%
\end{array}%
\right) \left( 
\begin{array}{c}
\phi \\ 
A_{\theta } \\ 
\pi%
\end{array}%
\right) =\mathcal{M}^{\lambda }\left( 
\begin{array}{c}
\phi \\ 
A_{\theta } \\ 
\pi%
\end{array}%
\right) =0.
\end{equation*}%
Again $\mathcal{M}^{\lambda }$ is formally self-adjoint. By studying the
difference of the signatures of $\mathcal{M}^{\lambda }$ at $0$ and at $%
\infty $, one can obtain sufficient conditions for linear instability of
general equilibria, which will generalize the instability criterion of the
monotone case. However we do not pursue the details here.

%%%%%%%%%%%%%%  Appendix  %%%%%%%%%%%%%%%%%%%%

\section{Appendix}

In this appendix, we list some common formulae in the cylindrical
coordinates. Assume $\psi=\psi\left( r,\theta,z \right) $ is a scalar
function and $\mathbf{A =}\left( A_{r},A_{\theta},A_{z}\right) $ is a vector
function.

\begin{equation*}
\nabla\psi=\frac{\partial\psi}{\partial r}\mathbf{e}_{r}+\frac{1}{r}\frac{%
\partial\psi}{\partial\theta}\mathbf{e}_{\theta}+\frac{\partial\psi }{%
\partial z}\mathbf{e}_{z},
\end{equation*}%
\begin{equation*}
\Delta\psi=\frac{1}{r}\frac{\partial}{\partial r}\left( r\frac{\partial\psi 
}{\partial r}\right) +\frac{1}{r^{2}}\frac{\partial^{2}\psi}{\partial
\theta^{2}}+\frac{\partial^{2}\psi}{\partial z^{2}},
\end{equation*}%
\begin{equation*}
\nabla \cdot\mathbf{A =}\frac{1}{r}\frac{\partial\left( rA_{r}\right) }{%
\partial r}+\frac{1}{r}\frac{\partial A_{\theta}}{\partial\theta}+\frac{%
\partial A_{z}}{\partial z},
\end{equation*}

\begin{equation*}
\nabla \times\mathbf{A =}\left( \frac{1}{r}\frac{\partial A_{z}}{%
\partial\theta}-\frac{\partial A_{\theta}}{\partial z}\right) \mathbf{e}%
_{r}+\left( \frac{\partial A_{r}}{\partial z}-\frac{\partial A_{z}}{\partial
r}\right) \mathbf{e}_{\theta}+\left( \frac{1}{r}\frac{\partial \left(
rA_{\theta}\right) }{\partial r}-\frac{1}{r}\frac{\partial A_{r}}{%
\partial\theta}\right) \mathbf{e}_{z}
\end{equation*}%
\begin{equation*}
\Delta\mathbf{A =}\left( \Delta A_{r}-\frac{1}{r^{2}}A_{r}-\frac {2}{r^{2}}%
\frac{\partial A_{\theta}}{\partial\theta}\right) \mathbf{e}_{r}+\left(
\Delta A_{\theta}-\frac{1}{r^{2}}A_{\theta}+\frac{2}{r^{2}}\frac{\partial
A_{r}}{\partial\theta}\right) \mathbf{e}_{\theta}+\Delta A_{z}\mathbf{e}_{z}.
\end{equation*}

We now present the derivation of (\ref{lin-3dvla}) in detail. The linearized
Vlasov equation can be written as 
\begin{equation*}
\partial _{t}f+Df=(\mathbf{E}+\hat{v}\times \mathbf{B})\cdot \nabla
_{v}f^{0}.
\end{equation*}%
Since $f^{0}=\mu \left( e,p\right) $, we have 
\begin{equation*}
\nabla _{v}f^{0}=\mu _{e}\hat{v}+\mu _{p}r\mathbf{e}_{\theta }.
\end{equation*}%
So 
\begin{eqnarray*}
\mathbf{E}\cdot \nabla _{v}f^{0} &=&(-\nabla _{x}\phi -\partial _{t}\mathbf{A%
})\cdot (\mu _{e}\hat{v}+\mu _{p}r\mathbf{e}_{\theta }) \\
&=&-\mu _{e}\hat{v}\cdot \nabla _{x}\phi -\mu _{e}\hat{v}\cdot \partial _{t}%
\mathbf{A}-\mu _{p}r\partial _{t}A_{\theta }
\end{eqnarray*}%
Moreover, 
\begin{eqnarray*}
\hat{v}\times \mathbf{B}\cdot \nabla _{v}f^{0} &=&\{\hat{v}\times (\nabla
_{x}\times \mathbf{A})\}\cdot \{\mu _{e}\hat{v}+\mu _{p}r\mathbf{e}_{\theta
}\} \\
&=&r\mu _{p}\{\hat{v}\times (\nabla _{x}\times \mathbf{A})\}\cdot \mathbf{e}%
_{\theta } \\
&=&-\mu _{p}\left( \hat{v}_{r}\partial _{r}\left( rA_{\theta }\right) +\hat{v%
}_{z}\partial _{z}\left( rA_{\theta }\right) \right) =-\mu _{p}D(rA_{\theta
}).
\end{eqnarray*}%
The last line is a consequence of the identity 
\begin{eqnarray*}
&&\hat{v}\times (\nabla _{x}\times \mathbf{A})\cdot \mathbf{e}_{\theta } \\
&=&\left\{ \left( \hat{v}_{r}\mathbf{e}_{r}+\hat{v}_{\theta }\mathbf{e}%
_{\theta }+\hat{v}_{z}\mathbf{e}_{z}\right) \times \left( -\frac{\partial
A_{\theta }}{\partial z}\mathbf{e}_{r}+\left( \frac{\partial A_{r}}{\partial
z}-\frac{\partial A_{z}}{\partial r}\right) \mathbf{e}_{\theta }+\frac{1}{r}%
\frac{\partial \left( rA_{\theta }\right) }{\partial r}\mathbf{e}_{z}\right)
\right\} \cdot \mathbf{e}_{\theta } \\
&=&-\hat{v}_{r}\frac{1}{r}\frac{\partial \left( rA_{\theta }\right) }{%
\partial r}-\hat{v}_{z}\frac{\partial A_{\theta }}{\partial z} = \frac1r
D(rA_\theta).
\end{eqnarray*}%
Combining the above computations, we obtain (\ref{lin-3dvla}). {\footnote{%
Our work was supported in part by NSF grants DMS-0405066 and DMS-0505460.}}

\end{document}